%% file: main.tex
\pdfoutput=1
\documentclass[conference]{IEEEtran}
\IEEEoverridecommandlockouts
\usepackage{cite}
\usepackage{amsmath,amssymb,amsfonts}
\usepackage{algorithmic}
\usepackage{graphicx}
\usepackage{textcomp}
\usepackage{xcolor}
\usepackage{multirow}
\usepackage{subcaption}
\usepackage{cleveref}
\usepackage{listings}
\usepackage{float}
\usepackage{makecell}
\usepackage{hhline}
\usepackage{listings}
\usepackage{adjustbox}
\usepackage{balance}
\usepackage{float}
\def\BibTeX{{\rm B\kern-.05em{\sc i\kern-.025em b}\kern-.08em
    T\kern-.1667em\lower.7ex\hbox{E}\kern-.125emX}}

\begin{document}

\title{AFrame: Extending DataFrames for Large-Scale Modern Data Analysis (Extended Version)\\
}

\author{\IEEEauthorblockN{Phanwadee Sinthong}
\IEEEauthorblockA{\textit{Dept. of Computer Science}\\
\textit{University of California, Irvine}\\
psinthon@uci.edu}
\and
\IEEEauthorblockN{Michael J. Carey}
\IEEEauthorblockA{\textit{Dept. of Computer Science}\\
\textit{University of California, Irvine}\\
mjcarey@ics.uci.edu}
}

\maketitle

\begin{abstract}
Analyzing the increasingly large volumes of data that are available today, possibly including the application of custom machine learning models, requires the utilization of distributed frameworks. This can result in serious productivity issues for ``normal" data scientists. This paper introduces AFrame, a new scalable data analysis package powered by a Big Data management system that extends the data scientists' familiar DataFrame operations to efficiently operate on managed data at scale. AFrame is implemented as a layer on top of Apache AsterixDB, transparently scaling out the execution of DataFrame operations and machine learning model invocation through a parallel, shared-nothing big data management system. AFrame incrementally constructs SQL++ queries and leverages AsterixDB's semistructured data management facilities, user-defined function support, and live data ingestion support. In order to evaluate the proposed approach, this paper also introduces an extensible micro-benchmark for use in evaluating DataFrame performance in both single-node and distributed settings via a collection of representative analytic operations. This paper presents the architecture of AFrame, describes the underlying capabilities of AsterixDB that efficiently support modern data analytic operations, and utilizes the proposed benchmark to evaluate and compare the performance and support for large-scale data analyses provided by alternative DataFrame libraries.
\end{abstract}

\begin{IEEEkeywords}
DataFrames, large-scale data management, data science, benchmark
\end{IEEEkeywords}

\section{Introduction}
In this era of big data, extracting useful patterns and intelligence for improved decision-making is becoming a standard practice for many businesses. Modern data increasingly has three main characteristics: the first characteristic is that much of it is generated and available on social media platforms. The rapid growth in the numbers of mobile devices and smartphones, Facebook users, and YouTube channels all combine to create a data-rich social media landscape. Information distribution through this landscape reaches a massive audience. As a result, social media is now used as a medium for advertisement, communication, and even political discourse. 

The second characteristic of modern data is the rapid rate at which the data is continuously being generated. In order to accommodate the rate and frequency at which modern data arrives, distributed data storage and management are required. Storing such massive data in a traditional file system is no longer an ideal solution because analysis often requires a complete file scan to retrieve even a modest subset of the data. In order to minimize time-to-insight, analyses need to be performed in close to real-time on the ever-arriving data. Database management systems are able to store, manage, and utilize indexes and query optimization to efficiently retrieve subsets of their data, enabling interactive data manipulation.

The third characteristic of modern data is the richness of the information encapsulated in the data.  Modern data is not only massive in size but is also often nested and loosely-structured. For example, Twitter~\cite{twitter} provides JSON data containing information related to each message along with information about the user who posted that message and their location details if available. Other social media sites such as Facebook and Instagram provide similar information through their web services. As a result, modern data enables analyses that go beyond interpreting content; one can also analyze the structure and relationships of the data, such as identifying communities. Information extraction from modern data requires complex custom algorithms and analyses using machine learning. 

The growing interest in collecting, monitoring, and interpreting large volumes of modern data for business advantages motivates the development of data analytic tools. The requirements that modern, at-scale data analysis impose on analytic tools are not met by a single current system. Instead, data scientists are typically required to integrate and maintain several separate platforms, such as HDFS~\cite{hdfs}, Spark~\cite{spark}, and TensorFlow~\cite{tensorflow}, which then demands systems expertise from analysts who should instead be focusing on data modeling, selection of machine learning techniques, and data exploration.

In this paper, we focus on providing a `scale-independent' user experience when moving from a local exploratory data analysis environment to a large-scale distributed workflow. We present AFrame, an Apache AsterixDB~\cite{asterixdb} based extension of DataFrame. AFrame is a data exploration library that provides a Pandas-like DataFrame~\cite{pandasdataframe} experience on top of a big data management platform that can support large-scale semi-structured data exploration and analysis. AFrame differs from other DataFrame libraries by leveraging a complete big data management system and its query processing capabilities to efficiently scale DataFrame operations and optimize data access on large distributed datasets.

The second contribution of this paper is a distributed DataFrame benchmark for general data analytics. The performance of a big data system is greatly affected by the characteristics of its workload. Understanding these characteristics and being able to compare various systems' performance on a set of related analytic tasks will lead to more effective tool selection. Various benchmarks~\cite{graysort, ycsb, bigbench, smartmeter, tpcds} have been developed for big data framework assessment, but these benchmarks are either SQL-oriented benchmarks for OLTP or OLAP operations or focus on end-to-end application-level performance. To our knowledge, there is no standard Data-Frame benchmark yet for large-scale data analytic use cases.

In order to evaluate the performance of our framework, we have designed a micro-benchmark to compare various distributed DataFrame libraries' performance by issuing a set of common analytic operations. Our DataFrame benchmark provides a detailed comparison of each analytic operation by separating the data preparation time (e.g., DataFrame creation) and expression execution time to give better insight into each system's performance and operation overheads. 

The rest of this paper is organized as follows: Section 2 discusses background and related work. Section 3 provides an overview of the AFrame system architecture, user model, and data analytic support. In Section 4, we describe the proposed DataFrame benchmark. Section 5 details our initial experiments and discusses their results. We discuss future improvements and conclude the paper in Section 6. 

\section{Background}
An important motivation for the AFrame project comes from the need to make the management of large-scale modern data available to the larger audience of the data science community by integrating the DataFrame user experience with a big data management system. Here we discuss the foundations of exploratory data analysis and some advantages and disadvantages of its standard evaluation strategy.
\subsection{Exploratory Data Analysis (EDA)} 
Exploratory Data Analysis (EDA)~\cite{eda} is an investigation process employed by analysts to extract information, identify anomalies, discover insights, and understand the underlying structures and characteristics of a dataset. The goal of EDA is to provide analysts with clues and a better understanding of the data in order to formulate reasonable hypotheses. Important applications of EDA include, but are not limited to, data exploration, cleaning, manipulation, and visualization.

Frameworks and technologies often used in these applications span across the fields of statistics and machine learning. There are a large number of prepackaged machine learning libraries that cover a wide variety of user requirements. However, not all of the machine learning frameworks that work out of the box are designed to work in a distributed environment. As a result, analysts have to resort to large-scale machine learning frameworks such as MLlib~\cite{mllib} because extensive effort is required to make locally constructed models operate on big data. Often times, these large-scale machine learning frameworks do not cover all types of analysis and models. 

Since EDA involves visualizing data, collecting statistics from the data, and is iterative in nature, most of the available tools targeting these types of analyses only accommodate smaller datasets and leave big data processing and the scaling out of the algorithmic process for data engineers to implement. In order to reduce the turn-around time and increase productivity for data analysts, several issues need to be addressed: 1) distributed application of custom machine learning models; 2) providing a seamless migration from a local workflow to a distributed environment; 3) having a scalable system that can acquire and operate on ever-changing incoming data.

\subsection{Eager vs. Lazy Evaluation} EDA frameworks such as Pandas target a local workstation environment and often rely on in-memory processing. These frameworks require data to be loaded into memory before any analysis operations can be performed on the data. Once the data is loaded into memory, analysis operations are evaluated eagerly, meaning as soon as they are initiated.  However, a similar evaluation strategy is not efficient on large-scale ever-arriving data, as processing every declared operation without any optimization would be expensive as it may result in repetitive scans over massive data.
  
Eager and lazy evaluation are strategies used in programming languages to determine when expressions should be evaluated~\cite{programing}. While eager evaluation causes programs to evaluate expressions as soon as they are assigned, lazy evaluation is the opposite and delays their evaluation until their values are required. With eager evaluation, programmers are responsible for ensuring code optimization to prevent performance degradation due to unnecessary operations over large datasets. Lazy evaluation, on the other hand, delays execution until values are required; it is employed to help with operation optimizations where multiple operations can be chained together, extended, and a single iteration over the source collection can be processed, e.g., as in LINQ~\cite{linq}. As a result, lazy evaluation is more suitable for exploratory operations on large-scale data. Its performance improvement becomes critical as the size of the data grows.



\subsection{Related Platforms}
We can compare and contrast existing systems in terms of Big data platforms and DataFrame technology. 

\subsubsection{\textbf{Big Data Platforms}} Here we consider frameworks that can operate on distributed data.

\textbf{Apache Spark}: Apache Spark~\cite{apachespark} is a general-purpose cluster computing system that provides in-memory parallel computation on a cluster with scalability and fault tolerance. SparkSQL~\cite{sparksql} is a module to simplify users' interactions with structured data. SparkSQL integrates relational processing with Spark's functional programming. MLlib~\cite{mllib}, which is built on top of Spark, provides the capability of constructing and running machine learning models on large-scale datasets. However, Spark does not provide data management and it requires the installation and configuration tuning of a distributed file system like HDFS.

\normalsize\textbf{Hive}:
Apache Hive~\cite{hive} is data warehouse software built on top of Apache Hadoop for providing data summary, query, and analysis capabilities. The introduction of Hive reduced the complexity of having to write pure MapReduce programs by providing a SQL-like interface and translating the input queries into MapReduce programs to be executed on the Hadoop platform. Now Hive also includes Apache Tez~\cite{tez} and Apache Spark~\cite{spark} as alternative query runtimes. However, to leverage Hive's processing power, knowledge of SQL is essential in addition to being able to install and appropriately configure and manage Hadoop and HDFS. 

\textbf{Apache AsterixDB}:
Apache AsterixDB~\cite{asterixweb},~\cite{asterixdb} is a parallel open source Big Data Management System (BDMS) that provides full distributed data management for large-scale, semi-structured data. AsterixDB utilizes a NoSQL style data model (ADM) which is a superset of JSON. Before storing data into AsterixDB, a user can create a Datatype, which describes known aspects of the data being stored, and a Dataset, which is a collection of objects of a Datatype. Datatypes are ``open" by default, in that the description of the data does not need to be complete prior to storing it; additional fields are permitted at runtime. This allows for uninterrupted ingestion of data with ever-changing data schemas. AsterixDB provides SQL++~\cite{sql}, a highly expressive semi-structured query language for users that are familiar with SQL, to explore stored NoSQL data.

Figure~\ref{fig:asterix_data} shows an example of creating an open datatype `Tweet' with only the field `id' being pre-defined and two datasets called `TrainingData' and `LiveTweets' which store records of this Tweet datatype. The TrainingData dataset is populated by reading data from a local file system. In this example, it is being populated using a labeled airline sentiment dataset. AsterixDB also provides support for user-defined functions (UDFs) and built-in live social media data acquisition through its data feed feature. The LiveTweets dataset is populated by connecting a data feed called `TwitterFeed' that continuously ingests Twitter data. (More details on how to create a live Twitter feed can be found in~\cite{asterixweb},~\cite{deem}). Figure~\ref{fig:asterix_data} also creates two indexes on the LiveTweets dataset.
\vspace{-1em}
\begin{figure}[h]
\footnotesize
\begin{lstlisting}[
           language=SQL,
           basicstyle=\ttfamily,
           showstringspaces=false,
           morekeywords={FEED, with, START, to, type, dataset, load},
           commentstyle=\color{gray},
           keywordstyle=\color{blue},
           frame=single
        ]
CREATE TYPE Tweet AS{id: int64};
CREATE DATASET TrainingData(Tweet);
CREATE DATASET LiveTweets(Tweet);
LOAD DATASET TrainingData USING localfs
 (("path"="1.1.1.1:///airline_data.json"),
  ("format"="adm"));
  
CREATE  FEED  TwitterFeed  WITH {...};
CONNECT  FEED  TwitterFeed  TO  LiveTweets;
START  FEED  TwitterFeed; 

CREATE PRIMARY INDEX ON LiveTweets;
CREATE INDEX coordIdx ON LiveTweets(coordinate); 

\end{lstlisting}
\vspace{-1em}
\caption{SQL++ queries}
\label{fig:asterix_data}
\vspace{-1.5em}
\end{figure}

\subsubsection{\textbf{DataFrames for Data Science}} Here we consider libraries that provide a DataFrame facility.

\textbf{Pandas}:
Pandas~\cite{pandas} is an open source data analysis tool that provides an easy-to-use data structure built specifically to support data wrangling in Python. Pandas reads data from various file formats (e.g., CSV, SQL databases, and Parquet) and creates a Python object, DataFrame, with rows and columns similar to a table in Excel. Pandas can be integrated with scientific visualization tools such as Jupyter notebooks~\cite{jupyter}; Jupyter notebooks provide a unified interface for organizing, executing code and visualizing results without referring to low-level systems' details. The rich set of features that are available in Pandas makes it one of today's most popular tools for data exploration. However, its limitation lies in scalability. Pandas does not provide either data storage or support for interacting with distributed data, as its focus has been on in-memory computation on a single node. Another well-known Pandas' limitation is its memory consumption. This is caused by the underlying internal memory requirements about which the Pandas creator, McKinney, advised: ``you should have 5 to 10 times as much RAM as the size of your dataset"~\cite{mckinney}.

\textbf{R Data Frames}:
R~\cite{r} is a language originally built for statistical computing and graphics. Since R is primarily used for statistical analysis, R has become one of the most popular languages in the data science community. R also provides Data Frame as a built-in native data structure, but working with data larger than memory in R still requires a distributed framework and data storage setup. For example, SparkR~\cite{sparkr} is an R package created by Apache Spark that supports distributed operations like R Data Frames but on large datasets.

\textbf{Spark DataFrames}:
Spark also provides a DataFrame API~\cite{sparkdf} to enable the wider audience of the data science community to leverage distributed data processing. This API is designed to support large-scale data science applications with inspirations from both the R DataFrame and Python Pandas. Spark employs the lazy evaluation technique to perform computations only when values are required. This is different from the eager evaluation strategies used in Python and R. Lazy evaluation is exploited by Spark's query optimizer, which understands the structure of the data and the operations. In order for Spark to determine the input data schema for unstructured data, a process called `schema inference' is required and can result in long wait times for data that does not fit in memory. 


\textbf{Pandas on Ray}:
Pandas on Ray, which recently become a part of the Modin project~\cite{modin}, is a recent attempt to make Pandas DataFrames work on big data by providing the Pandas syntax and transparently distributing the data and operations using Ray~\cite{ray}. Ray uses shared memory and employs a distributed scheduler to manage a system's resources. Pandas on Ray automatically utilizes all available cores on a machine or a cluster to execute operations in parallel. Since the Ray framework handles large data through shared memory, it requires a cluster with sufficient aggregate memory to hold the entire dataset. In addition, Pandas on Ray uses Pandas as a black box at its core, which does not address the high memory consumption issue of Pandas.

\section{AFrame System Architecture}
Exploratory tools such as Pandas work well against locally stored data that fits in the memory of a single machine, but this is not a solution for large-scale analysis. Still, Pandas is one of the most widely used libraries for data exploration due to the analyst-friendly characteristics of its data structure. As a result, we set out to integrate a Pandas-like user experience with big data management capabilities to provide analysts with a familiar environment while scaling out their analytic operations over a large data cluster to enable big data management and analysis. 

Our goal in the AFrame project is to create a unified system that can efficiently support all of the various stages~\cite{tdsp} in data science projects, from data understanding to model deployment and application, thus enabling very large-scale analysis and requiring little or no modification to analysts' existing local workflows. Instead of building such a system from scratch, we extend Apache AsterixDB with support for the use of machine learning libraries and with interactive data exploration capabilities. Here we describe the underlying architecture of AFrame, the relevant AsterixDB features, and illustrate AFrame's basic functionality through a small running example that shows how to perform a simple sentiment analysis on ever-growing Twitter data.

\subsection{Acquiring Data}

\begin{figure}[h]
    \includegraphics[width=8.5cm, height=1.5cm]{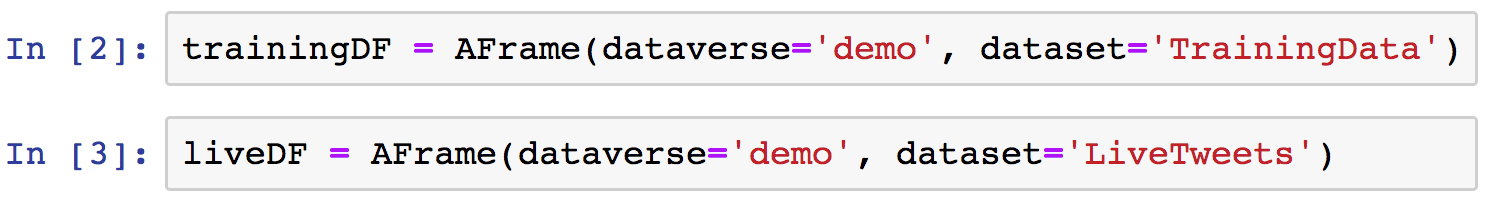}
    \caption{Initializing AFrame Objects}
    \label{fig:af_live_cnt}
    \vspace{-1em}
\end{figure}

AFrame is an API that provides a DataFrame syntax to interact with AsterixDB's datasets; it targets data scientists who are already familiar with Pandas DataFrames. AFrame works on distributed data by connecting to AsterixDB's webservice using its RESTful API. Figure~\ref{fig:af_live_cnt} shows how users can use AFrame in a Jupyter notebook to access datasets stored in AsterixDB. Input 2 (labeled ``In [2]") creates an AFrame object (trainingDF) from the TrainingData dataset initialized via the SQL++ statements in Figure~\ref{fig:asterix_data}. Input 3 creates another AFrame object (liveDF) from the LiveTweets dataset, which is connected to a data feed that continuously ingests data from Twitter. Building on top of AsterixDB allows AFrame to operate on such live data the same way as it does on a static dataset without requiring additional knowledge about how to setup a streaming engine. Since Figure~\ref{fig:asterix_data} created indexes on the LiveTweets dataset, the incoming data is also appropriately stored and indexed for efficient data access.
\vspace{-1em}

\subsection{Operating on Data}
%
As most EDA tools are designed to work with in-memory data, the eager evaluation strategy can suffice even when a session involves multiple scans over the entire dataset. However, multiple scans over a large distributed dataset would be very costly and have a negative effect on system performance.

\begin{figure}[h]
    \includegraphics[width=8.5cm, height=6.5cm]{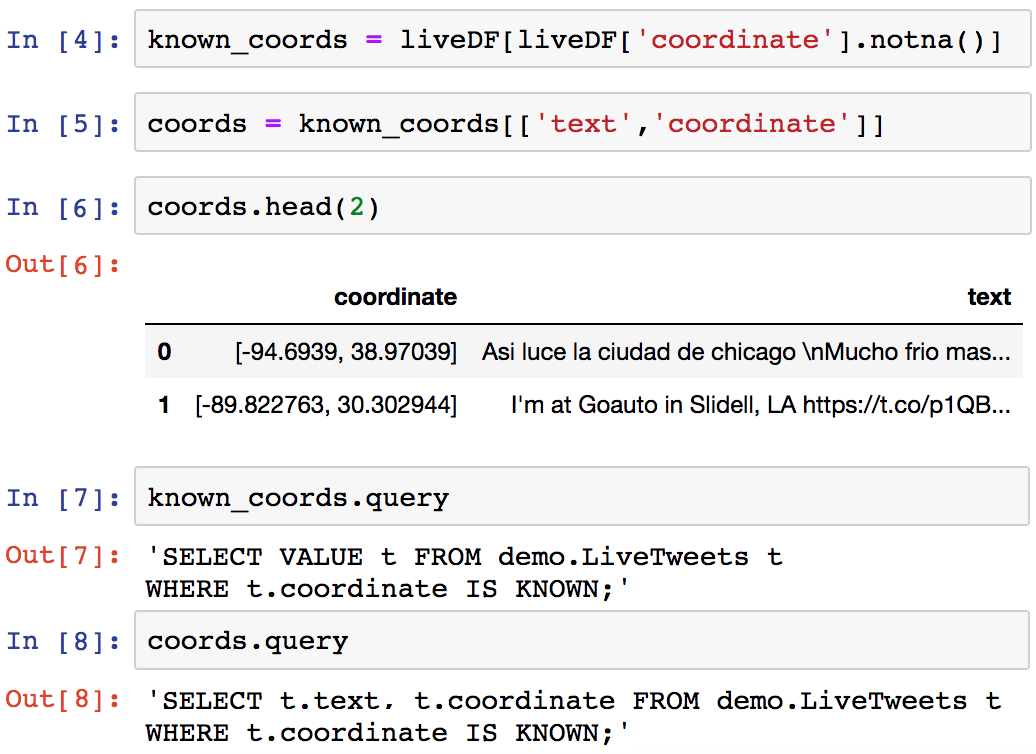}
    \caption{DataFrame expressions and underlying queries}
    \label{fig:af_queries}
    \vspace{-1em}
\end{figure}

AFrame leverages lazy evaluation. AFrame operations are incrementally translated into SQL++ queries that are sent to AsterixDB (via its RESTful API) only when final results are called for. Figure~\ref{fig:af_queries} shows an example of some expressions in AFrame when issuing Pandas-like DataFrame expressions. Input 4 (labeled In [4]) issues a selection predicate on the live dataset declared in Figure~\ref{fig:af_live_cnt}. Input 5 performs attribute projections. Neither inputs 4 or 5 trigger query evaluation; they only modify an underlying AFrame query. Input 6 performs an action that requests the actual output of two records, so AFrame takes the underlying query, appends a `LIMIT 2' clause to it, sends it to AsterixDB for evaluation, and displays the requested data. For debugging purposes, AFrame allows users to observe the underlying query resulting from the incremental query formation process. Input 7 prints the underlying query resulting from Input 4. Input 8 prints the underlying query of Input 5 (which adds projected attributes to the selection query). These are examples of queries that correspond to simple DataFrame operations. However, even complex DataFrame expressions that result in nested SQL++ queries are efficiently translated into optimized query plans in order to minimize data access. This is another benefit of operating on AsterixDB and utilizing its query optimizer.

In its early development stage, AFrame today covers essential Pandas’ operations for exploratory analyses that are suitable for large-scale unordered data. Currently, AFrame's supported operations include column selection and projection, statistical operations (e.g., describe), arithmetic operations (e.g., addition, subtraction, etc.), applying functions (both elementwise and tablewise), joining, categorizing data (sorting and ordering), grouping (group by and aggregation), and persisting data.
\vspace{-1em}

\subsection{Support for Machine Learning Models}
Following the data wrangling and hypothesis forming process, distributed systems are often required to accommodate the development and usage of customized machine learning models. The goal of the modeling step is to create an effective machine learning model that can make accurate predictions. With AFrame, analysts can apply either a prepackaged model or create a custom machine learning model from their local environment that can be applied to a distributed dataset directly from within a Jupyter notebook.

\begin{figure}
    \includegraphics[trim={0 1cm 0 0},width=8.2cm, height=2.5cm]{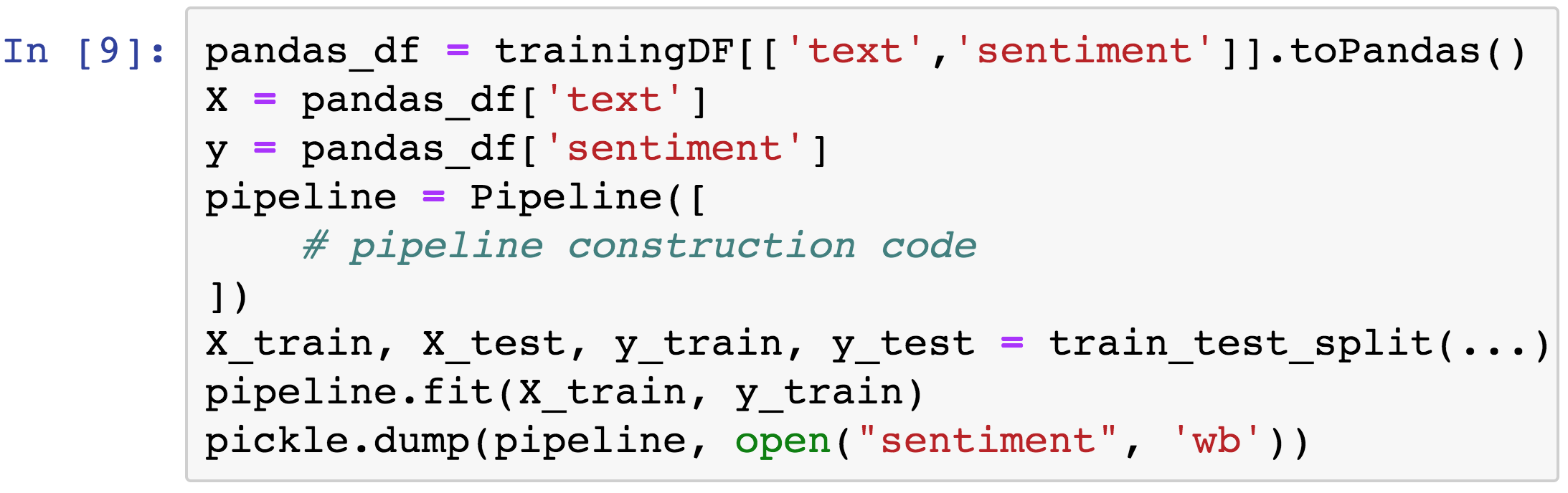}
    \caption{Training a Scikit-Learn Pipeline}
    \label{fig:ml_train}
    \vspace{-1em}
\end{figure}

Figure~\ref{fig:ml_train} illustrates a sentiment classifier training session using Python, Scikit-Learn~\cite{scikit}, Pandas, and AFrame. It trains a classifier on the training dataset from Figure~\ref{fig:af_live_cnt}. This is a dataset, publicly available on Kaggle~\cite{kaggle}, containing Twitter posts related to users' experiences with U.S. airlines released by CrowdFlower~\cite{crowdflower}. The dataset contains labeled tweet sentiments which are positive, negative, and neutral. The first step in Figure~\ref{fig:ml_train} selects a subset of attributes from the training dataset. Since the subsetted training data is small enough to fit in a single node's memory\footnote{Scikit-Learn's model training is required to take place on a single-node, but we are then able to utilize its trained models in a distributed setting.}, here we convert it to a Pandas DataFrame and use it to build and train a Scikit-Learn pipeline to classify sentiment values. The last step after training the model saves it as an executable which can then be dropped into AsterixDB and utilized as a UDF.

\begin{figure}[h]
    \includegraphics[trim={0 0.5cm 0 0},width=8.3cm, height=6.5cm]{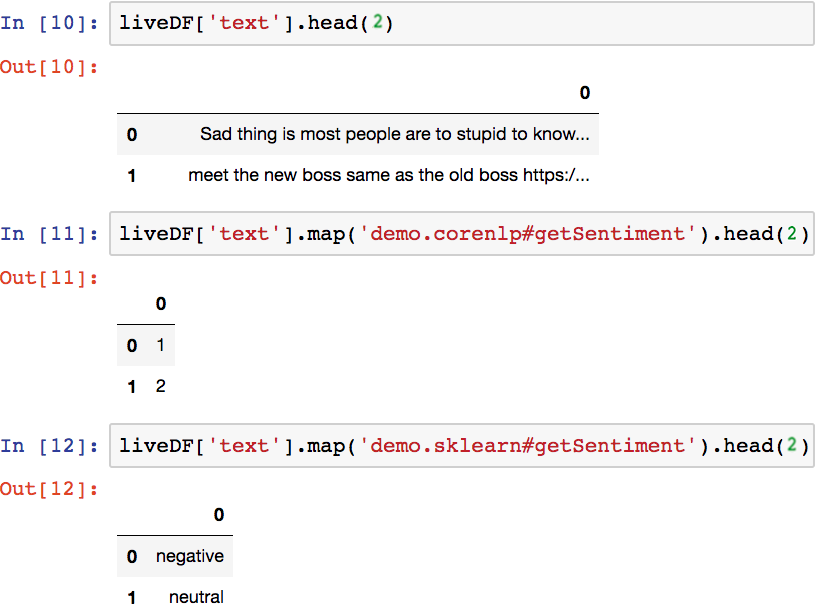}
    \caption{Applying CoreNLP and Scikit-Learn models}
    \label{fig:udfs}
\vspace{-1.5em}
\end{figure}

In Figure~\ref{fig:udfs}, we show sample code for applying machine learning models in AFrame. We first apply a pre-trained model (from Stanford CoreNLP) and then apply our custom Scikit-Learn sentiment analysis model (created in Figure~\ref{fig:ml_train}) using the Pandas-style map function syntax on the `text' column to get sentiment value predictions. Input 10 in the figure displays a sample of the text column from the liveDF dataset created in Figure~\ref{fig:af_live_cnt}. Input 11 applies the pre-trained Stanford CoreNLP sentiment analysis model~\cite{snlp} to the text column and displays two records. The CoreNLP sentiment annotator produces 5 sentiment classes ranging from very negative to very positive (0-4). Input 12 applies our custom Scikit-Learn sentiment analysis model to the same data. 

Under the hood, AFrame utilizes AsterixDB's UDF framework to enable users to import and then apply their own machine learning models written in popular programming languages (e.g., Java and Python) as functions. 
\vspace{-0.5em}

\subsection{Result Persistence}
\begin{figure}[h]
    \includegraphics[width=8.5cm,
    height=4.8cm]{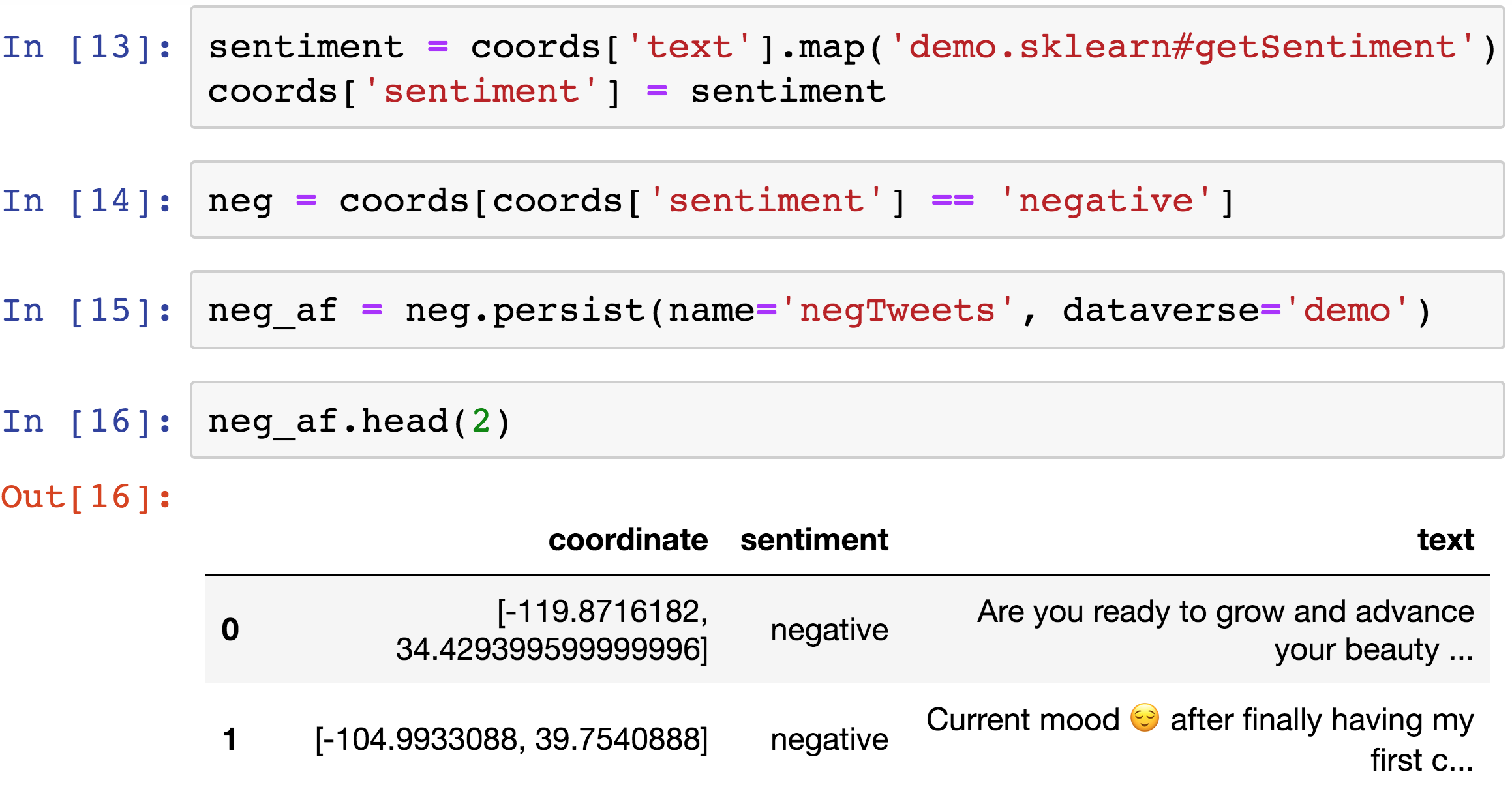}
    \caption{Persist Sentiment Analysis Results}
    \label{fig:persist}
    \vspace{-1em}
\end{figure}

After constructing a model, the next step would be to deploy the model and to apply it on real data. Input 13 in Figure~\ref{fig:persist} shows an example of how to apply the previously-constructed Scikit-Learn sentiment function to the `text' field of a queried subset (coords) of the live Twitter records resulting from the operations in Figure~\ref{fig:af_queries}. It then saves the sentiment prediction as a new field called `sentiment'. Input 14 selects only records with negative sentiment for future root cause analysis. In AFrame, the result of an AFrame operation can optionally be persisted as another dataset by issuing the {\lq{persist}\rq} command and providing a new dataset name, as shown by Input 15 in Figure~\ref{fig:persist}. Persisting an analysis result is efficient here, as the data has never left AsterixDB storage and the new dataset (demo.negTweets) can be accessed right away without having to wait for data re-loading or a file scan. Input 16 displays sampled records from the new dataset created using AFrame; their sentiment is negative and they only contain a subset of the attributes from the original dataset.
 \vspace{-1em}

\subsection{Summary}
We have demonstrated through an example how to use AFrame to acquire live Twitter data, manipulate the data, train and apply a custom Scikit-Learn model to get sentiments from the data, and save an analysis result for further investigation. AFrame provides a Pandas-like user experience without suffering from Pandas' single-node and in-memory requirements. AFrame does not load all data from a file or store its intermediate analysis results in memory. It can utilize database features to efficiently retrieve data and accelerate data manipulation on large-scale distributed data. By offloading data management to a distributed database system, AFrame remains a lightweight library that provides a scale-independent user experience to data scientists with any level of expertise.

\section{A DataFrame Benchmark}
In order to evaluate our AFrame implementation and compare its performance to that of other distributed DataFrame libraries, we have constructed a preliminary DataFrame benchmark. Inspired by the early Wisconsin Benchmark~\cite{wisconsin} from the relational world, we propose a benchmark that evaluates DataFrames in several key dimensions that are important to conducting large-scale data analyses. This is similar to how the Wisconsin Benchmark was used to assess early relational database system performance. We also aim to provide members of the data science community with a tool to help them select a framework that is best suited to their project. 

Our DataFrame Benchmark is designed to evaluate the performance of DataFrame libraries against data of various sizes in both local and distributed environments. As an initial set of evaluated systems, we selected the following DataFrame frameworks: Pandas, PySpark, Pandas on Ray (Modin), and AFrame. There are several factors that contributed to our framework selection. First, since our goal is to support DataFrame syntax on large-scale data, it is appropriate to compare how systems perform with regard to the original Pandas DataFrames in a single node environment. Second, Apache Spark is a popular framework for distributed processing of large-scale data, so comparing against Spark DataFrames gives us a good understanding and comparison to a commercial and well-maintained DataFrame project. Pandas on Ray is another project that is trying to solve the same data scientists' problem, but using a different approach, so we also include it in our initial set of platforms. (At the time of writing this paper, Pandas on Ray had not yet published its cluster installation details, so we have only run it on a single node.) 

\subsection{Benchmark Datasets}
In order to discover useful information from large volumes of modern data, most data science projects rely on data exploration. DataFrames are one of the most popular data structures used in data exploration and manipulation. A mature DataFrames library must be able to handle exploratory data manipulation operations on large volumes of data efficiently. The design of our DataFrame micro benchmark aims at reflecting these expectations in its workload.

\begin{figure}[h]
    \vspace{-0.5em}
    \includegraphics[width=8cm, height=6cm]{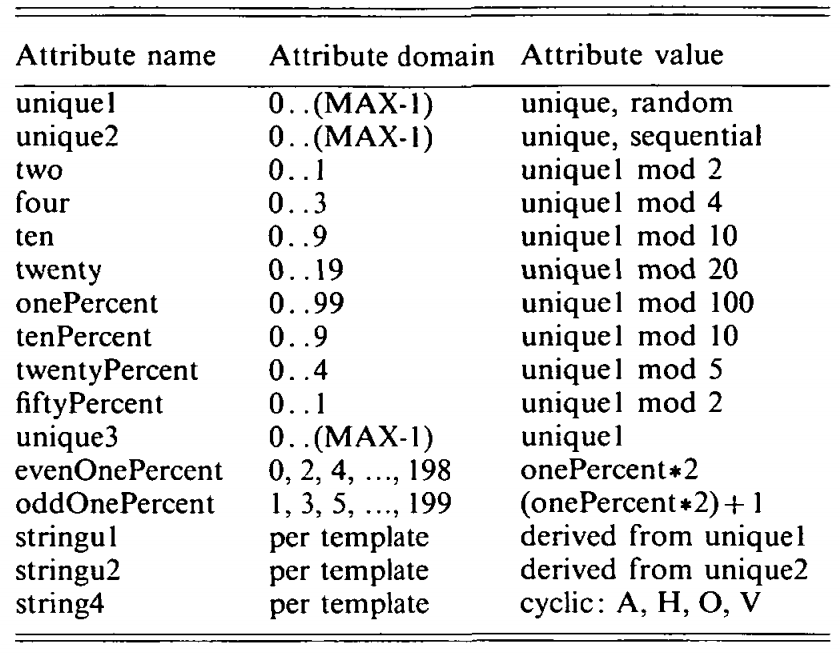}
    \caption{Scalable Wisconsin benchmark: attributes~\cite{wisconsin}}
    \label{fig:wisconsin_benchmark}
    \vspace{-1em}
\end{figure}

For our benchmark datasets, we have chosen to use a synthetically generated Wisconsin benchmark dataset instead of using data from social media sites to allow us to precisely control the selectivity percentages, to generate data with uniform value distributions, and to broadly represent data for general analysis use cases (not just social media). A specification of the attributes in the Wisconsin benchmark's dataset is displayed in Figure~\ref{fig:wisconsin_benchmark}. The unique2 attribute is a declared key and is ordered sequentially, while the unique1 attribute has 0 to (cardinality - 1) unique values that are randomly distributed.  The two, four, ten and twenty attributes have a random ordering of values which are derived by an appropriate mod of the unique1 values. The onePercent, tenPercent, twentyPercent, and fiftyPercent attributes are used to provide access to a known percentage of values in the dataset. The dataset also contains three string attributes: stringu1, stringu2, and string4. The stringu1 and stringu2 attributes derive their values from the unique1 and unique2 values respectively. The string 4 attribute takes on one of four unique values in a cyclic fashion; its unique values are constructed by forcing the first four positions of a string to have the same value chosen from a set of four letters: [A, H, O, V].

For our DataFrame benchmark, we used a JSON data generator to generate Wisconsin datasets of various sizes ranging from 1 GB (0.5 million records) to 40 GB (20 million records). In addition to JSON, we also evaluate systems using other widely used input formats, namely Parquet~\cite{parquet} and CSV.

\subsection{Benchmark Queries}
The essential characteristic that makes DataFrame an appealing choice for data scientists is its stepwise syntax for exploratory tasks and data manipulation. As a result, we have designed our benchmark queries to target a set of core exploratory operations and visualization tasks. Table~\ref{tab:operations} summarizes the details of our initial DataFrame benchmark expressions. All evaluated frameworks except Pandas on Ray provide support for all of our benchmark expressions. Pandas on Ray defaults back to Pandas if the given expression has not yet been implemented to take advantage of its parallel processing engine. (In our case, these expressions are expressions 4, 8, and 12 in Table~\ref{tab:operations}.) Our initial set of expressions consist of analysis operations that include selection, projection, grouping, sorting, aggregation, and join. For expressions 2, 5, 9, and 10, we only asked for sampling because loading the entire dataset into memory would not be desirable in an exploratory big data context. For the join expression, both datasets are of the same size with the same number of records ranging from 1 GB to 40 GB. When executing the benchmark, each expression is run 15 times, and the first five results were excluded from the calculation to account for any JVM warm-up overheads. The recorded results are averaged over 10 runs. Our DataFrame benchmark expressions are detailed in Table~\ref{tab:operations}. We randomly generated values for the expression predicates (e.g., df[`ten'] == \$x ) that fall within the tested attributes' range to reduce the effect of any in-memory caching between runs.

\lstdefinestyle{pythonNoFramestyle}
      {  language = python,
        showstringspaces=false,
        basicstyle=\ttfamily,
        keywordstyle=\color{blue},
        otherkeywords = {min,max,groupby,head,sort_values,map,agg,len,merge,read_json,read.json},
        morekeywords = [2]{min},
        morekeywords = [3]{groupby},
        morekeywords = [4]{head}
    }

\begin{table*}[h]

\begin{adjustbox}{width=\textwidth}
\begin{tabular}{|l|l|l|l|}
\hline
\textbf{ID} & \textbf{Operation} & \textbf{Description} & \textbf{DataFrame Expression} \\ \hline
1 & Total Count & Total count   &
\begin{lstlisting}[style=pythonNoFramestyle]
len(df)
\end{lstlisting} \\ \hline
2 & Project  & Project records on attributes two and four   &  
\begin{lstlisting}[style=pythonNoFramestyle]
df[[`two',`four']].head()
\end{lstlisting} \\ \hline
3 & Filter \& Count    & Count records that satisfy column conditions &  
\begin{lstlisting}[style=pythonNoFramestyle]
len(df[(df[`ten'] == x) & (df[`twentyPercent'] == y) & 
 (df[`two'] == z)])
\end{lstlisting}\\ \hline
4 & Group By& Count records with the same column value   &  
\begin{lstlisting}[style=pythonNoFramestyle]
df.groupby(`oddOnePercent').agg(`count')
\end{lstlisting} \\ \hline
5 & Map Function  & Apply a function to a column   &  
\begin{lstlisting}[style=pythonNoFramestyle]
df[`stringu1'].map(str.upper).head()
\end{lstlisting} \\ \hline
6 & Max     & Retrieve a max column value   &  
\begin{lstlisting}[style=pythonNoFramestyle]
df[`unique1'].max()
\end{lstlisting} \\ \hline
7 & Min     & Retrieve a min column value   &  
\begin{lstlisting}[style=pythonNoFramestyle]
df[`unique1'].min()
\end{lstlisting} \\ \hline
8 & Group By \& Max       & Retrieve the max column value for each group   &  
\begin{lstlisting}[style=pythonNoFramestyle]
df.groupby(`twenty')[`four'].agg(`max')
\end{lstlisting} \\ \hline
9 & Sort    & Order records based on a column   &  
\begin{lstlisting}[style=pythonNoFramestyle]
df.sort_values(`unique1', ascending=False).head()
\end{lstlisting} \\ \hline
10& Selection     & Retrieve some records that satisfy column conditions   &  
\begin{lstlisting}[style=pythonNoFramestyle]
df[(df[`ten'] == x)].head()
\end{lstlisting} \\ \hline
11& Range Selection     & Count records in a selected range   &  
\begin{lstlisting}[style=pythonNoFramestyle]
len(df[(df[`onePercent'] >= x) & (df[`onePercent'] <= y)])
\end{lstlisting} \\ \hline
12& Join \& Count    & Count records resulting from an inner join   &  
\begin{lstlisting}[style=pythonNoFramestyle]
len(pd.merge(df, df2, left_on=`unique1',
            right_on=`unique1',how=`inner'))
\end{lstlisting} \\ \hline
\end{tabular}
\end{adjustbox}
\caption{Benchmark Operations (df, df2 = DataFrame objects, x,y,z = variables representing random values within range)}
\label{tab:operations}
\vspace{-1em}
\end{table*}

\subsection{Evaluated System Details}
The details of each systems' setup are provided below.


\textbf{Pandas}: Pandas DataFrame only works on a single machine environment and on data that fits in memory. It is important to note that Pandas only utilizes a single core for processing and that we use it with its default settings (without any additional configuration). It is labeled ``Pandas" in the experimental results presented in this paper.

\textbf{Spark}: Spark indicates in its DataFrame API document that there is a significant difference in its DataFrame creation time when reading from JSON files if a data schema is provided. This performance benefit comes from eliminating its initial schema inference step. As a result, a dataset schema was also included in our benchmark. For single node experiments, we used Spark in its local standalone operating mode. In the distributed environment, we configured HDFS as its distributed storage and used its standalone cluster manager. We evaluated Spark's DataFrame on both JSON and Parquet data using the default setup configurations. The three evaluated Spark variations are labeled ``Spark JSON", ``Spark JSON Schema", and ``Spark Parquet" in the experimental results section.

\textbf{AFrame}: In order to evaluate AFrame, the benchmark datasets are expected to be resident in AsterixDB (as opposed, e.g., to HDFS) when running the operations. Similar to the Wisconsin benchmark queries, some of the expressions can benefit from indexes, so we executed the queries on both indexed and non-indexed data. Also, even though AsterixDB's default data typing is open, there is some benefit when a data schema is provided. Since we also provided Spark with a schema, we decided to also evaluate AFrame on a closed data type with the same pre-defined schema. AFrame's translated SQL++ queries for all benchmark expressions are presented in section~\ref{sec:translated_queries} of the appendix. The three evaluated AFrame variations are labeled ``AFrame", ``AFrame Schema", and ``AFrame Index" in the experiments presented here.

\textbf{Pandas on Ray}: When we began evaluating the systems, Pandas on Ray had not yet provided cluster installation instructions, so we executed the DataFrame benchmark only on its single node setup. Notably, Pandas on Ray has implemented an impressive number of Pandas' operations to utilize all of the available cores in the given system. (For functions that have not been parallelized, it defaults back to using the original Pandas' operations.) When we did a preliminary run of the benchmark to check supported expressions, we noticed that Pandas on Ray had not yet parallelized Pandas' load\_json method, so we decided to evaluate Pandas on Ray using CSV files instead. Pandas on Ray is based on a shared, in-memory architecture; its strength lies in in-memory computation. However, it is worth mentioning that the project has started to implement support for large datasets using disk as an overflow for in-memory DataFrames.

\subsection{Experimental Setup}
Our DataFrame benchmark provides a set of configurable parameters to enable both single-node and cluster performance evaluations. The same suite of benchmark queries were applied to both settings. Each evaluated framework handles DataFrame creation differently, and some utilize an eager evaluation strategy while the others employ lazy evaluation. On top of that, depending on the flow of an analysis session, data might or might not already be available in memory, resulting in additional time to create a DataFrame before issuing analytic operations. Sometimes, when only a small subset of the data is needed, DataFrame creation time can dominate the overall actual operation time. As a result, we separately consider expression-only run times and total run times (which include both the DataFrame creation time and the DataFrame expression execution time). Each system's timing point is provided in section~\ref{sec:timingAppendix} of the appendix.

In order to provide a reproducible environment for evaluating these systems, we set all of the evaluated systems up and executed our benchmark on Amazon EC2 instances. For each node, we selected the m4.large instance type with the Linux 16.04 operating system, 2 cores, 8 GB of memory, and 100 GB of SSD.

\subsubsection{\textbf{Single-Node Setup}}
We generated the Wisconsin benchmark as JSON data in various sizes ranging from 1 GB (0.5 million records) to 10 GB (5 millions records). The Parquet and CSV datasets were created by converting the JSON files; they contained the exact same logical records as the JSON datasets. Table~\ref{tab:onenode} shows the numbers of records and the byte sizes of each dataset for all file formats. The sizes of the Parquet files are significantly smaller due to its compression and its internal data representation. The JSON structure is based on key-value pairs. Each JSON record contains all of the necessary information about its content, and in principle each record could contain different fields in different orders. CSV is more compact than JSON due to the facts that its schema is only declared once for the whole file and that each record has an identical list of fields in the exact same order. Parquet is a column-oriented binary file that contains metadata about its content. Parquet is the most compact file format among the three formats tested.


\begin{table}[h]
\resizebox{0.48\textwidth}{!}{%
\begin{tabular}{l|l|l|l|l|l|}
\cline{2-6}
 & \multicolumn{5}{c|}{\textbf{Dataset Name}} \\ \cline{2-6} 
 & \multicolumn{1}{c|}{\textbf{XS}} & \multicolumn{1}{c|}{\textbf{S}} & \multicolumn{1}{c|}{\textbf{M}} & \multicolumn{1}{c|}{\textbf{L}} & \multicolumn{1}{c|}{\textbf{XL}} \\ \hline
\multicolumn{1}{|l|}{Number of Records} & 0.5 mil  & 1.25 mil  & 2.5 mil  & 3.75 mil & 5 mil \\ \hline
\multicolumn{1}{|l|}{JSON File Size} & 1 GB  & 2.5 GB  & 5 GB  & 7.5 GB & 10 GB \\ \hline
\multicolumn{1}{|l|}{Parquet File Size} & 43 MB & 110 MB  & 217 MB & 317 MB & 426 MB \\ \hline
\multicolumn{1}{|l|}{CSV File Size} & 715 MB & 2.3 GB  & 4.6 GB & 6.8 GB  & 9.3 GB \\ \hline
\end{tabular}%
}
\caption{Dataset Summary (mil = million)}
\label{tab:onenode}
\vspace{-1em}
\end{table}

\subsubsection{\textbf{Multi-Node Setup}}
For the multi-node setting, we only evaluated Spark and AFrame. The evaluated cluster size ranged from 2-4 nodes, where each node is a worker except for one node that is also a master. Speedup and scaleup are the two preferred and widely used metrics to evaluate the processing performance of distributed systems, so we evaluated the two systems using these two metrics.

\textbf{Speedup Experiment}: Ideal speedup is when increasing resources by a certain factor to operate on a fixed amount of data results in the overall task processing time being reduced by the same factor. As a result, speedup reduces the response time, which also makes resources available sooner for other tasks. Linear speedup is not always achievable due to reasons such as start up cost and system interference between parallel processes accessing shared resources. 

For our DataFrame benchmark, we conducted speedup experiments using a fixed size dataset while increasing the number of machines from one up to four. The details are summarized in Table~\ref{tab:speedup}, where aggregate memory is the sum of all of the available memory in the cluster.

\begin{table}[h]
\resizebox{0.48\textwidth}{!}{%
\begin{tabular}{l|l|l|l|l|}
\cline{2-5}
 & \textbf{1 node} & \textbf{2 nodes} & \textbf{3 nodes} & \textbf{4 nodes} \\ \hline
\multicolumn{1}{|l|}{\textbf{Aggregate Memory}} & 8 GB & 16 GB & 24 GB & 32 GB \\ \hline
\multicolumn{1}{|l|}{\textbf{JSON File Size}} & 10 GB & 10 GB & 10GB & 10 GB \\ \hline
\multicolumn{1}{|l|}{\textbf{Parquet File Size}} & 426 MB & 426 MB & 426 MB & 426 MB \\ \hline
\end{tabular}%
}
\caption{Speedup Experiment Setup}
\label{tab:speedup}
\end{table}

\textbf{Scaleup Experiment}
Ideal scaleup is the system's ability to maintain the same response time when both the system resources and work (data) increase by the same factor.

For the scaleup experiments, we increased both the number of machines and the amount of data proportionally, as summarized in Table~\ref{tab:scaleup}, to measure each system's performance.

\begin{table}[h]
\resizebox{0.48\textwidth}{!}{%
\begin{tabular}{l|l|l|l|l|}
\cline{2-5}
 & \textbf{1 node} & \textbf{2 nodes} & \textbf{3 nodes} & \textbf{4 nodes} \\ \hline
\multicolumn{1}{|l|}{\textbf{Aggregated Memory}} & 8 GB & 16 GB & 24 GB & 32 GB \\ \hline
\multicolumn{1}{|l|}{\textbf{JSON File Size}} & 10 GB & 20 GB & 30GB & 40 GB \\ \hline
\multicolumn{1}{|l|}{\textbf{Parquet File Size}} & 426 MB & 818 MB & 1.33 GB & 1.75 GB \\ \hline
\end{tabular}%
}
\caption{Scaleup Experiment Setup}
\label{tab:scaleup}
\end{table}
\vspace{-1em}

\section{Initial Benchmark Results}
In this section, we present the initial experimental results from both the single-node and cluster environments.
\input{single_node.tex}

\input{single_node_images.tex}

\input{multinode_speedup.tex} 
\input{multinode_scaleup.tex}

\input{discussion.tex}
\input{conclusion.tex}

\balance
\bibliographystyle{abbrv}
\bibliography{references.bib}
\vspace{12pt}
\newpage
\appendix

\lstdefinestyle{pythonstyle}
{
    language = python,
    showstringspaces=false,
    basicstyle=\ttfamily,
    keywordstyle=\color{blue},
    commentstyle=\color{gray}\ttfamily,
    otherkeywords = {min, groupby,head,max,sort_values,map,agg,len, merge,count, read_json},
    morekeywords = [2]{min},
    morekeywords = [3]{groupby},
    morekeywords = [4]{head},
    frame=single
}

\lstdefinestyle{sqlstyle}
{
    language = sql,
    showstringspaces=false,
    basicstyle=\ttfamily,
    keywordstyle=\color{blue},
    otherkeywords = {value, count, group by, START, FEED, TO, WITH, TYPE, CLOSED, DATASET},
    morekeywords = [2]{value},
    morekeywords = [3]{group by},
    morekeywords = [4]{from},
    frame=single
}

\subsection{AsterixDB Twitter Feed}

\begin{figure}[h]
\footnotesize
\begin{lstlisting}[style=sqlstyle]
CREATE FEED TwitterFeed with {
   "adapter-name" : "push_twitter",
   "type-name" : "Tweet",
   "format" : "twitter-status",
   "consumer.key" : "***",
   "access.token" : "***",
   "access.token.secret" : "***"
 };
CONNECT FEED TwitterFeed TO LiveTweets;
START FEED TwitterFeed;
\end{lstlisting}
\caption{Create a Twitter feed to collect tweets.}
\label{fig:twitter_feed}
\end{figure}

\subsection{AsterixDB DDL}
\small
\begin{figure}[h]
\footnotesize
\begin{lstlisting}[style=sqlstyle]
CREATE TYPE ClosedType AS CLOSED{
    unique1: int64,
    unique2: int64,
    unique3: int64,
    two: int64,
    four: int64,
    ten: int64,
    twenty: int64,
    onePercent: int64,
    tenPercent: int64,
    twentyPercent: int64,
    fiftyPercent: int64,
    evenOnePercent: int64,
    oddOnePercent: int64,
    stringu1: string,
    stringu2: string,
    string4: string
};

CREATE TYPE OpenType AS {unique2: int64};

CREATE DATASET ClosedData(ClosedType)
PRIMARY KEY unique2
WITH {"storage-block-compression": 
    {"scheme": "snappy"}};

CREATE DATASET OpenData(OpenType)
PRIMARY KEY unique2
WITH {"storage-block-compression": 
    {"scheme": "snappy"}};
        
\end{lstlisting}
\caption{Create datatypes and datasets to use for benchmarking}
\label{fig:asterix_ddls}
\end{figure}
\vspace{-0.5cm}

\subsection{Benchmark Translated SQL++ Queries}
\label{sec:translated_queries}
\footnotesize
\begin{lstlisting}[style=sqlstyle]
1.  SELECT VALUE COUNT(*) FROM Data;
2.  SELECT t.two, t.four FROM Data t LIMIT 5;
3.  SELECT VALUE COUNT(*) FROM Data t 
        WHERE t.ten = x 
        AND t.twentyPercent = y 
        AND t.two = z;
4.  SELECT grp_id, COUNT(*) AS cnt 
        FROM Data t 
        GROUP BY t.oddOnePercent AS grp_id; 
5.  SELECT VALUE UPPER(t.stringu1) 
        FROM Data t LIMIT 5;
6.  SELECT MAX(t.unique1) FROM Data t;
7.  SELECT MIN(t.unique1) FROM Data t;
8.  SELECT grp_id, MAX(t.four) AS max 
        FROM Data t 
        GROUP BY t.twenty AS grp_id;
9.  SELECT VALUE t 
        FROM Data t 
        ORDER BY t.unique1 DESC LIMIT 5;
10. SELECT VALUE t 
        FROM Data t 
        WHERE t.ten = x LIMIT 5;
11. SELECT VALUE t 
    	FROM Data t 
    	WHERE t.onePercent >= x 
    	AND t.onePercent <= y;
12. SELECT VALUE COUNT(*) 
        FROM (SELECT l,r 
              FROM leftData l JOIN rightData r 
              ON l.unique1 = r.unique1) t;
\end{lstlisting}

\subsection{Benchmark Timing Points}
\label{sec:timingAppendix}
\begin{itemize}

\item\textbf{Pandas \& Pandas on Ray Timing}
\vspace{-0.5em}
\begin{lstlisting}[style=pythonstyle]
# DataFrame creation time
df = pd.read_json(file_path)
# Expression-only time
df.head()
\end{lstlisting}

\item\textbf{Spark Timing}
\begin{lstlisting}[style=pythonstyle]
# DataFrame creation time
df = sparkContext.read.json(file_path) 
# Expression-only time
df.head(5) 
\end{lstlisting}

\item\textbf{AFrame Timing}
\begin{lstlisting}[style=pythonstyle]
# DataFrame creation time
df = AFrame(dataverse, dataset) 
# Expression-only time
df.head() 
\end{lstlisting}

\end{itemize}

\begin{table*}[ht]
\resizebox{\textwidth}{!}{%
\begin{tabular}{|l||l|l|l|l|l|l|l|l||l|l|l|l|l|l|l|l|}
\hline
\multirow{2}{*}{\begin{tabular}[c]{@{}l@{}}\textbf{Expression: \ Time}\end{tabular}} & \multicolumn{8}{c||}{\textbf{1GB}} & \multicolumn{8}{c|}{\textbf{2.5GB}} \\ \cline{2-17} 
 & \multicolumn{1}{c|}{\rotatebox[origin=c]{90}{\textbf{Pandas}}} & \multicolumn{1}{c|}{\rotatebox[origin=c]{90}{\textbf{Pandas on Ray}}} & \multicolumn{1}{c|}{\rotatebox[origin=c]{90}{\textbf{AFrame}}} & \multicolumn{1}{c|}{\rotatebox[origin=c]{90}{\textbf{AFrame Schema}}} & \multicolumn{1}{c|}{\rotatebox[origin=c]{90}{\textbf{AFrame Index}}} & \multicolumn{1}{c|}{\rotatebox[origin=c]{90}{\textbf{Spark JSON}}} & \multicolumn{1}{c|}{\rotatebox[origin=c]{90}{\textbf{Spark JSON Schema}}} & \multicolumn{1}{c||}{\rotatebox[origin=c]{90}{\textbf{Spark Parquet}}} & \multicolumn{1}{c|}{\rotatebox[origin=c]{90}{\textbf{Pandas}}} & \multicolumn{1}{c|}{\rotatebox[origin=c]{90}{\textbf{Pandas on Ray}}} & \multicolumn{1}{c|}{\rotatebox[origin=c]{90}{\textbf{AFrame}}} & \multicolumn{1}{c|}{\rotatebox[origin=c]{90}{\textbf{AFrame Schema}}} & \multicolumn{1}{c|}{\rotatebox[origin=c]{90}{\textbf{AFrame Index}}} & \multicolumn{1}{c|}{\rotatebox[origin=c]{90}{\textbf{Spark JSON}}}& \multicolumn{1}{c|}{\rotatebox[origin=c]{90}{\textbf{Spark JSON Schema}}} & \multicolumn{1}{c|}{\rotatebox[origin=c]{90}{\textbf{Spark Parquet}}} \\ \hhline{=================}
Exp. 1: Total & 23.55 & 12.19 & 1.828 & 0.452 & 0.092 & 22.09 & 7.52 & 3.97 & 41.46 & 27.999 & 2.884 & 2.628 & 0.193 & 25.44 & 9.11 & 4.08 \\ \hline
Exp. 2: Total & 23.560 & 13.61 & 0.036 & 0.029 & - & 17.790 & 3.227 & 3.908 & 41.473 & 29.521 & 0.039 & 0.031 & - & 19.042 & 3.011 & 4.024 \\ \hline
Exp. 3: Total & 23.635 & 17.551 & 2.686 & 1.004 & - & 23.216 & 8.375 & 4.056 & 41.556 & 35.028 & 3.673 & 3.113 & - & 25.707 & 9.494 & 4.186 \\ \hline
Exp. 4: Total & 23.579 & 16.652 & 3.45 & 2.548 & - & 23.751 & 10.630 & 4.747 & 41.494 & 33.015 & 6.697 & 4.226 & - & 27.443 & 11.578 & 4.864 \\ \hline
Exp. 5: Total & 24.132 & 16.052 & 0.022 & 0.02 & - & 17.777 & 3.208 & 3.907 & 42.876 & 35.688 & 0.023 & 0.022 & - & 19.026 & 2.995 & 4.058 \\ \hline
Exp. 6: Total & 23.561 & 13.204 & 3.048 & 2.304 & - & 22.985 & 8.208 & 3.952 & 41.474 & 29.419 & 3.924 & 3.469 & - & 25.144 & 8.95 & 4.063 \\ \hline
Exp. 7: Total & 23.561 & 13.201 & 3.027 & 1.319 & - & 22.98 & 8.258 & 3.953 & 41.474 & 29.44 & 3.893 & 3.446 & - & 25.087 & 8.952 & 4.067 \\ \hline
Exp. 8: Total & 23.568 & 16.636 & 6.167 & 2.674 & - & 24.355 & 9.185 & 4.518 & 41.489 & 32.844 & 6.495 & 4.311 & - & 25.95 & 11.189 & 4.685 \\ \hline
Exp. 9: Total & 25.763 & 24.651 & 2.968 & 1.247 & - & 27.699 & 12.571 & 5.175 & 43.944 & 56.086 & 3.848 & 3.318 & - & 32.986 & 15.152 & 6.421 \\ \hline
Exp. 10: Total & 23.626 & 15.718 & 0.025 & 0.023 & - & 17.88 & 3.321 & 4.128 & 41.554 & 32.338 & 0.029 & 0.025 & - & 19.136 & 3.104 & 4.268 \\ \hline
Exp. 11: Total & 23.572  & 13.907 & 1.677  & 0.625  & 0.12  & 20.278  & 5.514  & 4.081  & 41.518  & 32.125  & 4.131  & 3.549  & 0.141  & 24.77  & 8.517  & 4.165  \\ \hline
Exp. 12: Total & 38.899  & 60.367  & 3.406  & 3.004 & 3.149  & 23.230  & 9.428  & 4.543  & -  & -  & 9.716  & 8.749  & 9.213  & 43.437  & 16.997  & 5.069 \\ \hhline{=================}
Exp. 1: Expression-only & 3.34E-06 & 3.34E-06 & 1.828 & 0.452 & 0.195 & 4.365 & 4.355 & 0.121 & 3.45E-06 & 3.47E-06 & 2.884 & 2.628 & 0.206 & 6.474 & 6.175 & 0.124 \\ \hline
Exp. 2: Expression-only & 0.008 & 1.418 & 0.036 & 0.029 & - & 0.068 & 0.066 & 0.061 & 0.009 & 1.522 & 0.039 & 0.031 & - & 0.076 & 0.074 & 0.073 \\ \hline
Exp. 3: Expression-only & 0.083 & 5.359 & 2.686 & 1.004 & - & 5.494 & 5.214 & 0.209 & 0.092 & 7.029 & 3.673 & 3.113 & - & 6.741 & 6.557 & 0.235 \\ \hline
Exp. 4: Expression-only & 0.027 & 4.46 & 3.45 & 2.548 & - & 6.029 & 7.469 & 0.9 & 0.030 & 5.016 & 6.697 & 4.226 & - & 8.477 & 8.641 & 0.913 \\ \hline
Exp. 5: Expression-only & 0.580 & 3.86 & 0.022 & 0.02 & - & 0.055 & 0.047 & 0.06 & 1.412 & 7.689 & 0.023 & 0.022 & - & 0.06 & 0.058 & 0.107 \\ \hline
Exp. 6: Expression-only & 0.009 & 1.012 & 3.048 & 2.304 & - & 5.263 & 5.047 & 0.105 & 0.01 & 1.42 & 3.924 & 3.469 & - & 6.178 & 6.013 & 0.112 \\ \hline
Exp. 7: Expression-only & 0.009 & 1.009 & 3.027 & 1.319 & - & 5.258 & 5.097 & 0.106 & 0.01 & 1.441 & 3.893 & 3.446 & - & 6.121 & 6.015 & 0.116 \\ \hline
Exp. 8: Expression-only & 0.016 & 4.444 & 6.167 & 2.674 & - & 6.633 & 6.024 & 0.671 & 0.025 & 4.845 & 6.495 & 4.311 & - & 6.984 & 8.252 & 0.734 \\ \hline
Exp. 9: Expression-only & 2.211 & 12.459 & 2.968 & 1.247 & - & 9.977 & 9.41 & 1.328 & 2.48 & 28.087 & 3.848 & 3.318 & - & 14.02 & 12.215 & 2.47 \\ \hline
Exp. 10: Expression-only & 0.074 & 3.526 & 0.025 & 0.023 & - & 0.162 & 0.161 & 0.281 & 0.09 & 4.339 & 0.029 & 0.025 & - & 0.166 & 0.164 & 0.318 \\ \hline
Exp. 11: Expression-only & 0.020  & 1.715  & 1.677  & 0.625  & 0.12  & 2.556  & 2.353  & 0.234  & 0.054  & 4.126  & 4.131  & 3.549  & 0.141  & 5.804  & 5.58  & 0.214  \\ \hline
Exp. 12: Expression-only & 0.627  & 26.356  & 3.406  & 3.004 & 3.149  & 7.074  & 6.93  & 0.696  & -  & -  & 9.716  & 8.749  & 9.213  & 14.352  & 14.06  & 1.118 \\ \hline
\end{tabular}%
}
\caption{Single Node Average Elapsed Times (in sec.)}
\label{tab:combined-single-node}
\end{table*}

\begin{table*}[h]
\resizebox{\textwidth}{!}{%
\begin{tabular}{|l||l|l|l|l|l|l||l|l|l|l|l|l||l|l|l|l|l|l|}
\hline
\multirow{2}{*}{\begin{tabular}[c]{@{}l@{}}\textbf{Expression: Time}\end{tabular}} & \multicolumn{6}{c||}{\textbf{5GB}} & \multicolumn{6}{c||}{\textbf{7.5GB}} & \multicolumn{6}{c|}{\textbf{10GB}} \\ \cline{2-19} 
 & \rotatebox[origin=c]{90}{\textbf{AFrame}} & \rotatebox[origin=c]{90}{\textbf{AFrame Schema}} & \rotatebox[origin=c]{90}{\textbf{AFrame Index}} & \rotatebox[origin=c]{90}{\textbf{Spark JSON}} & \rotatebox[origin=c]{90}{\textbf{Spark JSON Schema}} & \rotatebox[origin=c]{90}{\textbf{Spark Parquet}} & \rotatebox[origin=c]{90}{\textbf{AFrame}} & \rotatebox[origin=c]{90}{\textbf{AFrame Schema}} &\rotatebox[origin=c]{90}{\textbf{AFrame Index}} & \rotatebox[origin=c]{90}{\textbf{Spark JSON}} & \rotatebox[origin=c]{90}{\textbf{Spark JSON Schema}} & \rotatebox[origin=c]{90}{\textbf{Spark Parquet}} & \rotatebox[origin=c]{90}{\textbf{AFrame}} & \rotatebox[origin=c]{90}{\textbf{AFrame Schema}} & \rotatebox[origin=c]{90}{\textbf{AFrame Index}} & \rotatebox[origin=c]{90}{\textbf{Spark JSON}} & \rotatebox[origin=c]{90}{\textbf{Spark JSON Schema}} & \rotatebox[origin=c]{90}{\textbf{Spark Parquet}} 
 
 \\ \hhline{===================}
Exp. 1: Total & 5.663 & 5.088 & 0.391 & 42.61 & 13.52 & 4.17 & 8.855 & 7.966 & 0.524 & 261.79 & 136.91 & 4.40 & 11.103 & 10.186 & 0.705 & 321.46  & 190.24 & 4.45 \\ \hline
Exp. 2: Total & 0.043 & 0.034 & - & 31.905 & 3.099 & 4.116 & 0.043 & 0.034 & - & 127.493 & 3.833 & 4.244 & 0.054 & 0.053 & - & 134.580  & 3.993 & 4.287 \\ \hline
Exp. 3: Total & 7.057 & 6.21 & - & 46.185 & 15.183 & 4.289 & 10.73 & 9.518 & - & 260.588 & 136.747 & 4.456 & 13.909 & 12.003 & - & 314.543  & 183.860 & 4.531 \\ \hline
Exp. 4: Total & 12.254 & 8.765 & - & 45.050 & 18.338 & 4.970 & 18.368 & 12.781 & - & 259.536 & 136.845 & 5.230 & 24.29 & 16.454 & - & 317.081  & 187.809 & 5.305 \\ \hline
Exp. 5: Total & 0.027 & 0.027 & - & 31.879 & 3.078 & 4.154 & 0.03 & 0.028 & - & 127.458 & 3.798 & 4.286 & 0.033 & 0.031 & - & 134.516  & 3.930 & 4.321 \\ \hline
Exp. 6: Total & 7.744 & 6.822 & - & 44.101 & 14.109 & 4.169 & 11.698 & 10.242 & - & 259.556 & 132.956 & 4.324 & 15.413 & 13.286 & - & 314.517  & 185.830 & 4.399 \\ \hline
Exp. 7: Total & 7.728 & 6.771 & - & 43.912 & 14.781 & 4.162 & 11.842 & 10.266 & - & 260.563 & 132.907 & 4.311 & 15.356 & 13.213 & - & 314.611  & 185.802 & 4.356 \\ \hline
Exp. 8: Total & 11.947 & 8.645 & - & 47.593 & 18.176 & 4.828 & 17.486 & 12.932 & - & 260.541 & 132.838 & 4.993 & 23.572 & 16.72 & - & 324.423  & 185.249 & 5.069 \\ \hline
Exp. 9: Total & 7.687 & 6.552 & - & 58.836 & 26.126 & 8.668 & 11.53 & 9.957 & - & 260.553 & 133.61 & 11.513 & 15.148 & 12.518 & - & 318.329  & 184.363 & 13.409 \\ \hline
Exp. 10: Total & 0.03 & 0.027 & - & 31.981 & 3.189 & 4.372 & 0.036 & 0.03 & - & 127.588 & 3.929 & 4.525 & 0.039 & 0.034 & - & 134.672  & 4.074 & 4.588 \\ \hline

Exp. 11: Total & 8.192  & 6.893  & 0.19  & 42.915  & 13.654  & 4.269  & 12.421  & 10.459  & 0.238  & 258.098  & 135.861  & 4.446  & 16.143  & 13.726  & 0.267  & 315.402  & 184.810  & 4.72   \\ \hline
Exp. 12: Total & 19.817  & 17.524  & 19.306  & 272.46  & 137.245  & 8.053  & 29.602  & 25.428  & 29.261  & 484.181  & 264.376  & 9.556  & 38.934  & 34.645  & 40.241  & 732.945  & 364.429  & 10.694  \\ \hhline{===================}

Exp. 1: Expression-only & 5.663 & 5.088 & 0.391 & 10.798 & 10.502 & 0.134 & 8.855 & 7.966 & 0.524 & 134.403 & 133.174 & 0.247 & 11.103 & 10.186 & 0.705 & 187.027  & 186.388 & 0.267 \\ \hline
Exp. 2: Expression-only & 0.043 & 0.034 & - & 0.094 & 0.083 & 0.076 & 0.05 & 0.042 & - & 0.109 & 0.099 & 0.09 & 0.054 & 0.053 & - & 0.144 & 0.142 & 0.102 \\ \hline
Exp. 3: Expression-only & 7.057 & 6.21 & - & 14.374 & 12.167 & 0.249 & 10.73 & 9.518 & - & 133.204 & 133.013 & 0.302 & 13.909 & 12.003 & - & 180.107  & 180.009 & 0.346 \\ \hline
Exp. 4: Expression-only & 12.254 & 8.765 & - & 13.239 & 15.322 & 0.93 & 18.368 & 12.781 & - & 132.152 & 133.111 & 1.076 & 24.29 & 16.454 & - & 182.645  & 183.958 & 1.12 \\ \hline
Exp. 5: Expression-only & 0.027 & 0.027 & - & 0.068 & 0.062 & 0.114 & 0.03 & 0.028 & - & 0.074 & 0.064 & 0.132 & 0.033 & 0.031 & - & 0.08 & 0.079 & 0.136 \\ \hline
Exp. 6: Expression-only & 7.744 & 6.822 & - & 12.29 & 11.093 & 0.129 & 11.698 & 10.242 & - & 132.172 & 129.222 & 0.17 & 15.413 & 13.286 & - & 180.081  & 181.979 & 0.214 \\ \hline
Exp. 7: Expression-only & 7.728 & 6.771 & - & 12.101 & 11.765 & 0.122 & 11.842 & 10.266 & - & 133.179 & 129.173 & 0.157 & 15.356 & 13.213 & - & 180.175  & 181.951 & 0.171 \\ \hline
Exp. 8: Expression-only & 11.947 & 8.645 & - & 15.782 & 15.16 & 0.788 & 17.486 & 12.932 & - & 133.157 & 129.104 & 0.839 & 23.572 & 16.72 & - & 189.987  & 181.398 & 0.884 \\ \hline
Exp. 9: Expression-only & 7.687 & 6.552 & - & 27.025 & 23.11 & 4.628 & 11.53 & 9.957 & - & 133.169 & 129.876 & 7.359 & 15.148 & 12.518 & - & 183.893  & 180.512 & 9.224 \\ \hline
Exp. 10: Expression-only & 0.03 & 0.027 & - & 0.171 & 0.169 & 0.332 & 0.036 & 0.03 & - & 0.208 & 0.199 & 0.375 & 0.039 & 0.034 & - & 0.236 & 0.223 & 0.398 \\ \hline
Exp. 11: Expression-only & 8.192  & 6.893  & 0.19  & 11.104  & 10.638  & 0.229  & 12.421  & 10.459  & 0.238  & 130.714  & 132.127  & 0.292  & 16.143  & 13.726  & 0.267  & 180.966  & 180.959  & 0.535\\ \hline
Exp. 12: Expression-only & 19.817  & 17.524  & 19.306  & 136.877  & 134.229  & 4.013  & 28.970  & 25.428  & 29.261  & 263.3  & 260.642  & 5.402  & 38.934  & 34.645  & 40.241  & 367.073  & 360.578  & 6.509\\ \hline
\end{tabular}%
}
\caption{Single Node Average Elapsed Times (in sec.) continued}
\label{tab:combined-single-node2}
\end{table*}

\begin{table*}[h]
\resizebox{\textwidth}{!}{%
\begin{tabular}{|l||l|l|l|l|l|l||l|l|l|l|l|l||l|l|l|l|l|l|}
\hline
\multirow{2}{*}{\begin{tabular}[c]{@{}l@{}}\textbf{Expression: Time}\end{tabular}} & \multicolumn{6}{c||}{\textbf{2 Nodes}} & \multicolumn{6}{c||}{\textbf{3 Nodes}} & \multicolumn{6}{c|}{\textbf{4 Nodes}} \\ \cline{2-19} 
 & \rotatebox[origin=c]{90}{\textbf{AFrame}} & \rotatebox[origin=c]{90}{\textbf{AFrame Schema}} & \rotatebox[origin=c]{90}{\textbf{AFrame Index}} & \rotatebox[origin=c]{90}{\textbf{Spark JSON}} & \rotatebox[origin=c]{90}{\textbf{Spark JSON Schema}} & \rotatebox[origin=c]{90}{\textbf{Spark Parquet}} & \rotatebox[origin=c]{90}{\textbf{AFrame}} & \rotatebox[origin=c]{90}{\textbf{AFrame Schema}} &\rotatebox[origin=c]{90}{\textbf{AFrame Index}} & \rotatebox[origin=c]{90}{\textbf{Spark JSON}} & \rotatebox[origin=c]{90}{\textbf{Spark JSON Schema}} & \rotatebox[origin=c]{90}{\textbf{Spark Parquet}} & \rotatebox[origin=c]{90}{\textbf{AFrame}} & \rotatebox[origin=c]{90}{\textbf{AFrame Schema}} & \rotatebox[origin=c]{90}{\textbf{AFrame Index}} & \rotatebox[origin=c]{90}{\textbf{Spark JSON}} & \rotatebox[origin=c]{90}{\textbf{Spark JSON Schema}} & \rotatebox[origin=c]{90}{\textbf{Spark Parquet}} 
  \\ \hhline{===================}
Exp. 1: Total & 7.865 & 7.743 & 0.830 & 54.724 & 17.157 & 3.627 & 4.814 & 4.379 & 0.524 & 41.964 & 13.491 & 3.51 & 3.837 & 3.449 & 0.309 & 33.697 & 11.490 & 3.258 \\ \hline
Exp. 2: Total & 0.051 & 0.039 & - & 37.007 & 1.612 & 3.411 & 0.043 & 0.034 & - & 28.485 & 1.583 & 3.312 & 0.035 & 0.031 & - & 22.741 & 1.317 & 3.074 \\ \hline
Exp. 3: Total & 7.925 & 7.493 & - & 55.391 & 19.415 & 3.668 & 5.657 & 5.548 & - & 44.224 & 16.136 & 3.557 & 3.533 & 3.355 & - & 36.649 & 12.222 & 3.297 \\ \hline
Exp. 4: Total & 9.298 & 7.441 & - & 56.485 & 19.284 & 4.565 & 8.083 & 6.510 & - & 43.904 & 16.563 & 4.116 & 6.059 & 3.717 & - & 37.308 & 15.271 & 3.824 \\ \hline
Exp. 5: Total & 0.031 & 0.029 & - & 37.030 & 1.626 & 3.491 & 0.029 & 0.028 & - & 28.508 & 1.61 & 3.385 & 0.028 & 0.0250 & - & 22.747 & 1.328 & 3.134 \\ \hline
Exp. 6: Total & 7.100 & 6.692 & - & 56.229 & 18.560 & 3.589 & 5.064 & 4.417 & - & 42.645 & 14.462 & 3.479 & 3.941 & 3.363 & - & 33.394 & 10.356 & 3.228 \\ \hline
Exp. 7: Total & 7.775 & 6.701 & - & 56.239 & 18.843 & 3.633 & 5.081 & 4.422 & - & 40.990 & 13.571 & 3.504 & 3.915 & 3.370 & - & 33.271 & 11.199 & 3.221 \\ \hline
Exp. 8: Total & 11.684 & 8.447 & - & 56.681 & 19.365 & 4.526 & 7.814 & 5.548 & - & 42.208 & 14.238 & 4.17 & 5.874 & 4.236 & - & 34.547 & 12.950 & 3.848 \\ \hline
Exp. 9: Total & 7.068 & 6.500 & - & 65.105 & 25.903 & 8.773 & 5.040 & 4.318 & - & 52.005 & 21.563 & 7.506 & 3.852 & 3.295 & - & 43.567 & 18.559 & 6.009 \\ \hline
Exp. 10: Total & 0.038 & 0.036 & - & 37.084 & 1.684 & 3.715 & 0.036 & 0.034 & - & 28.345 & 1.625 & 3.614 & 0.033 & 0.032 & - & 22.777 & 1.353 & 3.361 \\ \hline
Exp. 11: Total & 8.086  & 6.908  & 0.193  & 55.914  & 20.485  & 3.836  & 5.478  & 4.615  & 0.169  & 41.035  & 13.854  & 3.704  & 4.147  & 3.570  & 0.122  & 32.716  & 11.138  & 3.445   \\ \hline
Exp. 12: Total & 19.507  & 17.768  & 22.863  & 273.822  & 145.211  & 7.239  & 13.786  & 12.380  & 15.078  & 162.117  & 76.788  & 6.316  & 10.774  & 9.794  & 11.731  & 63.404  & 21.957  & 5.685  \\ \hhline{===================}

Exp. 1: Expression-only & 7.865 & 7.743 & 0.830 & 17.813 & 15.640 & 0.252 & 4.814 & 4.379 & 0.524 & 13.569 & 11.989 & 0.23 & 2.837 & 2.549 & 0.309 & 11.037 & 10.251 & 0.209 \\ \hline
Exp. 2: Expression-only & 0.051 & 0.039 & - & 0.096 & 0.095 & 0.036 & 0.043 & 0.034 & - & 0.090 & 0.081 & 0.032 & 0.035 & 0.031 & - & 0.081 & 0.078 & 0.025 \\ \hline
Exp. 3: Expression-only & 7.925 & 7.493 & - & 18.480 & 17.898 & 0.293 & 5.657 & 5.548 & - & 15.829 & 14.634 & 0.277 & 3.533 & 3.355 & - & 13.989 & 10.983 & 0.248 \\ \hline
Exp. 4: Expression-only & 9.298 & 7.441 & - & 19.574 & 17.767 & 1.19 & 8.083 & 6.510 & - & 15.509 & 15.061 & 0.836 & 6.059 & 3.717 & - & 14.648 & 14.032 & 0.775 \\ \hline
Exp. 5: Expression-only & 0.031 & 0.029 & - & 0.079 & 0.076 & 0.116 & 0.029 & 0.028 & - & 0.073 & 0.072 & 0.105 & 0.028 & 0.0250 & - & 0.067 & 0.066 & 0.085 \\ \hline
Exp. 6: Expression-only & 7.100 & 6.692 & - & 19.318 & 17.043 & 0.204 & 5.064 & 4.417 & - & 13.250 & 12.960 & 0.199 & 3.941 & 3.363 & - & 10.734 & 9.117 & 0.179 \\ \hline
Exp. 7: Expression-only & 7.775 & 6.701 & - & 19.328 & 17.326 & 0.158 & 5.081 & 4.422 & - & 12.595 & 12.069 & 0.124 & 3.915 & 3.370 & - & 10.611 & 9.960 & 0.072 \\ \hline
Exp. 8: Expression-only & 11.684 & 8.447 & - & 19.770 & 17.848 & 0.595 & 7.814 & 5.548 & - & 13.813 & 12.736 & 0.49 & 5.874 & 4.236 & - & 11.887 & 11.711 & 0.399 \\ \hline
Exp. 9: Expression-only & 7.068 & 6.500 & - & 28.194 & 24.386 & 5.398 & 5.040 & 4.318 & - & 23.761 & 20.061 & 4.226 & 3.852 & 3.295 & - & 20.907 & 17.320 & 2.946 \\ \hline
Exp. 10: Expression-only & 0.038 & 0.036 & - & 0.173 & 0.167 & 0.340 & 0.036 & 0.034 & - & 0.130 & 0.123 & 0.334 & 0.033 & 0.032 & - & 0.117 & 0.114 & 0.312 \\ \hline

Exp. 11: Expression-only & 8.086  & 6.908  & 0.193  & 19.003  & 18.968  & 0.461  & 5.478  & 4.615  & 0.169  & 12.640  & 12.352  & 0.424  & 4.147  & 3.570  & 0.122  & 10.056  & 9.899  & 0.396\\ \hline
Exp. 12: Expression-only & 19.507  & 17.768  & 22.863  & 143.726  & 143.694  & 3.864  & 13.786  & 12.380  & 15.078  & 75.841  & 75.286  & 3.036  & 10.774  & 9.794  & 11.731  & 22.216  & 20.718  & 2.636\\ \hline
\end{tabular}%
}
\caption{Multi-node Speedup for Average Elapsed Times (in sec.)}
\label{tab:combined-speedup}
\end{table*}

\begin{table*}[h]
\resizebox{\textwidth}{!}{%
\begin{tabular}{|l||l|l|l|l|l|l||l|l|l|l|l|l||l|l|l|l|l|l|}
\hline
\multirow{2}{*}{\begin{tabular}[c]{@{}l@{}}\textbf{Expression: Time}\end{tabular}} & \multicolumn{6}{c||}{\textbf{2 Nodes}} & \multicolumn{6}{c||}{\textbf{3 Nodes}} & \multicolumn{6}{c|}{\textbf{4 Nodes}} \\ \cline{2-19} 
 & \rotatebox[origin=c]{90}{\textbf{AFrame}} & \rotatebox[origin=c]{90}{\textbf{AFrame Schema}} & \rotatebox[origin=c]{90}{\textbf{AFrame Index}} & \rotatebox[origin=c]{90}{\textbf{Spark JSON}} & \rotatebox[origin=c]{90}{\textbf{Spark JSON Schema}} & \rotatebox[origin=c]{90}{\textbf{Spark Parquet}} & \rotatebox[origin=c]{90}{\textbf{AFrame}} & \rotatebox[origin=c]{90}{\textbf{AFrame Schema}} &\rotatebox[origin=c]{90}{\textbf{AFrame Index}} & \rotatebox[origin=c]{90}{\textbf{Spark JSON}} & \rotatebox[origin=c]{90}{\textbf{Spark JSON Schema}} & \rotatebox[origin=c]{90}{\textbf{Spark Parquet}} & \rotatebox[origin=c]{90}{\textbf{AFrame}} & \rotatebox[origin=c]{90}{\textbf{AFrame Schema}} & \rotatebox[origin=c]{90}{\textbf{AFrame Index}} & \rotatebox[origin=c]{90}{\textbf{Spark JSON}} & \rotatebox[origin=c]{90}{\textbf{Spark JSON Schema}} & \rotatebox[origin=c]{90}{\textbf{Spark Parquet}} 
  \\ \hhline{===================}
Exp. 1: Total & 11.38  & 10.092  & 0.7  & 323.498  & 182.316  & 3.537  & 11.539  & 10.383  & 0.701  & 324.481  & 185.514  & 3.403  & 11.699  & 10.507  & 0.716  & 320.086  & 185.439  & 3.521 \\ \hline
Exp. 2: Total & 0.062  & 0.054  & -  & 139.251  & 1.350  & 3.383  & 0.052  & 0.053  & -  & 139.932  & 1.7  & 3.224  & 0.055  & 0.049  & -  & 135.494  & 1.516  & 3.336 \\ \hline
Exp. 3: Total & 13.855  & 12.145  & -  & 326.53  & 183.042  & 3.890  & 14.180  & 12.263  & -  & 324.088  & 185.431  & 3.715  & 14.318  & 12.637  & -  & 319.648  & 185.515  & 3.898 \\ \hline
Exp. 4: Total & 24.646  & 16.948  & -  & 323.959  & 185.701  & 4.516  & 24.790  & 16.973  & -  & 324.778  & 185.985  & 4.398  & 25.118  & 16.957  & -  & 320.333  & 186.094  & 4.497 \\ \hline
Exp. 5: Total & 0.036  & 0.035  & -  & 139.188  & 1.286  & 3.404  & 0.039  & 0.037  & -  & 139.878  & 1.373  & 3.264  & 0.036  & 0.032  & -  & 135.438  & 1.453  & 3.366 \\ \hline
Exp. 6: Total & 15.411  & 13.335  & -  & 323.467  & 182.316  & 3.537  & 15.571  & 13.357  & -  & 324.132  & 183.834  & 3.388  & 15.927  & 13.754  & -  & 319.707  & 185.018  & 3.429 \\ \hline
Exp. 7: Total & 15.438  & 12.661  & -  & 326.544  & 189.142  & 3.532  & 15.563  & 13.395  & -  & 324.105  & 185.633  & 3.325  & 15.915  & 13.742  & -  & 319.581  & 185.422  & 3.493 \\ \hline
Exp. 8: Total & 24.312  & 16.998  & -  & 326.625  & 188.657  & 4.233  & 24.17  & 16.816  & -  & 324.103  & 185.666  & 4.11  & 24.914  & 17.325  & -  & 319.682  & 185.518  & 4.227 \\ \hline
Exp. 9: Total & 15.876  & 13.053  & -  & 326.597  & 188.703  & 12.422  & 15.431  & 13.078  & -  & 324.38  & 185.613  & 13.069  & 15.67  & 13.135  & -  & 319.778  & 185.575  & 13.256 \\ \hline
Exp. 10: Total & 0.031  & 0.027  & -  & 139.214  & 1.299  & 3.562  & 0.03  & 0.029  & -  & 139.898  & 1.382  & 3.387  & 0.035  & 0.034  & -  & 135.456  & 1.468  & 3.518 \\ \hline

Exp. 11: Total & 16.046  & 13.714  & 0.264  & 324.854  & 186.914  & 3.908  & 16.31  & 13.756  & 0.258  & 322.176  & 183.637  & 3.792  & 16.425  & 13.729  & 0.281  & 317.508  & 183.917  & 3.915   \\ \hline
Exp. 12: Total & 38.289  & 34.168  & 47.297  & 506.986  & 369.924  & 9.153  & 38.232  & 34.081  & 46.533  & 507.827  & 369.465  & 9.492  & 38.718  & 35.236  & 47.676  & 503.444  & 369.942  & 9.443  \\ \hhline{===================}

Exp. 1: Expression-only & 11.38  & 10.092  & 0.700  & 184.392  & 181.108  & 0.263  & 11.539  & 10.383  & 0.701  & 184.69  & 184.224  & 0.289  & 11.699  & 10.507  & 0.716  & 184.735  & 184.065  & 0.293 \\ \hline
Exp. 2: Expression-only & 0.062  & 0.054  & -  & 0.145  & 0.142  & 0.109  & 0.052  & 0.053  & -  & 0.141  & 0.14  & 0.11  & 0.055  & 0.049  & -  & 0.143  & 0.142  & 0.108 \\ \hline
Exp. 3: Expression-only & 13.855  & 12.145  & -  & 187.424  & 181.834  & 0.616  & 14.18  & 12.263  & -  & 184.297  & 184.141  & 0.601  & 14.18  & 12.263  & -  & 184.297  & 184.141  & 0.601 \\ \hline
Exp. 4: Expression-only & 24.646  & 16.948  & -  & 184.853  & 184.493  & 1.242  & 24.79  & 16.973  & -  & 184.987  & 184.695  & 1.284  & 25.118  & 16.957  & -  & 184.982  & 184.72  & 1.269 \\ \hline
Exp. 5: Expression-only & 0.036  & 0.035  & -  & 0.082  & 0.078  & 0.13  & 0.039  & 0.037  & -  & 0.087  & 0.083  & 0.15  & 0.036  & 0.032  & -  & 0.087  & 0.079  & 0.138 \\ \hline
Exp. 6: Expression-only & 15.411  & 13.335  & -  & 184.361  & 181.108  & 0.263  & 15.571  & 13.357  & -  & 184.341  & 182.544  & 0.274  & 15.927  & 13.754  & -  & 184.356  & 183.644  & 0.201 \\ \hline
Exp. 7: Expression-only & 15.438  & 12.661  & -  & 187.438  & 187.934  & 0.258  & 15.563  & 13.395  & -  & 184.314  & 184.343  & 0.211  & 15.915  & 13.742  & -  & 184.23  & 184.048  & 0.265 \\ \hline
Exp. 8: Expression-only & 24.312  & 16.998  & -  & 187.519  & 187.449  & 0.959  & 24.17  & 16.816  & -  & 184.312  & 184.376  & 0.996  & 24.914  & 17.325  & -  & 184.331  & 184.144  & 0.999 \\ \hline
Exp. 9: Expression-only & 15.876  & 13.053  & -  & 187.491  & 187.495  & 9.148  & 15.431  & 13.078  & -  & 184.589  & 184.323  & 9.955  & 15.67  & 13.135  & -  & 184.427  & 184.201  & 10.028 \\ \hline
Exp. 10: Expression-only & 0.031  & 0.027  & -  & 0.198  & 0.191  & 0.388  & 0.03  & 0.029  & -  & 0.197  & 0.192  & 0.373  & 0.035  & 0.034  & -  & 0.195  & 0.194  & 0.39 \\ \hline
Exp. 11: Expression-only & 16.046  & 13.714  & 0.264  & 185.748  & 185.706  & 0.634  & 16.31  & 13.756  & 0.258  & 182.385  & 182.347  & 0.678  & 16.425  & 13.729  & 0.281  & 182.157  & 182.543  & 0.687 \\ \hline
Exp. 12: Expression-only & 38.289  & 34.168  & 47.297  & 367.88  & 368.716  & 5.879  & 38.232  & 34.081  & 46.533  & 368.036  & 368.175  & 6.378  & 38.718  & 35.236  & 47.676  & 368.093  & 368.568  & 6.215\\ \hline
\end{tabular}%
}
\caption{Multi-node Scaleup for Average Elapsed Times (in sec.)}
\label{tab:combined-scaleup}
\end{table*}

\end{document}

%% file: single_node.tex
\setcounter{topnumber}{7}

\begin{figure*}[!ht]
     \centering

   \begin{subfigure}[t]{0.68\textwidth}
        \includegraphics[trim=0.5cm 1.5 0 2,width=\textwidth,height=1cm]{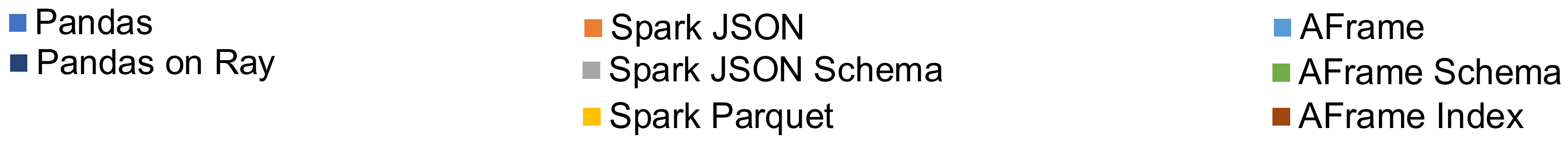}%
    \end{subfigure}
    \begin{subfigure}[t]{0.43\textwidth}
        \includegraphics[trim=1.5 1.5 0cm 1.5,width=\textwidth,height=3.7cm]{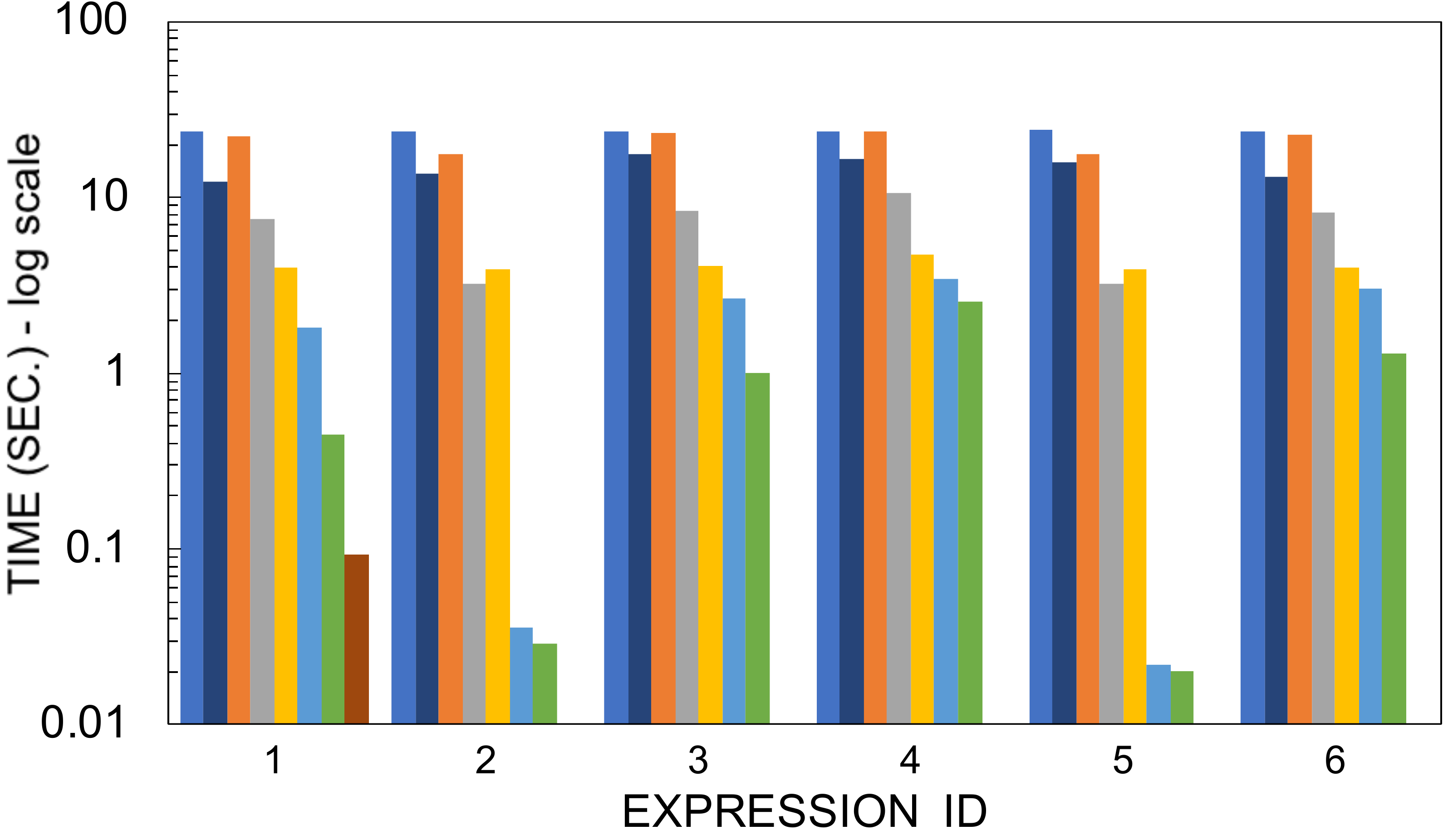}
        \caption{Expression 1-6 total times}
        \label{fig:1-5total}
    \end{subfigure}
   \hspace{0.2cm}
    \begin{subfigure}[t]{0.43\textwidth}
        \includegraphics[trim=0.5cm 1.5 0.3cm 1.5,width=\textwidth,height=3.75cm]{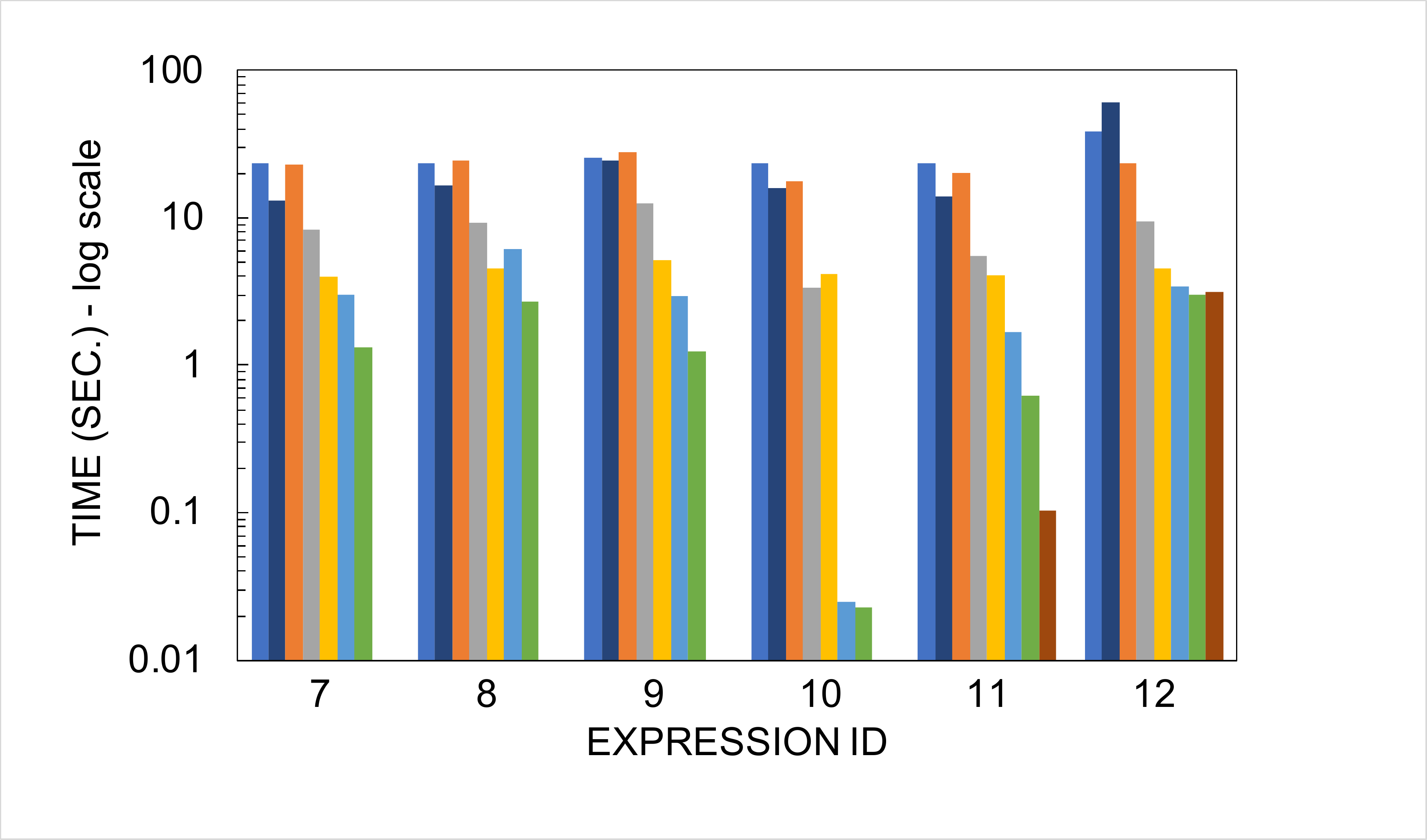}
        \caption{Expression 7-12 total times}
        \label{fig:6-10total}
    \end{subfigure}
    \hfill
    \begin{subfigure}[t]{0.43\textwidth}
        \includegraphics[trim=0.9cm 1.5 0.5cm 1.5,width=\textwidth,height=3.7cm]{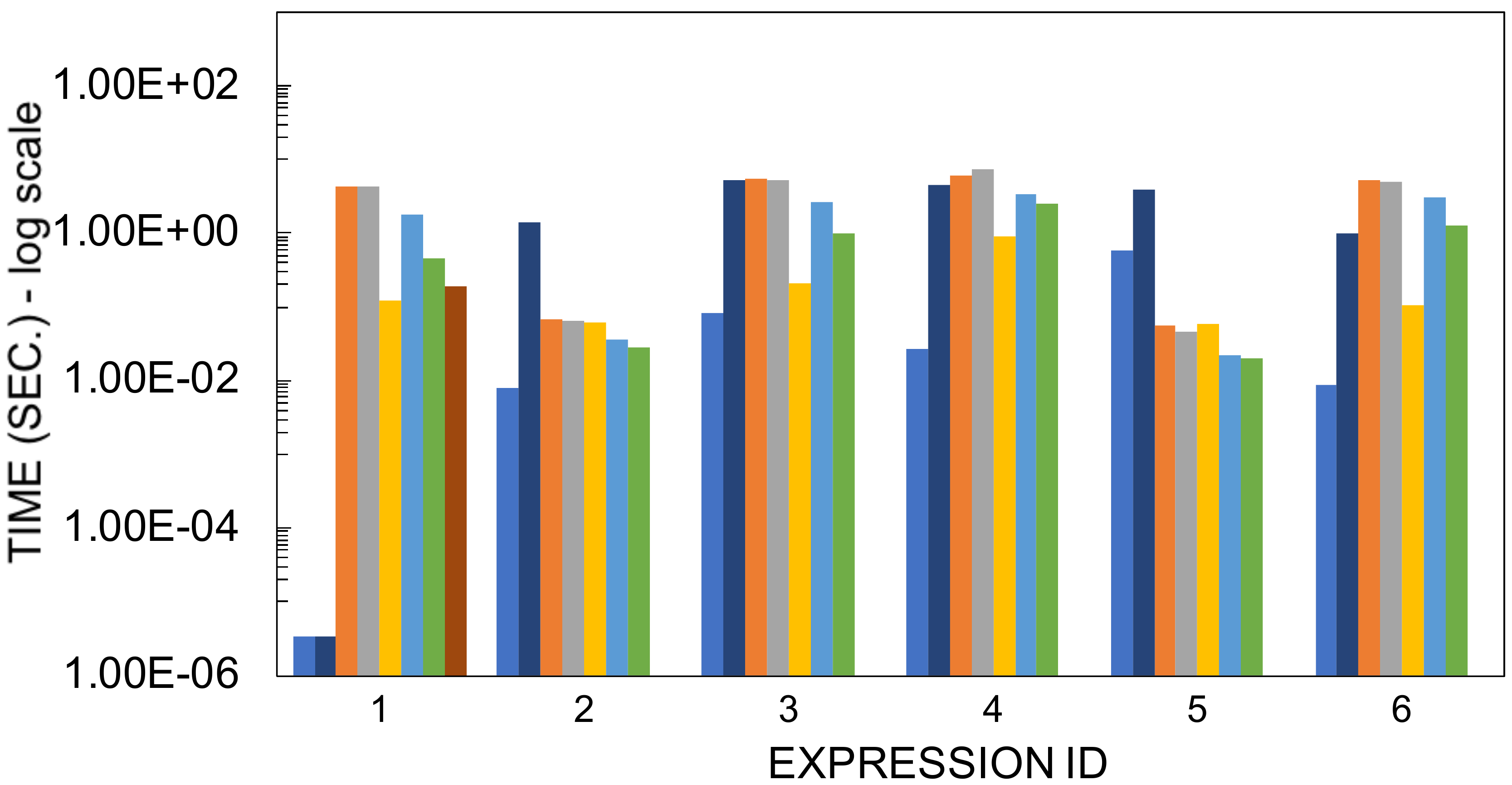}
        \caption{Expression 1-6 expression-only times}
        \label{fig:1-5task}
    \end{subfigure}
    \hspace{0.2cm}
    \begin{subfigure}[t]{0.43\textwidth}
        \includegraphics[trim=0cm 1.5 0.5cm 1.5,width=\textwidth,height=3.7cm]{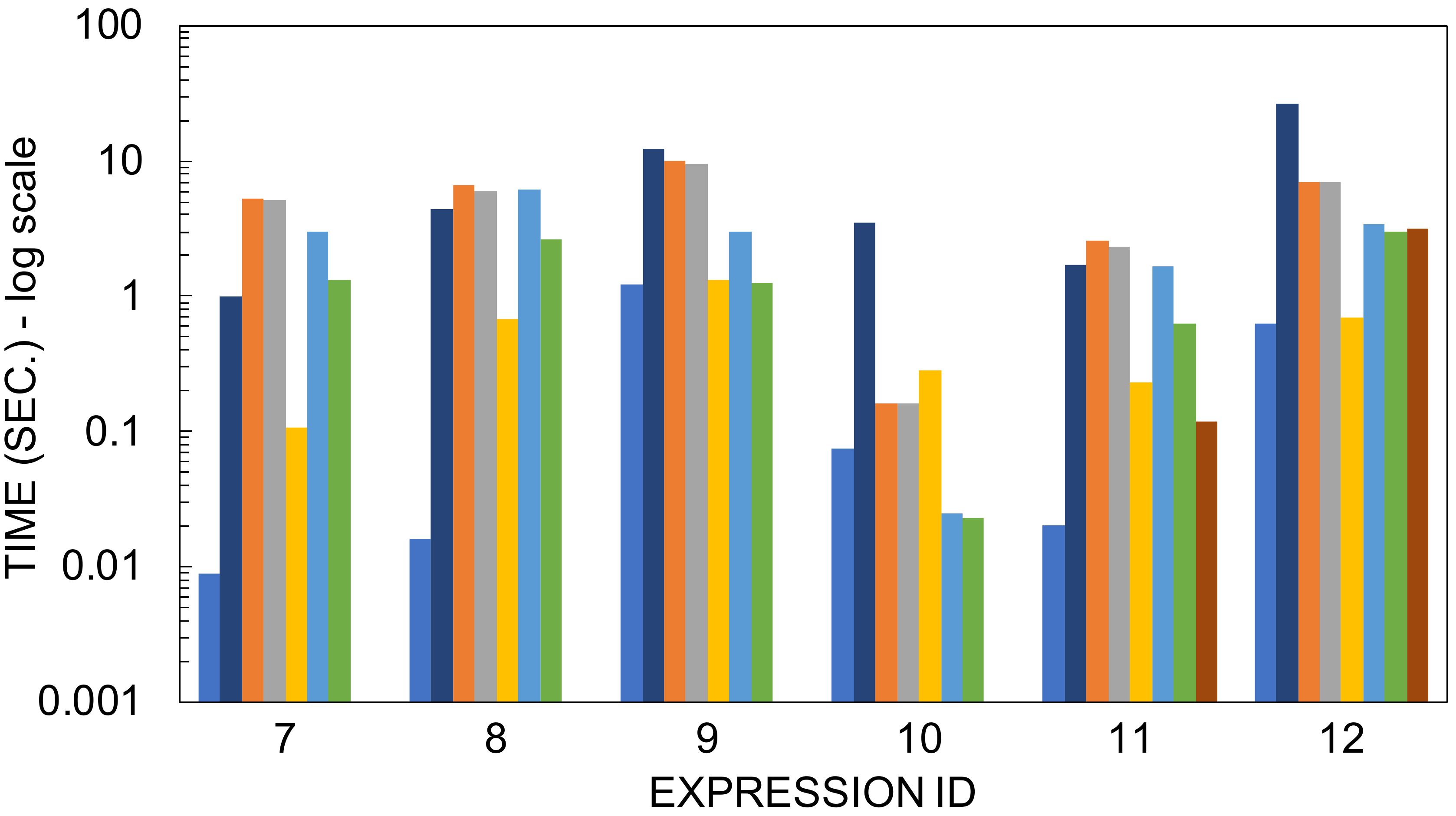}
        \caption{Expression 7-12 expression-only times}
        \label{fig:6-10task}
    \end{subfigure}
    \caption{XS Results of Single Node Evaluation}
    \vspace{-1em}
    \label{fig:single_node_1}
\end{figure*}

\subsection{Single Node Results}
For the single-node evaluations, we ran the test suite first on the XS Wisconsin dataset as a preliminary test to determine the level of feature support in each framework and to observe their relative performance across all twelve expressions. The XS results are displayed in Figure~\ref{fig:single_node_1}. After the first round, we ran the benchmark on four other dataset sizes, S, M, L and XL to evaluate the data scalability of each framework on a single node. As mentioned in the experimental setup section, we present both the expression run times and the total run times (which include the DataFrame creation times). The raw averaged run times for single node results are also included in the appendix.

\subsubsection{\textbf{Baseline Performance Results}}

The XS results are presented in Figure~\ref{fig:single_node_1}. \Cref{fig:1-5total,fig:6-10total} show the total run times (including DataFrame creation). \Cref{fig:1-5total} displays expression 1-6's results and Figure \ref{fig:6-10total} displays expression 7-12's results. \Cref{fig:1-5task,fig:6-10task} show the expression-only execution times. Figure~\ref{fig:1-5task} displays expression 1-6's results, and Figure~\ref{fig:6-10task} displays expression 7-12's results. The differences between the total times and expression-only times indicate that the DataFrame creation process can significantly impact system performance.

Pandas requires data to be loaded into memory before its operation evaluations. Since it was not designed for parallel processing, the total run time including DataFrame creation was high for Pandas in all of the test cases. However, once the data was loaded into memory, as shown in Figures~\ref{fig:1-5task} and \ref{fig:6-10task}, Pandas performed the best in 10 of the 12 expressions. The two cases where Pandas was not the fastest were operations 5 and 10, and the reason was Pandas' eager evaluation strategy. Expression 5 applies a function to a string column, while expression 10 selects rows that satisfy a column predicate. However, in the end, both expressions 5 and 10 require only a small subset (head()) of rows from the dataset. The strict nature of Pandas' eager evaluation caused both the function and the predicate to be applied to the whole dataset before selecting only a few samples to return. On the other hand, with lazy evaluation, the expressions can be applied to just the subset of data needed to fulfill the result's required size. 

Pandas on Ray leverages parallel processing by utilizing all available cores in a system to load and process the data. However, there are overheads associated with distributing a DataFrame, as we can see from Figures~\ref{fig:1-5task} and \ref{fig:6-10task}, where Pandas outperformed Pandas on Ray on all but one expression. However, Pandas on Ray's total run time was better than Pandas' due to parallel data loading. As the size of the data grows, so does the time taken to process the data. Pandas on Ray outperforms Pandas when the task processing time dominates its work distribution overheads.

Among the three Spark DataFrames, the Parquet-based DataFrame (Spark Parquet) outperformed the JSON-based DataFrame (Spark JSON) and the JSON-based DataFrame with a pre-defined schema (Spark JSON Schema) in most of the tested cases for both the total and expression-only evaluation metrics. Spark produces different runtime plans for the JSON-based DataFrame and the Parquet-based DataFrame, resulting in the difference in their task execution times even after the schema inferencing step. 


AFrame was the fastest in terms of the total-time evaluation since its DataFrame creation process does not involve first loading data into memory from a file. In addition, AFrame also benefits from the presence of database indexes. Its performance results for the datasets with indexes are an order of magnitude faster, as seen for Expressions 1, 11, and 12. Even in terms of just the expression-only time, AFrame with an index on the range attribute performed better than Spark Parquet on expression 11 (see Figure~\ref{fig:6-10task}).


%% file: single_node_images.tex
\subsubsection{\textbf{Scalability}}
\begin{figure*}[!ht]
    \centering
    \begin{subfigure}[t]{0.65\textwidth}
        \includegraphics[trim=0.5cm 1.5 0 2,width=\textwidth,height=1cm]{figures/single_legend_2.pdf}%
    \end{subfigure}
    \begin{subfigure}[t]{0.42\textwidth}
        \includegraphics[trim=0.5cm 1.5 0 2,width=\textwidth,height=4cm]{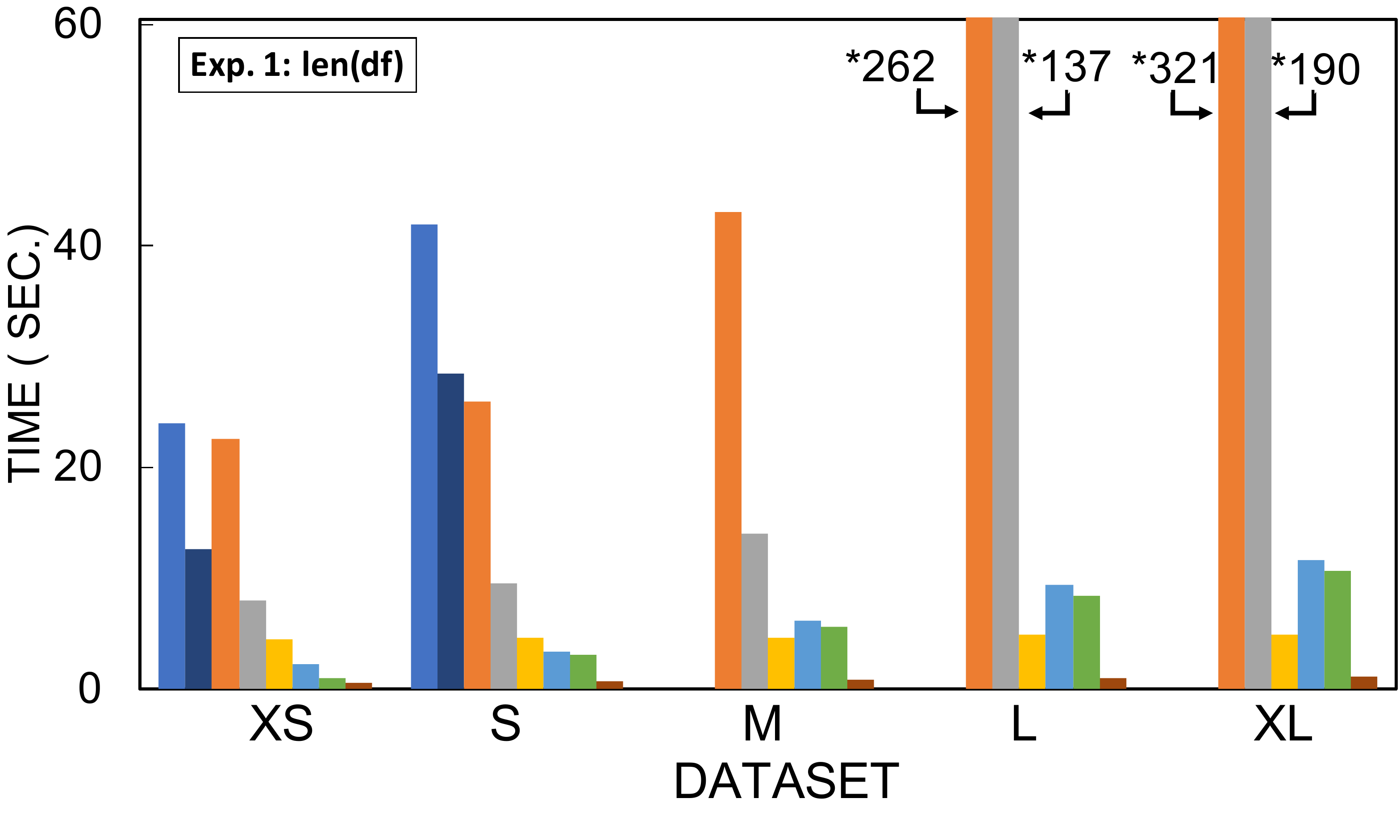}%
        \vspace{-1em}
        \caption{Expression 1 total times}
        \label{fig:q1total}
    \end{subfigure}
    \hspace{0.2cm}
    \begin{subfigure}[t]{0.42\textwidth}
        \includegraphics[trim=1cm 1.5 0 1.5,width=\textwidth,height=4cm]{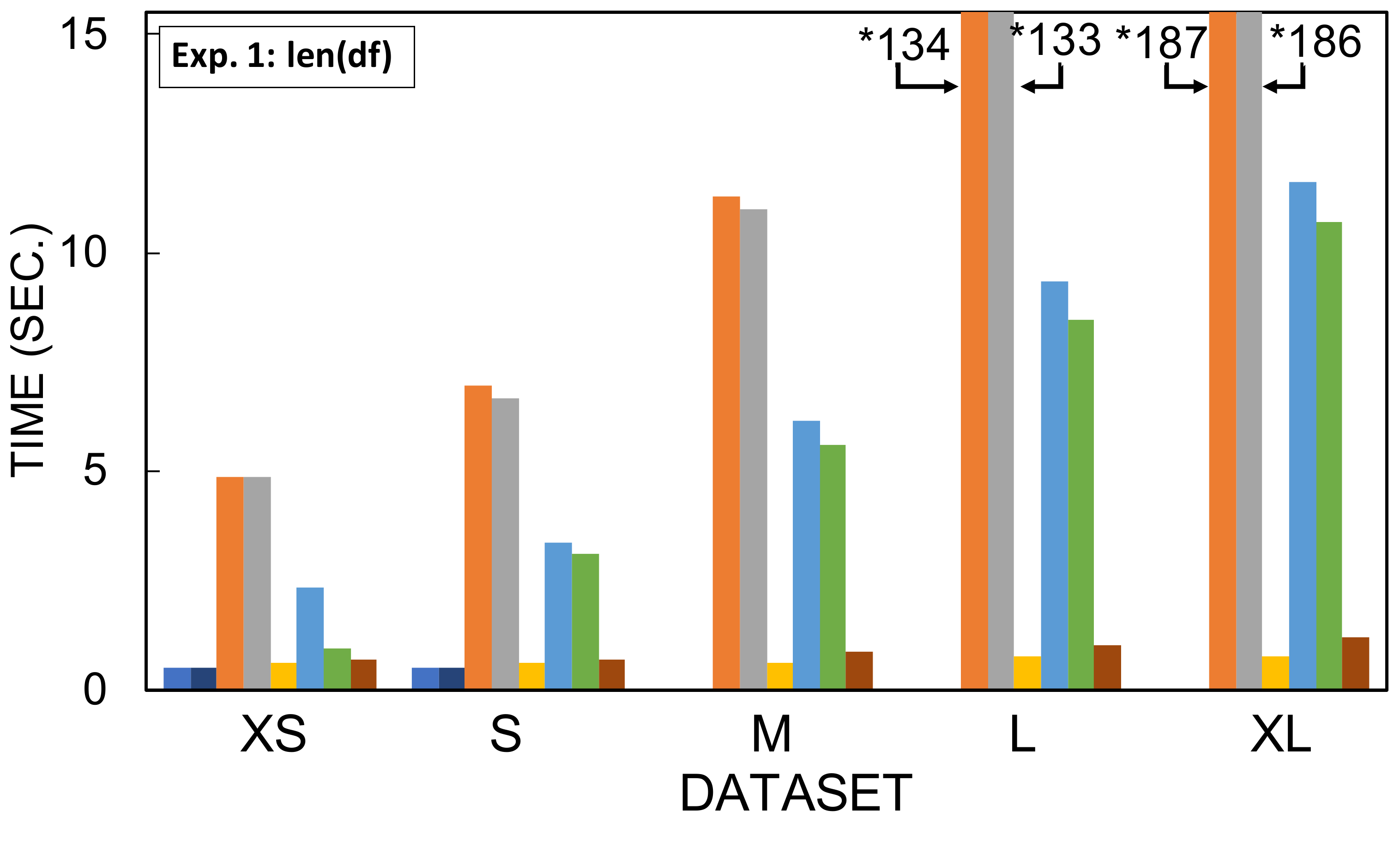}%
        \vspace{-1em}
        \caption{Expression 1 expression-only times}
        \label{fig:q1task}
    \end{subfigure}
    \begin{subfigure}[t]{0.42\textwidth}
        \includegraphics[trim=0.5cm 1.5 0 2,width=\textwidth,height=4cm]{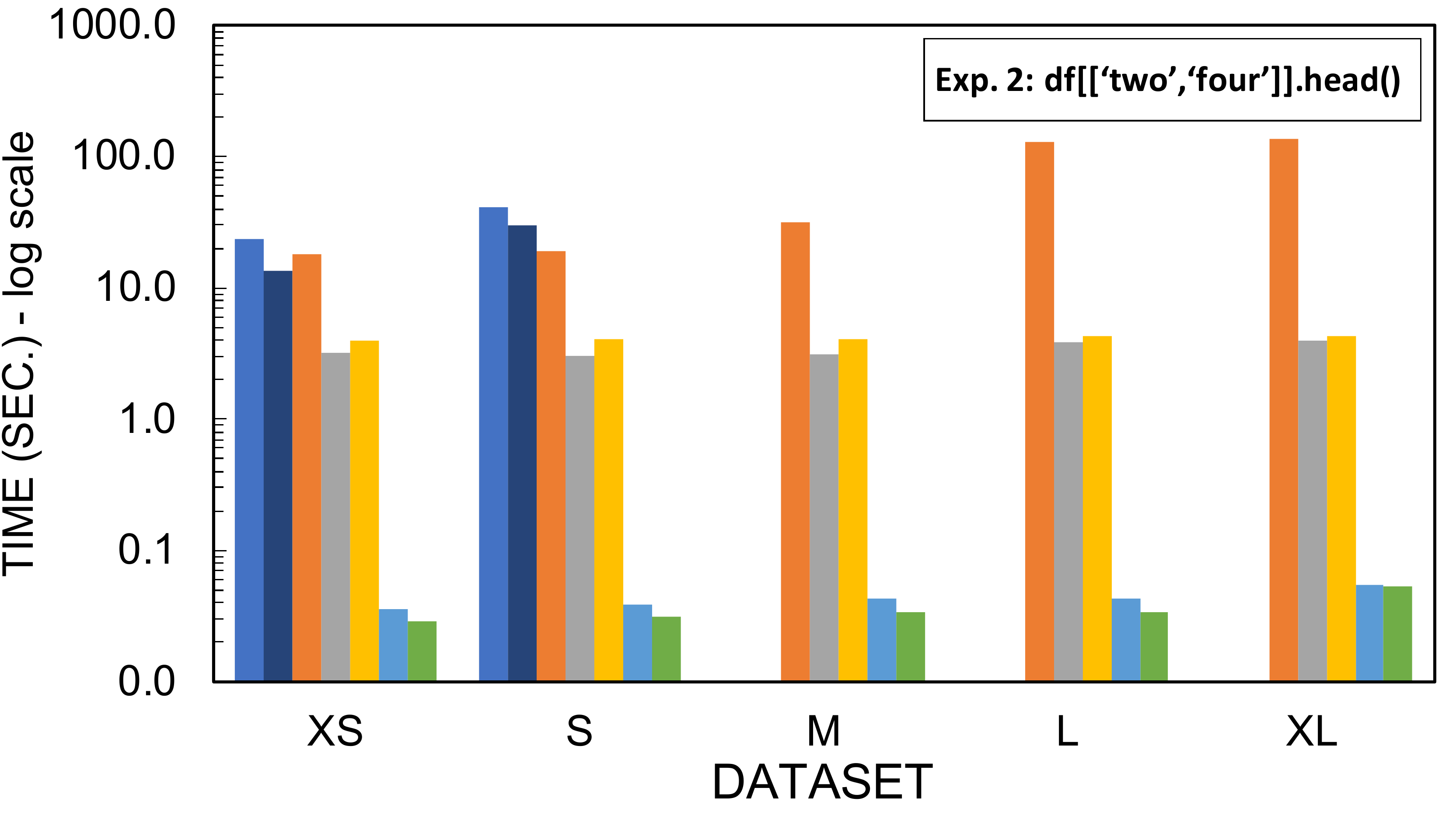}%
        \vspace{-1em}
        \caption{Expression 2 total times}
        \label{fig:q2total}
    \end{subfigure}
    \hspace{0.2cm}
    \begin{subfigure}[t]{0.42\textwidth}
        \includegraphics[trim=1cm 1.5 0 1.5,width=\textwidth,height=4cm]{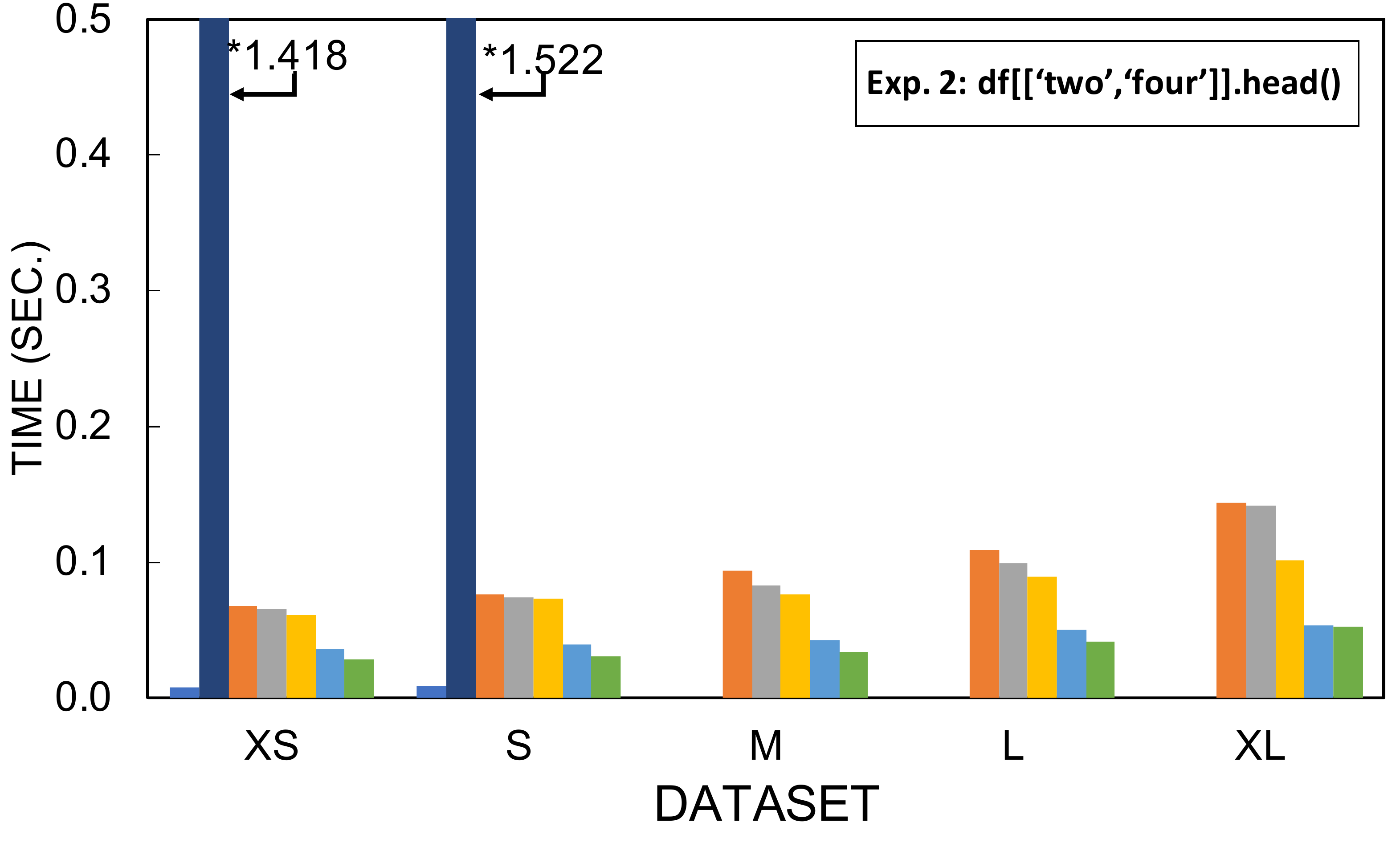}%
        \vspace{-1em}
        \caption{Expression 2 expression-only times}
        \label{fig:q2task}
    \end{subfigure}
    \hfill
    \begin{subfigure}[t]{0.42\textwidth}
        \includegraphics[trim=0.5cm 1.5 0 2,width=\textwidth,height=4cm]{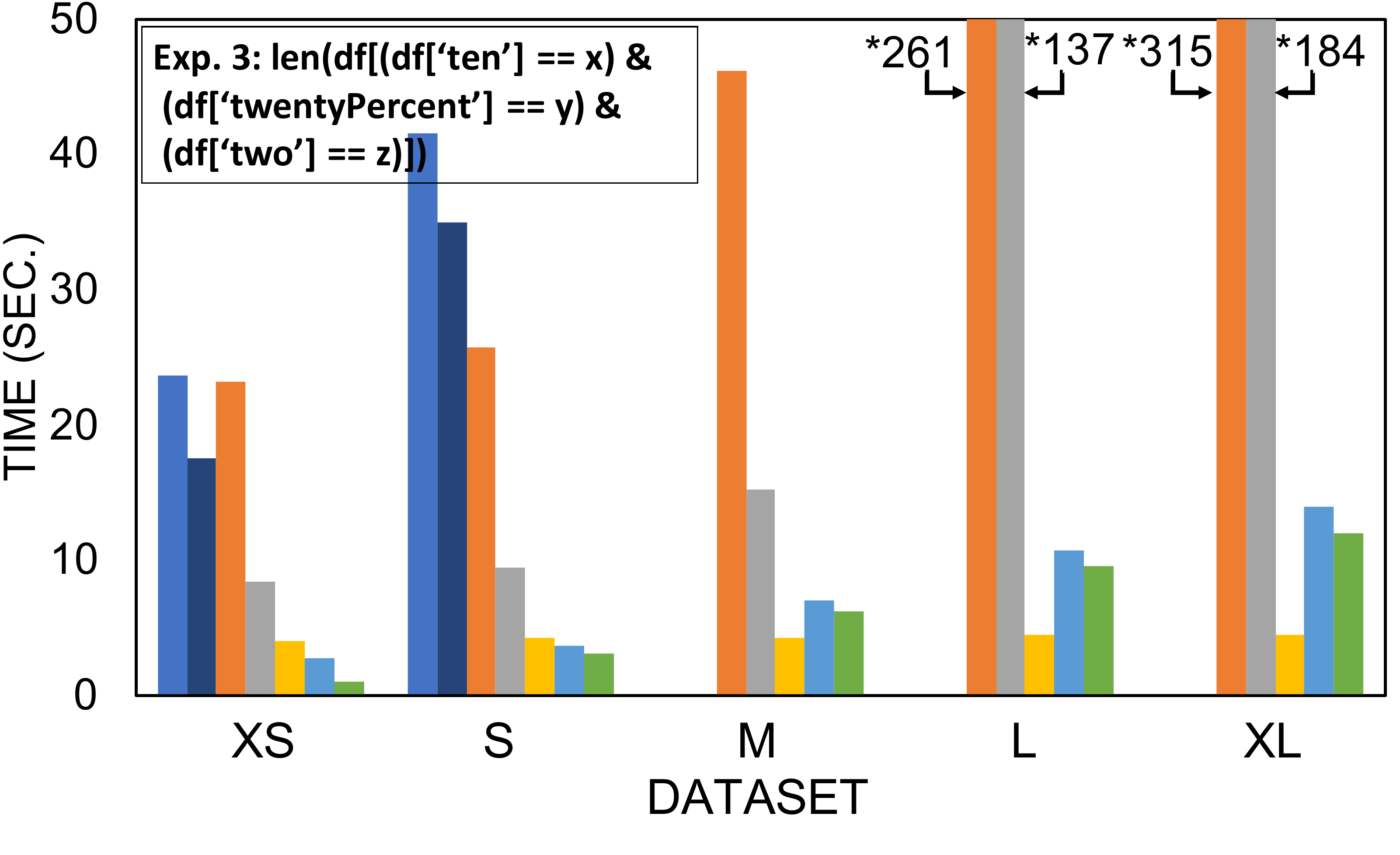}%
        \vspace{-1em}
        \caption{Expression 3 total times}
        \label{fig:q3total}
    \end{subfigure}
    \hspace{0.2cm}
    \begin{subfigure}[t]{0.42\textwidth}
        \includegraphics[trim=1cm 1.5 0 1.5,width=\textwidth,height=4cm]{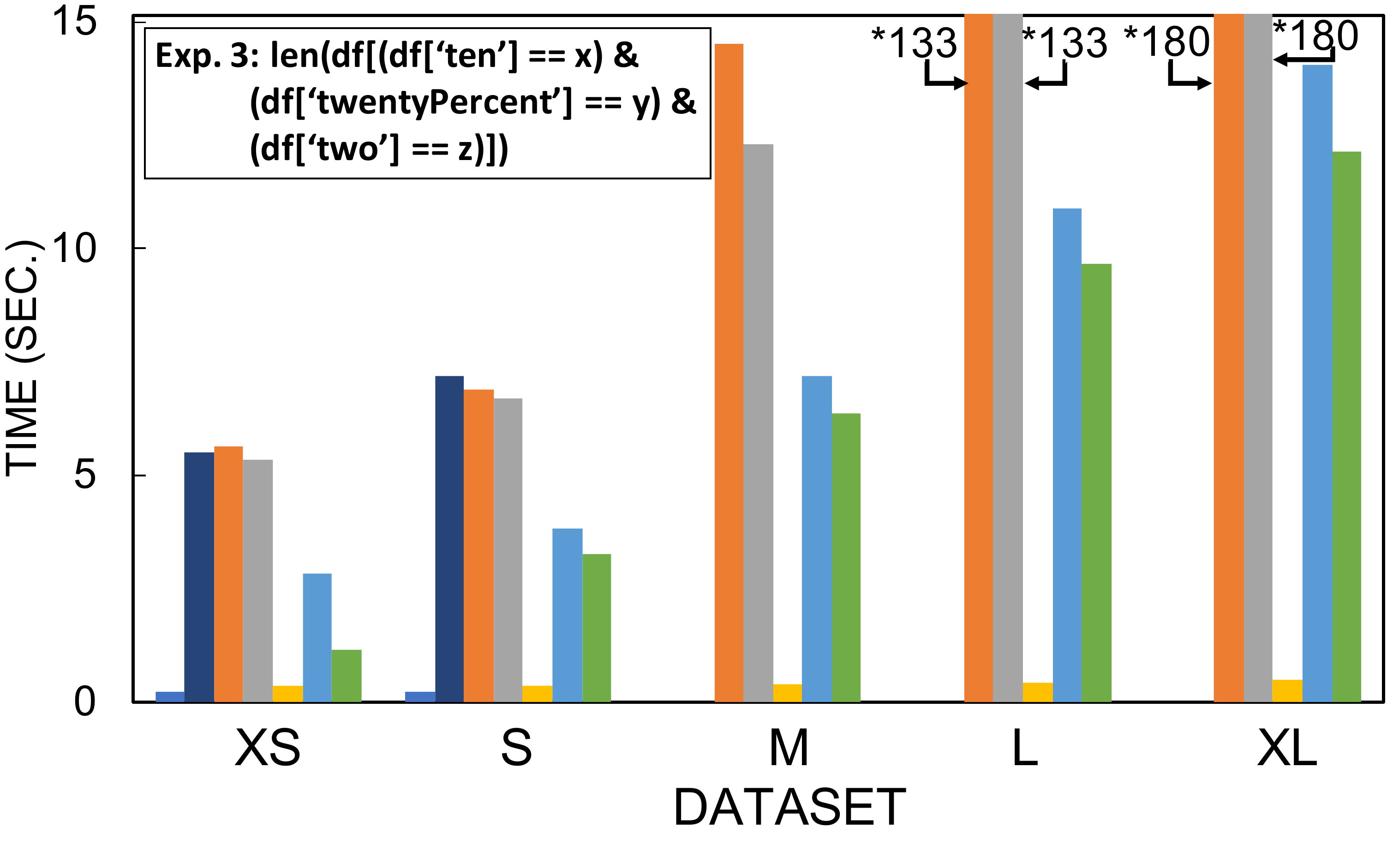}%
        \vspace{-1em}
        \caption{Expression 3 expression-only times}
        \label{fig:q3task}
    \end{subfigure}
    \begin{subfigure}[t]{0.42\textwidth}
        \includegraphics[trim=0.5cm 1.5 0 2,width=\textwidth,height=4cm]{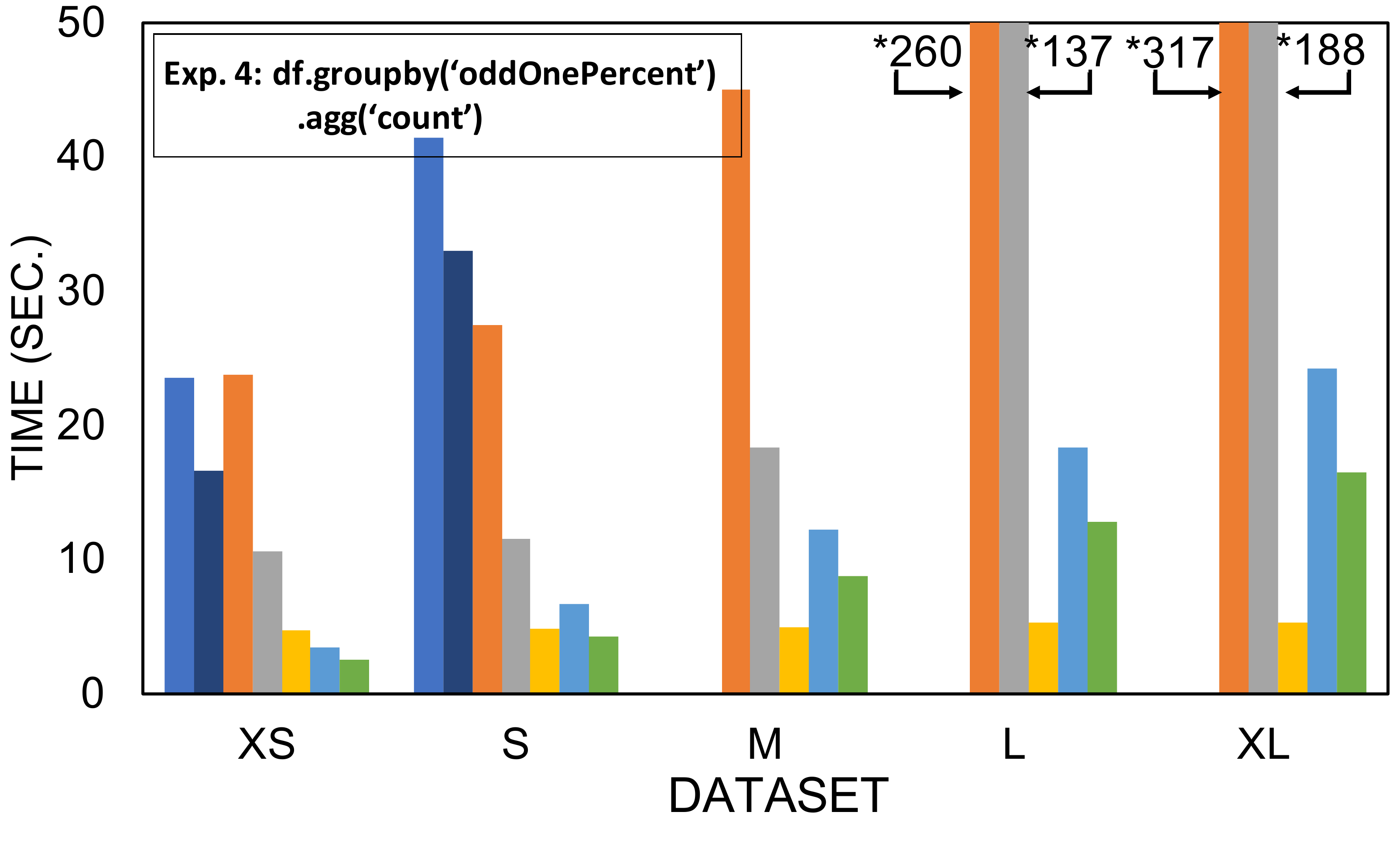}%
        \vspace{-1em}
        \caption{Expression 4 total times}
        \label{fig:q4total}
    \end{subfigure}
    \hspace{0.2cm}
    \begin{subfigure}[t]{0.42\textwidth}
        \includegraphics[trim=1cm 1.5 0 1.5,width=\textwidth,height=4cm]{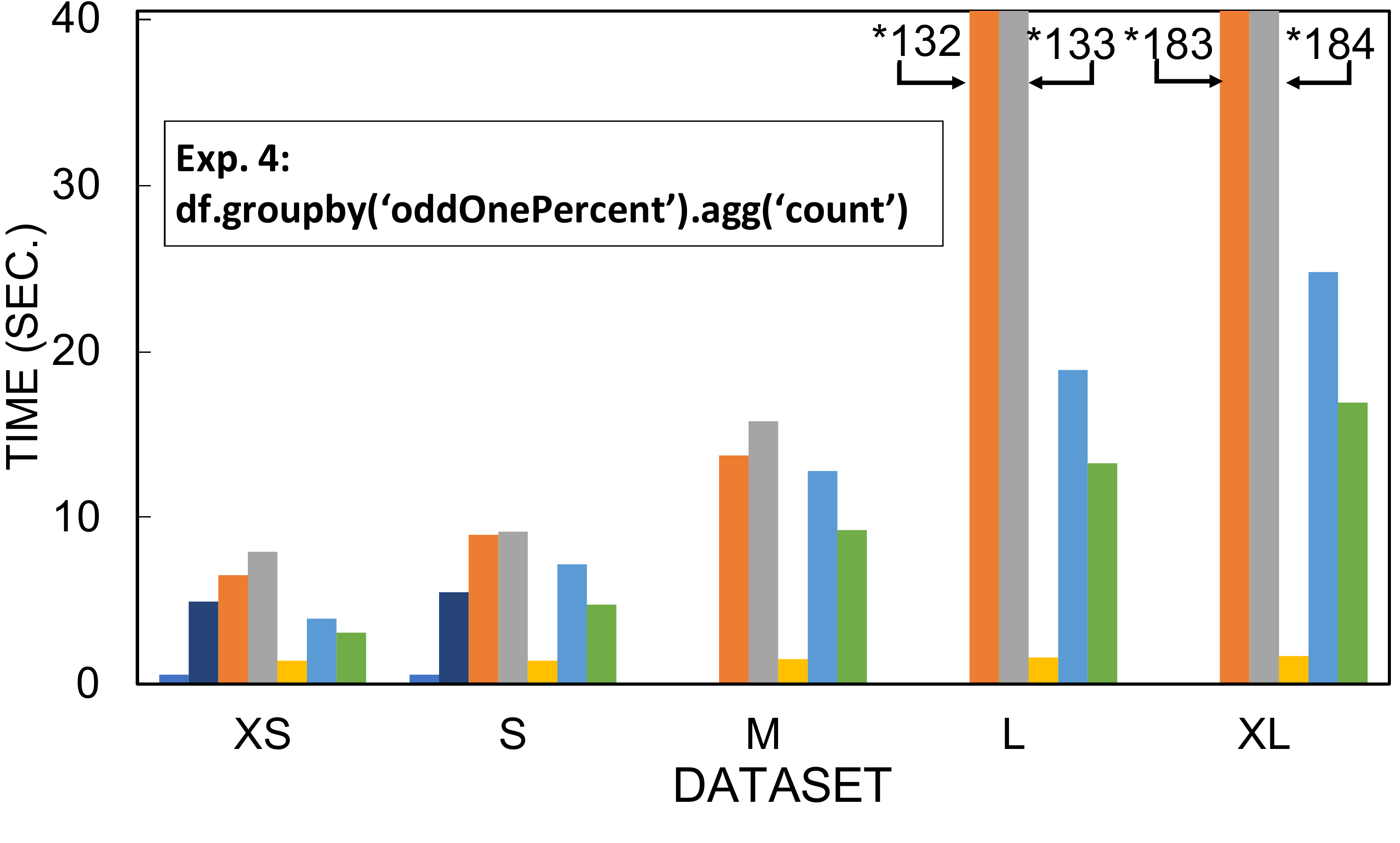}%
        \vspace{-1em}
        \caption{Expression 4 expression-only times}
        \label{fig:q4task}
    \end{subfigure}
    \begin{subfigure}[t]{0.42\textwidth}
        \includegraphics[trim=0.5cm 1.5 0 2,width=\textwidth,height=4cm]{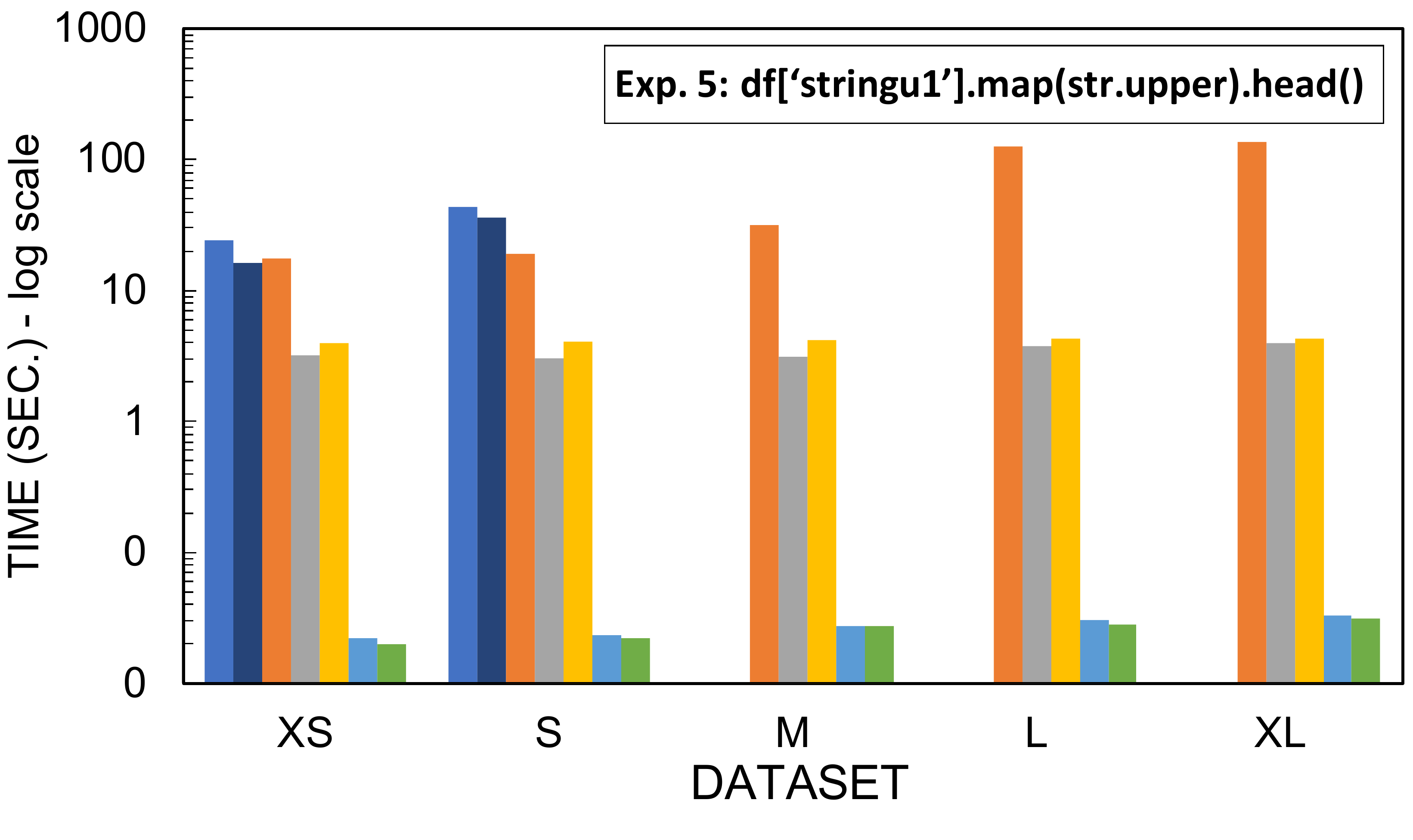}%
        \vspace{-1em}
        \caption{Expression 5 total times}
        \label{fig:q5total}
    \end{subfigure}
    \hspace{0.2cm}
    \begin{subfigure}[t]{0.42\textwidth}
        \includegraphics[trim=1cm 1.5 0 1.5,width=\textwidth,height=4cm]{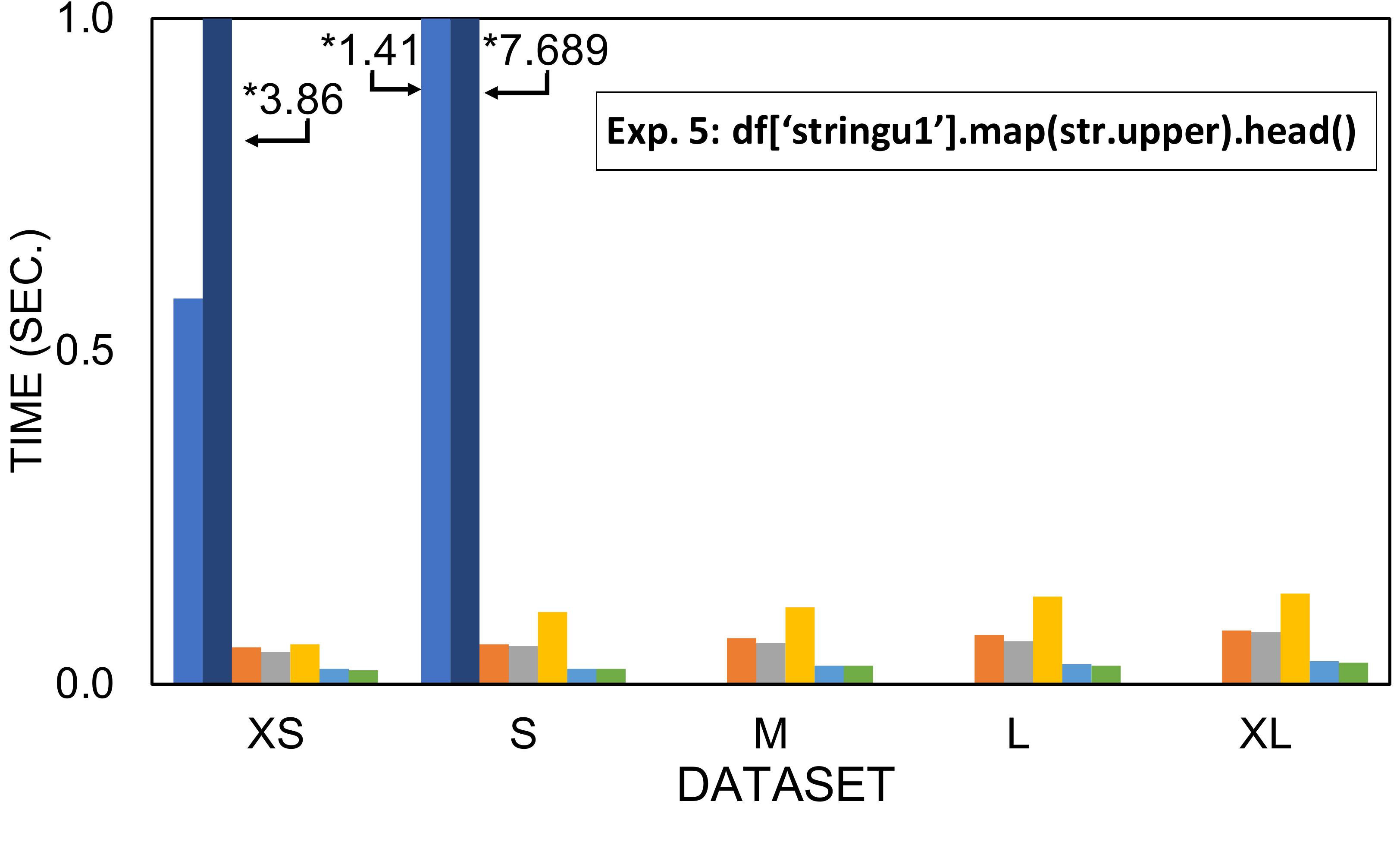}%
        \vspace{-1em}
        \caption{Expression 5 expression-only times}
        \label{fig:q5task}
    \end{subfigure}

    \caption{Single Node Evaluation: Expression 1-5 Results  (* = value where the bar ends)}
    \vspace{-1em}
    \label{fig:single_node_2}
\end{figure*}

\begin{figure*}[!ht]
    \centering
    \begin{subfigure}[t]{0.65\textwidth}
        \includegraphics[trim=0.5cm 1.5 0 2,width=\textwidth,height=1cm]{figures/single_legend_2.pdf}%
    \end{subfigure}
    \begin{subfigure}[t]{0.42\textwidth}
        \includegraphics[trim=0.5cm 1.5 0 2,width=\textwidth,height=4cm]{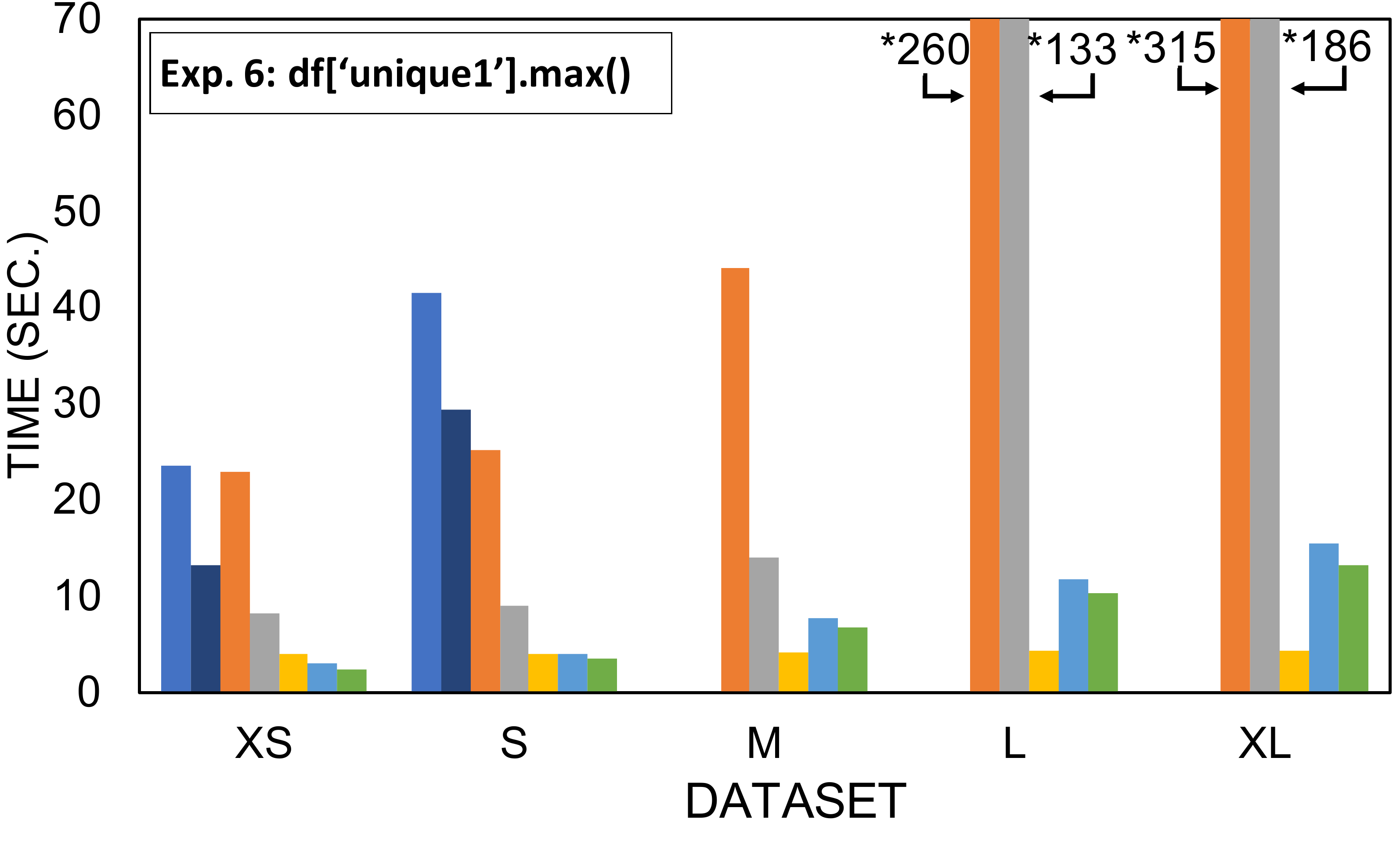}%
        \vspace{-1em}
        \caption{Expression 6 total times}
        \label{fig:q6total}
    \end{subfigure}
    \hspace{0.2cm}
    \begin{subfigure}[t]{0.42\textwidth}
        \includegraphics[trim=1cm 1.5 0 1.5,width=\textwidth,height=4cm]{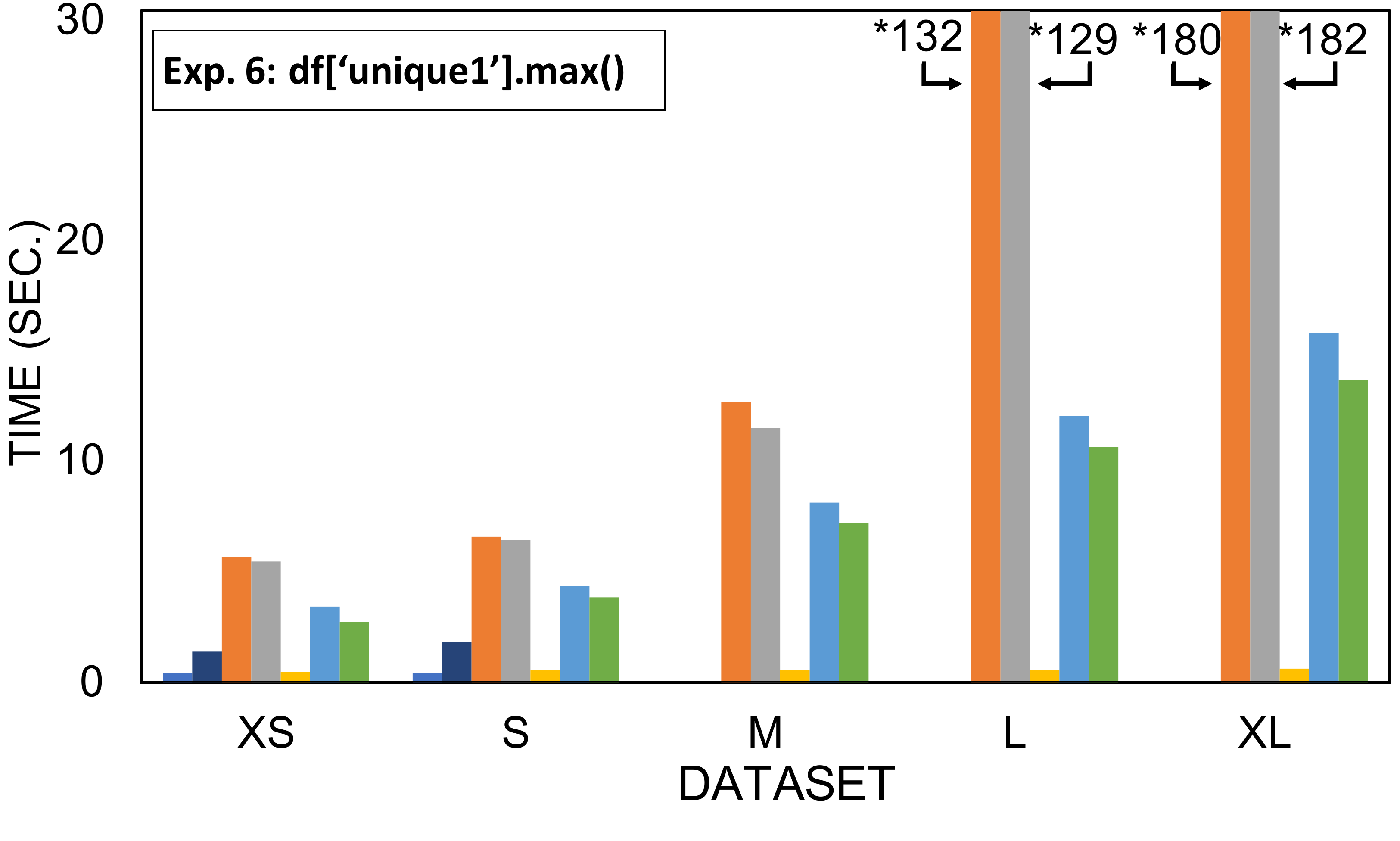}%
        \vspace{-1em}
        \caption{Expression 6 expression-only times}
        \label{fig:q6task}
    \end{subfigure}
    \begin{subfigure}[t]{0.42\textwidth}
        \includegraphics[trim=0.5cm 1.5 0 2,width=\textwidth,height=4cm]{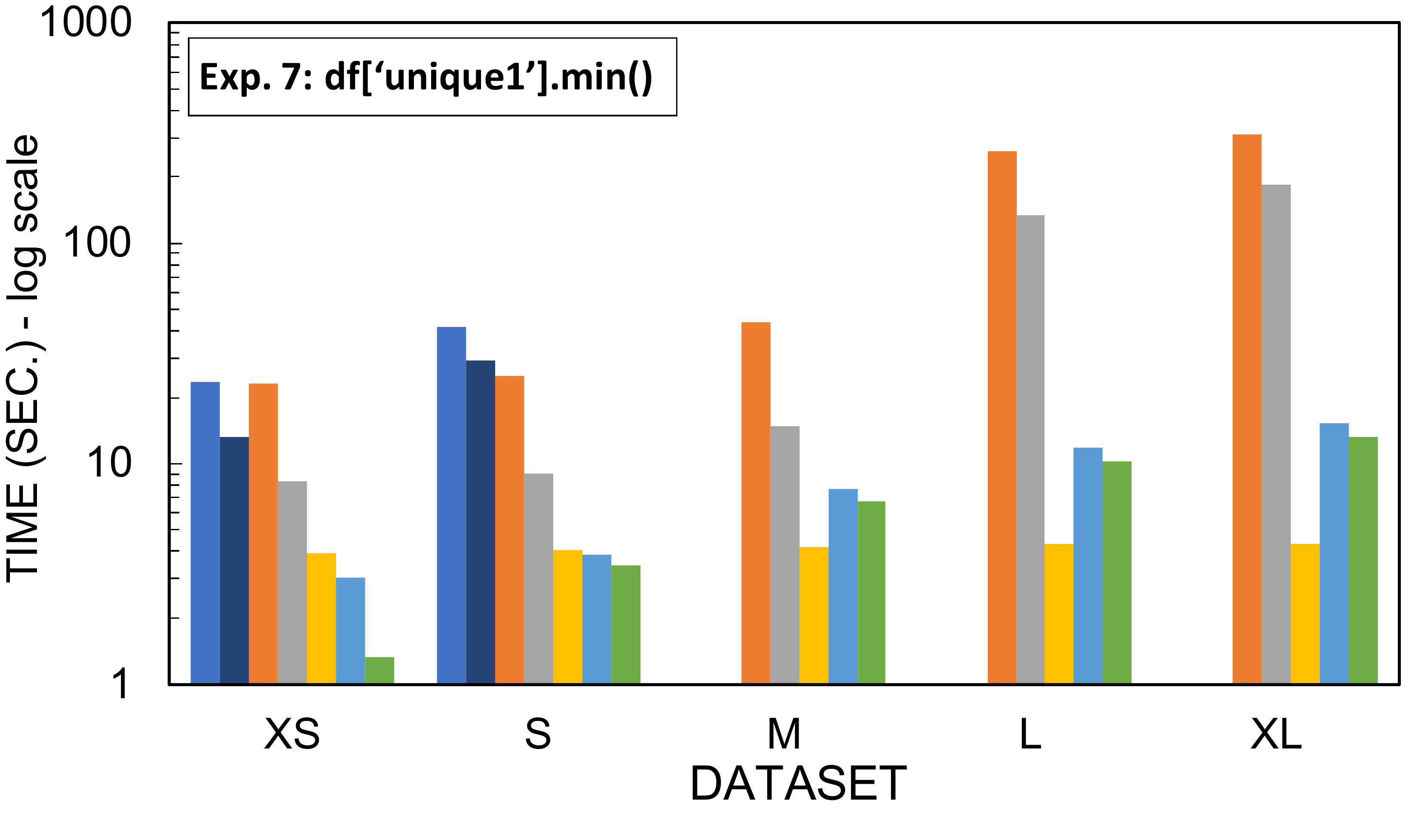}%
        \vspace{-1em}
        \caption{Expression 7 total times}
        \label{fig:q7total}
    \end{subfigure}
    \hspace{0.2cm}
    \begin{subfigure}[t]{0.42\textwidth}
        \includegraphics[trim=1cm 1.5 0 1.5,width=\textwidth,height=4cm]{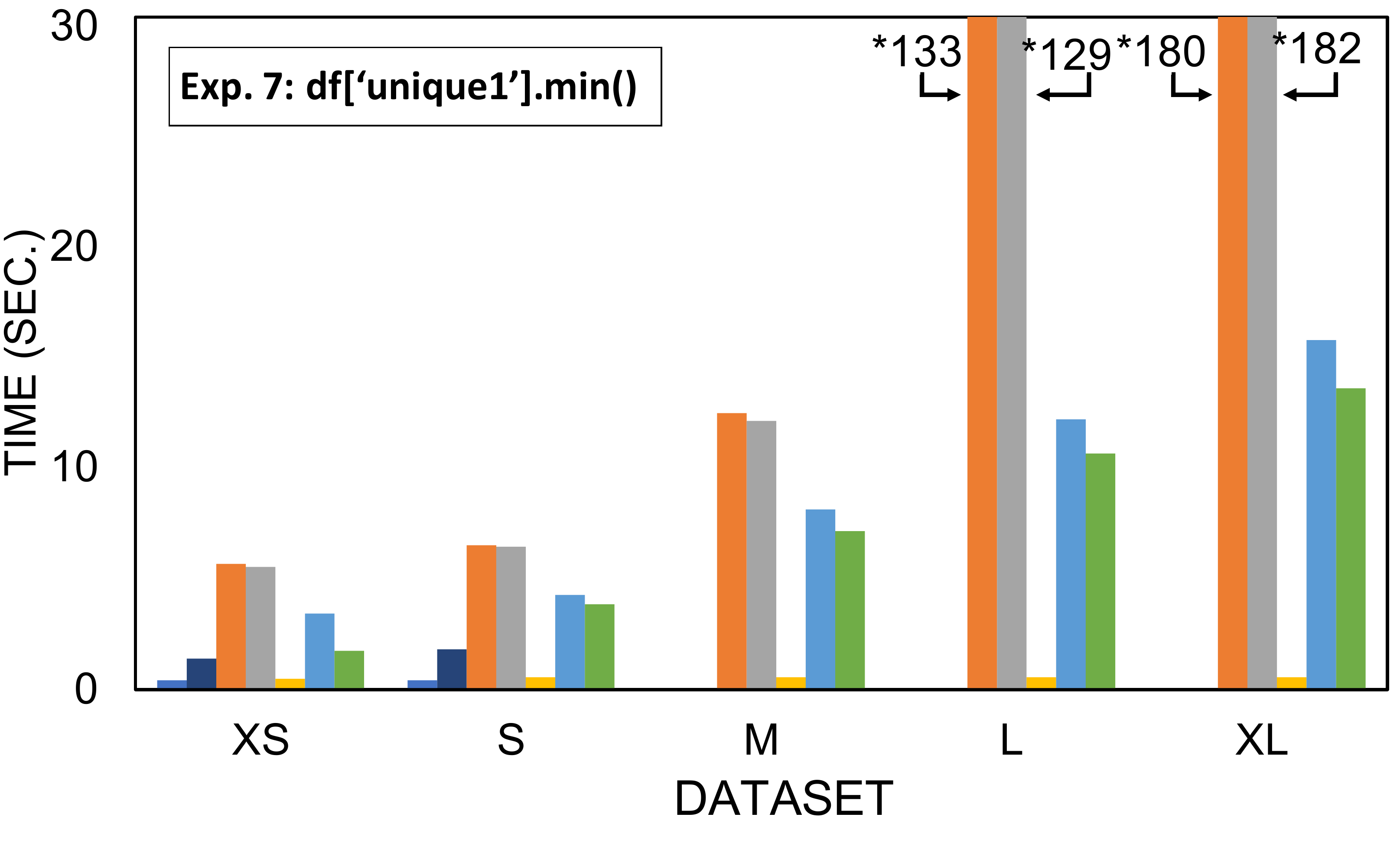}%
        \vspace{-1em}
        \caption{Expression 7 expression-only times}
        \label{fig:q7task}
    \end{subfigure}
    \hfill
    \begin{subfigure}[t]{0.42\textwidth}
        \includegraphics[trim=0.5cm 1.5 0 2,width=\textwidth,height=4cm]{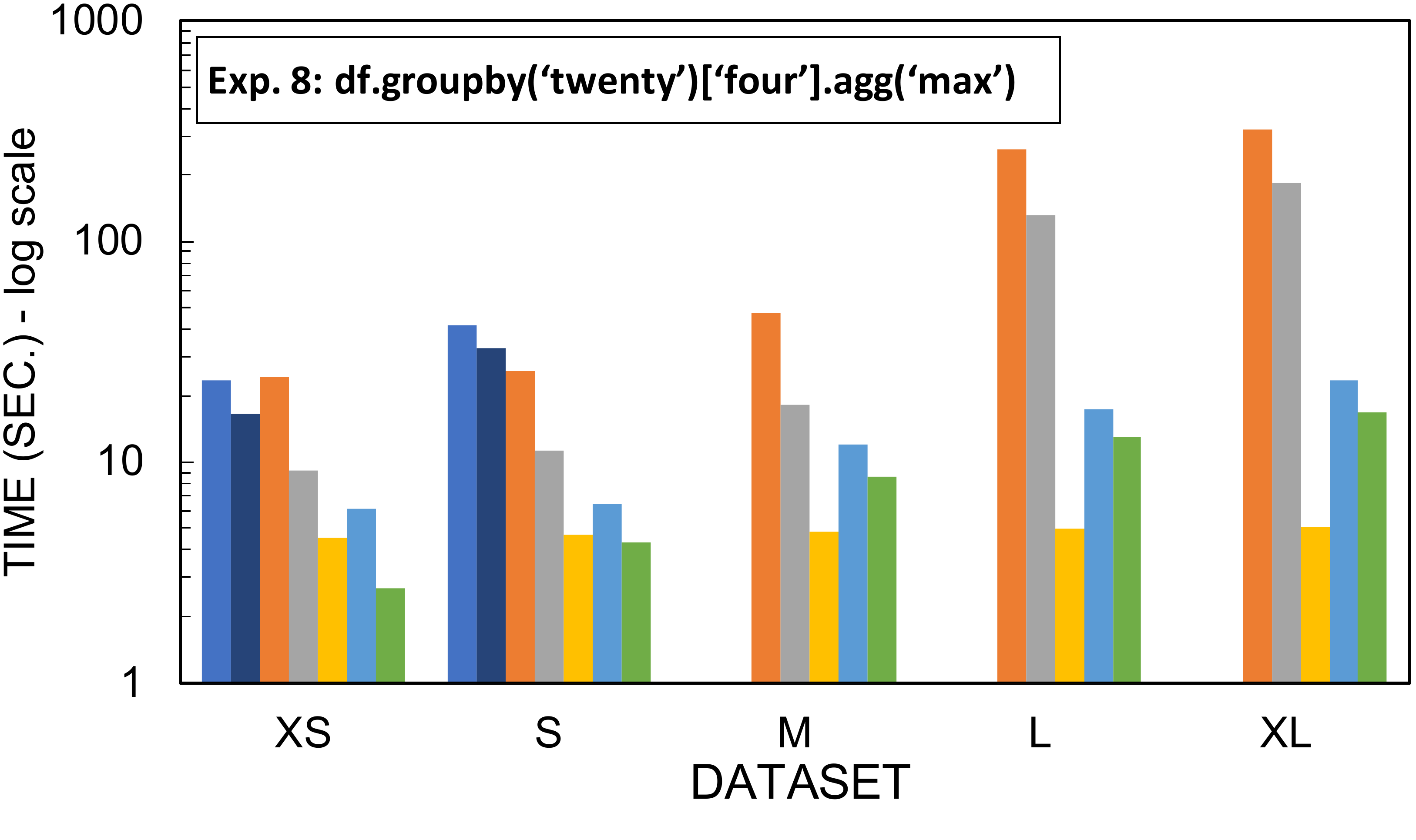}%
        \vspace{-1em}
        \caption{Expression 8 total times}
        \label{fig:q8total}
    \end{subfigure}
    \hspace{0.2cm}
    \begin{subfigure}[t]{0.42\textwidth}
        \includegraphics[trim=1cm 1.5 0 1.5,width=\textwidth,height=4cm]{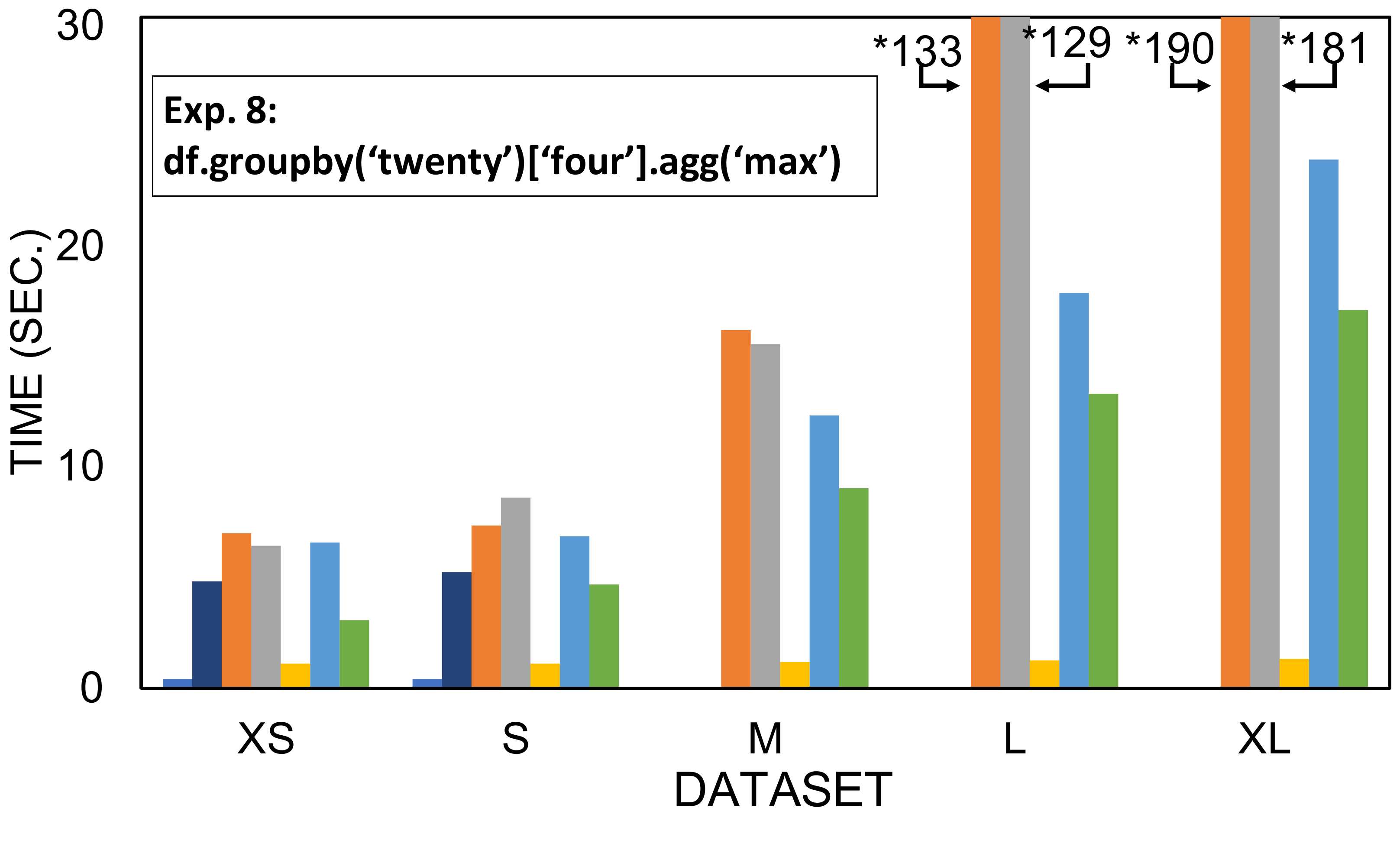}%
        \vspace{-1em}
        \caption{Expression 8 expression-only times}
        \label{fig:q8task}
    \end{subfigure}
    \begin{subfigure}[t]{0.42\textwidth}
        \includegraphics[trim=0.5cm 1.5 0 2,width=\textwidth,height=4cm]{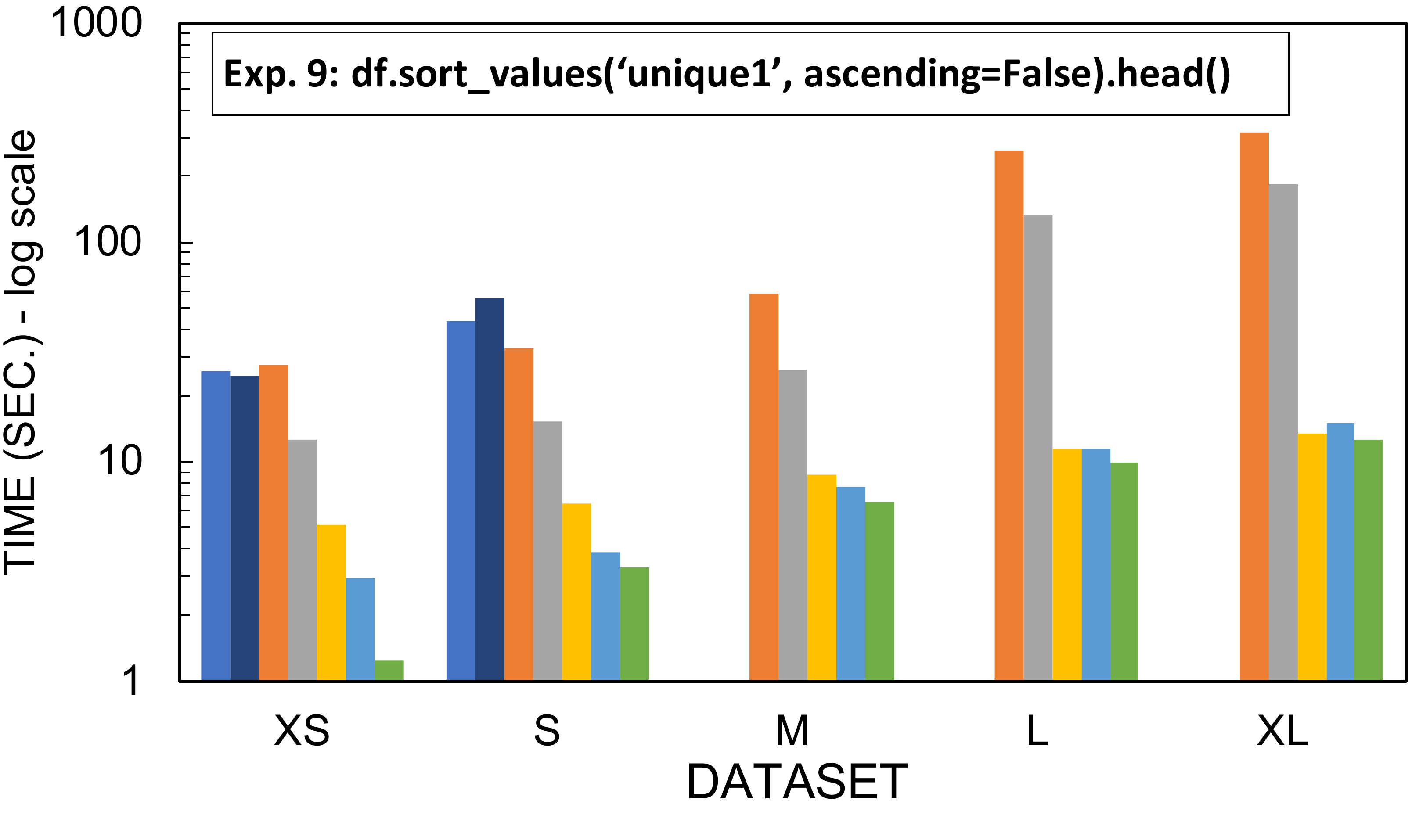}%
        \vspace{-1em}
        \caption{Expression 9 total times}
        \label{fig:q9total}
    \end{subfigure}
    \hspace{0.2cm}
    \begin{subfigure}[t]{0.42\textwidth}
        \includegraphics[trim=1cm 1.5 0 1.5,width=\textwidth,height=4cm]{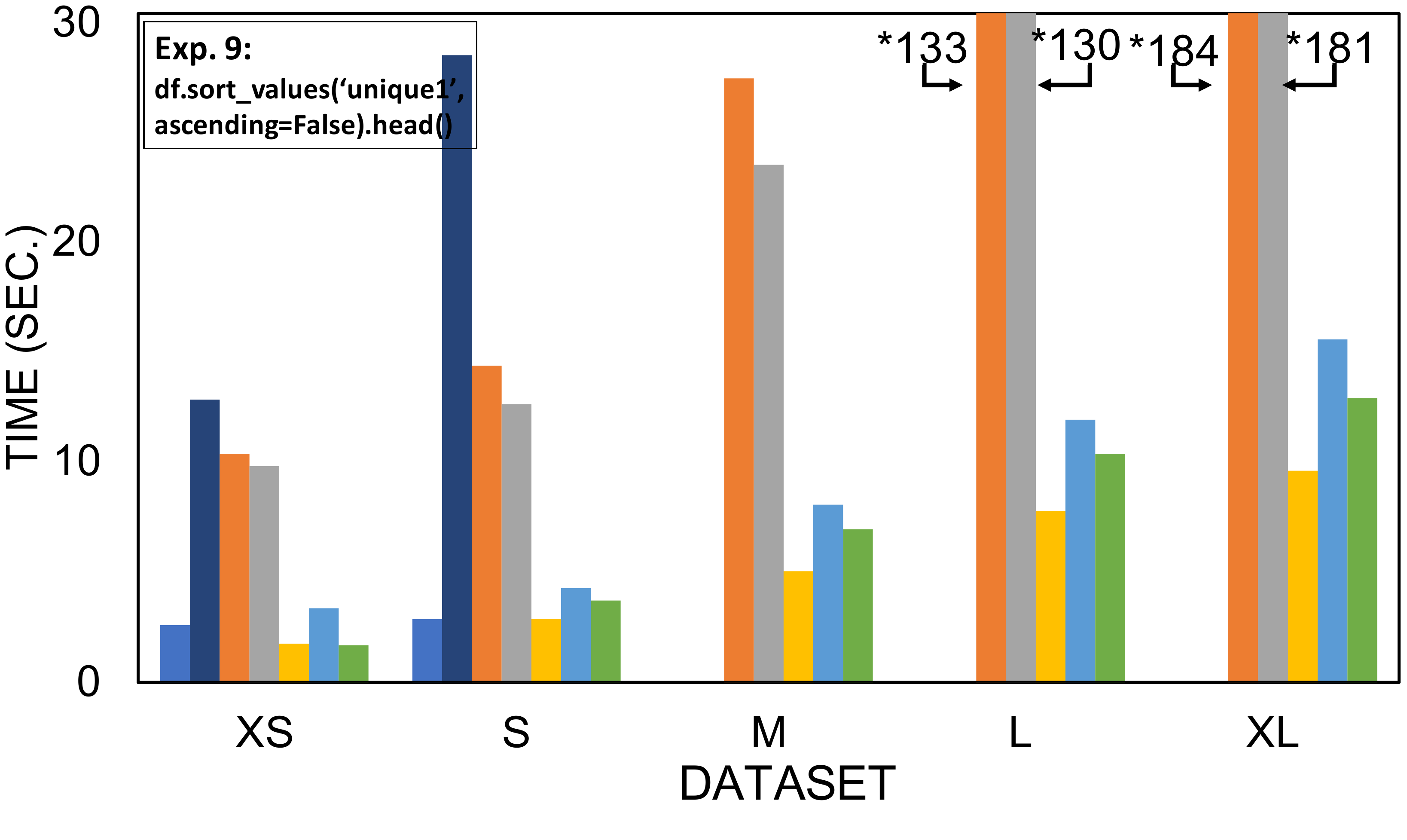}%
        \vspace{-1em}
        \caption{Expression 9 expression-only times}
        \label{fig:q9task}
    \end{subfigure}
    \begin{subfigure}[t]{0.42\textwidth}
        \includegraphics[trim=0.5cm 1.5 0 2,width=\textwidth,height=4cm]{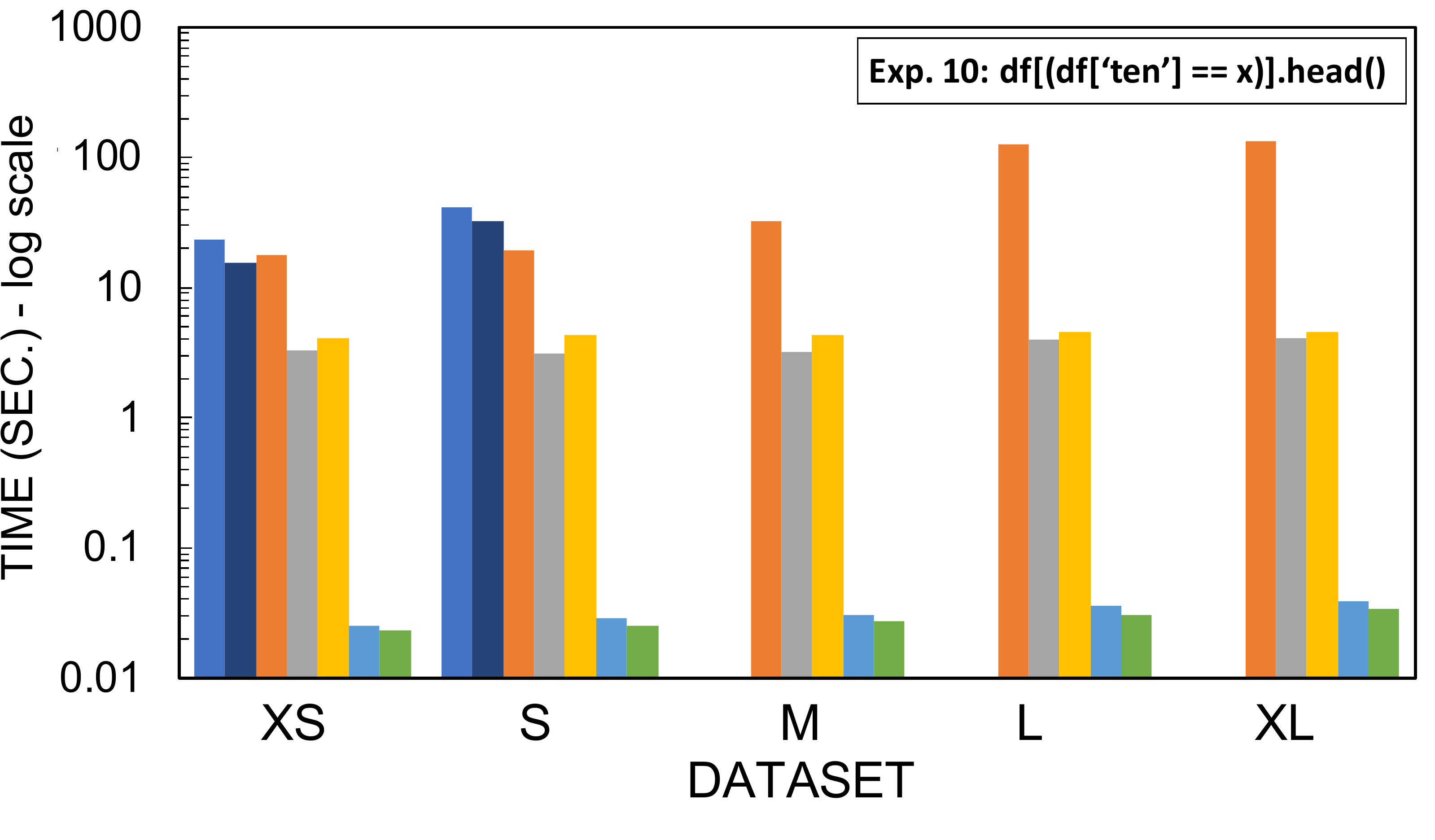}%
        \vspace{-1em}
        \caption{Expression 10 total times}
        \label{fig:q10total}
    \end{subfigure}
    \hspace{0.2cm}
    \begin{subfigure}[t]{0.42\textwidth}
        \includegraphics[trim=1cm 1.5 0 1.5,width=\textwidth,height=4cm]{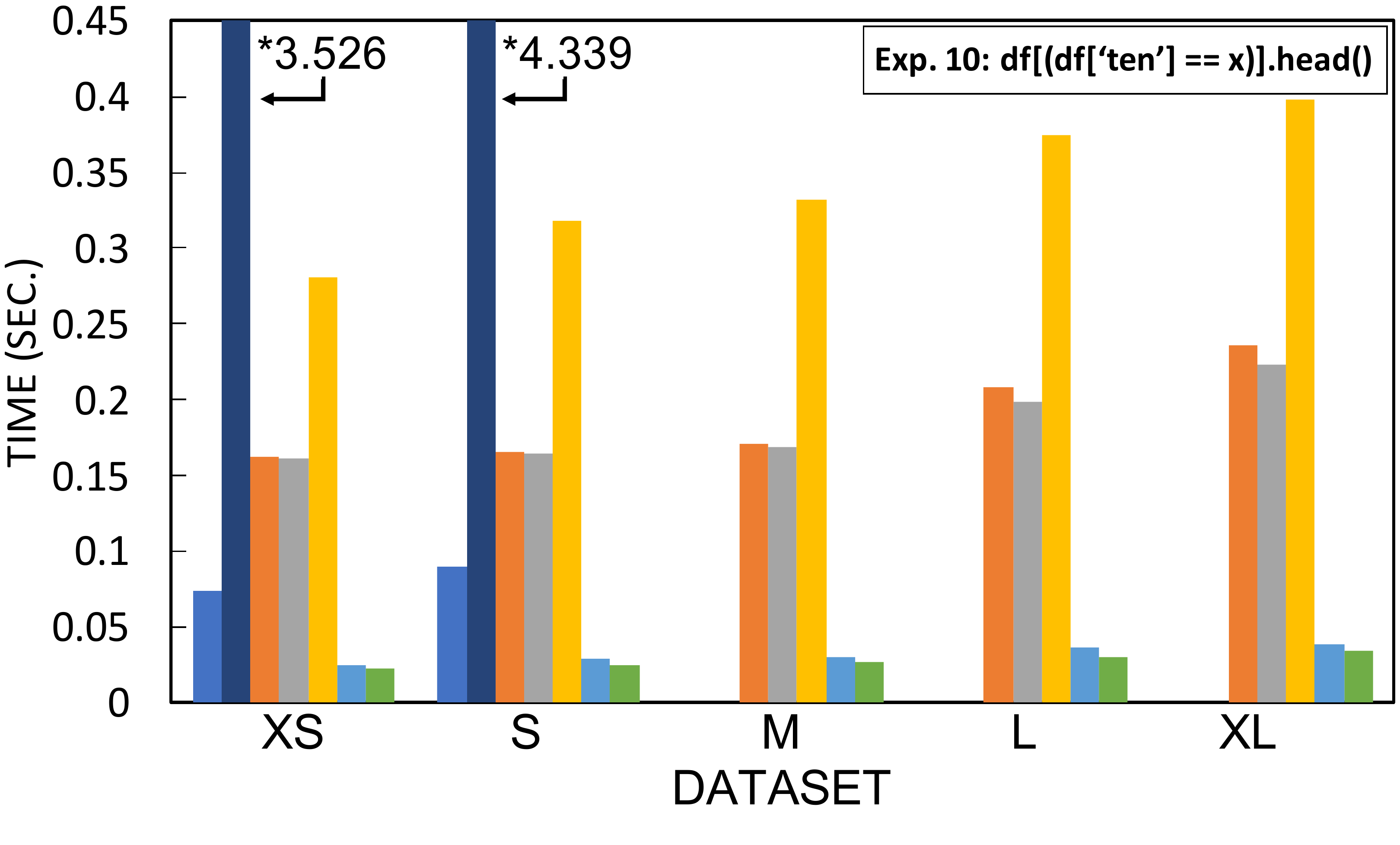}%
        \vspace{-1em}
        \caption{Expression 10 expression-only times}
        \label{fig:q10task}
    \end{subfigure}

    \caption{Single Node Evaluation: Expression 6-10 Results (* = value where the bar ends)}
    \vspace{-1em}
    \label{fig:single_node_6-10}
\end{figure*}

\begin{figure*}[!ht]
    \centering
    \begin{subfigure}[t]{0.65\textwidth}
        \includegraphics[trim=0.5cm 1.5 0 2,width=\textwidth,height=1cm]{figures/single_legend_2.pdf}%
    \end{subfigure}
    \begin{subfigure}[t]{0.42\textwidth}
        \includegraphics[trim=0.5cm 1.5 0 2,width=\textwidth,height=4cm]{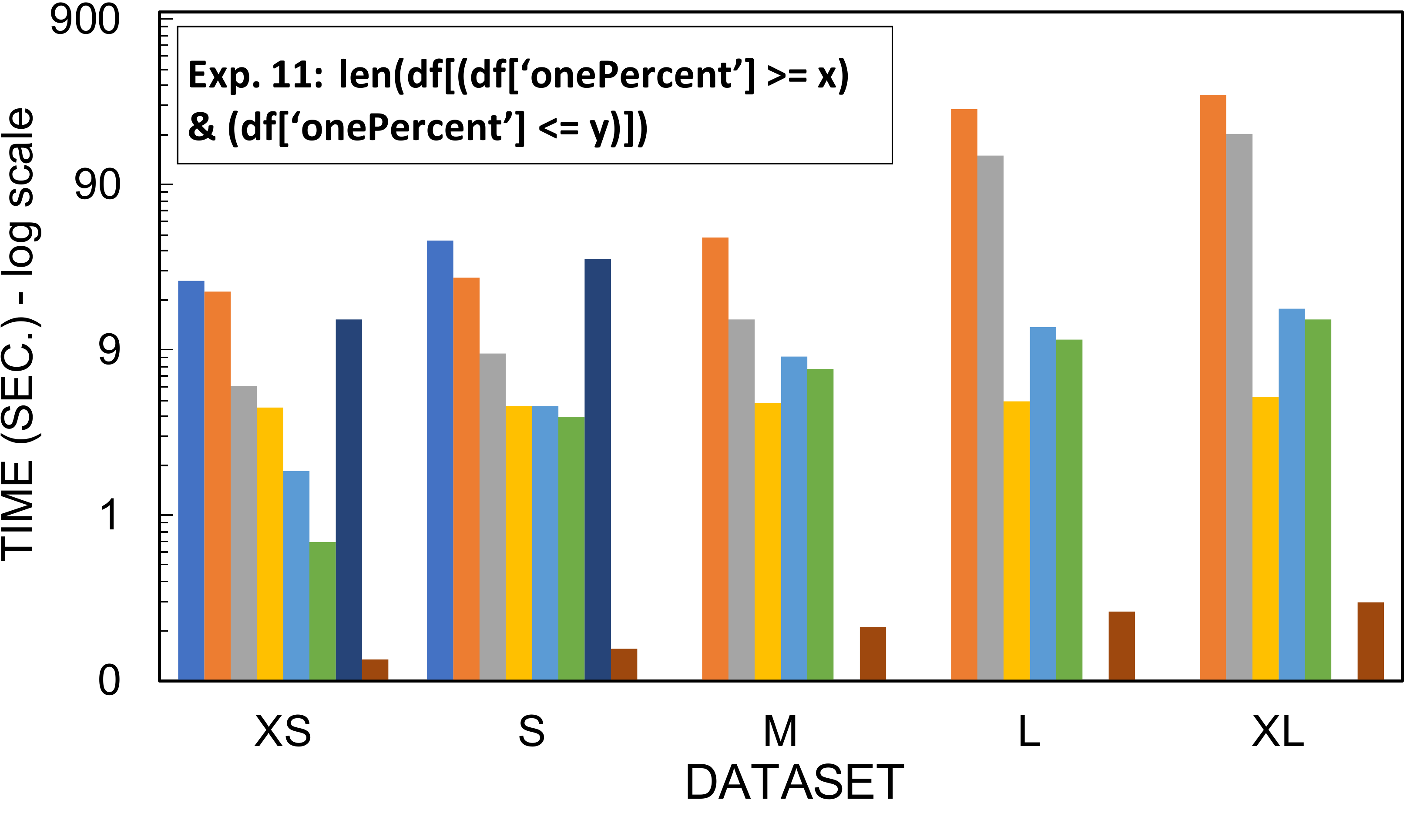}%
        \vspace{-1em}
        \caption{Expression 11 total times}
        \label{fig:q11total}
    \end{subfigure}
    \hspace{0.2cm}
    \begin{subfigure}[t]{0.42\textwidth}
        \includegraphics[trim=1cm 1.5 0 1.5,width=\textwidth,height=4cm]{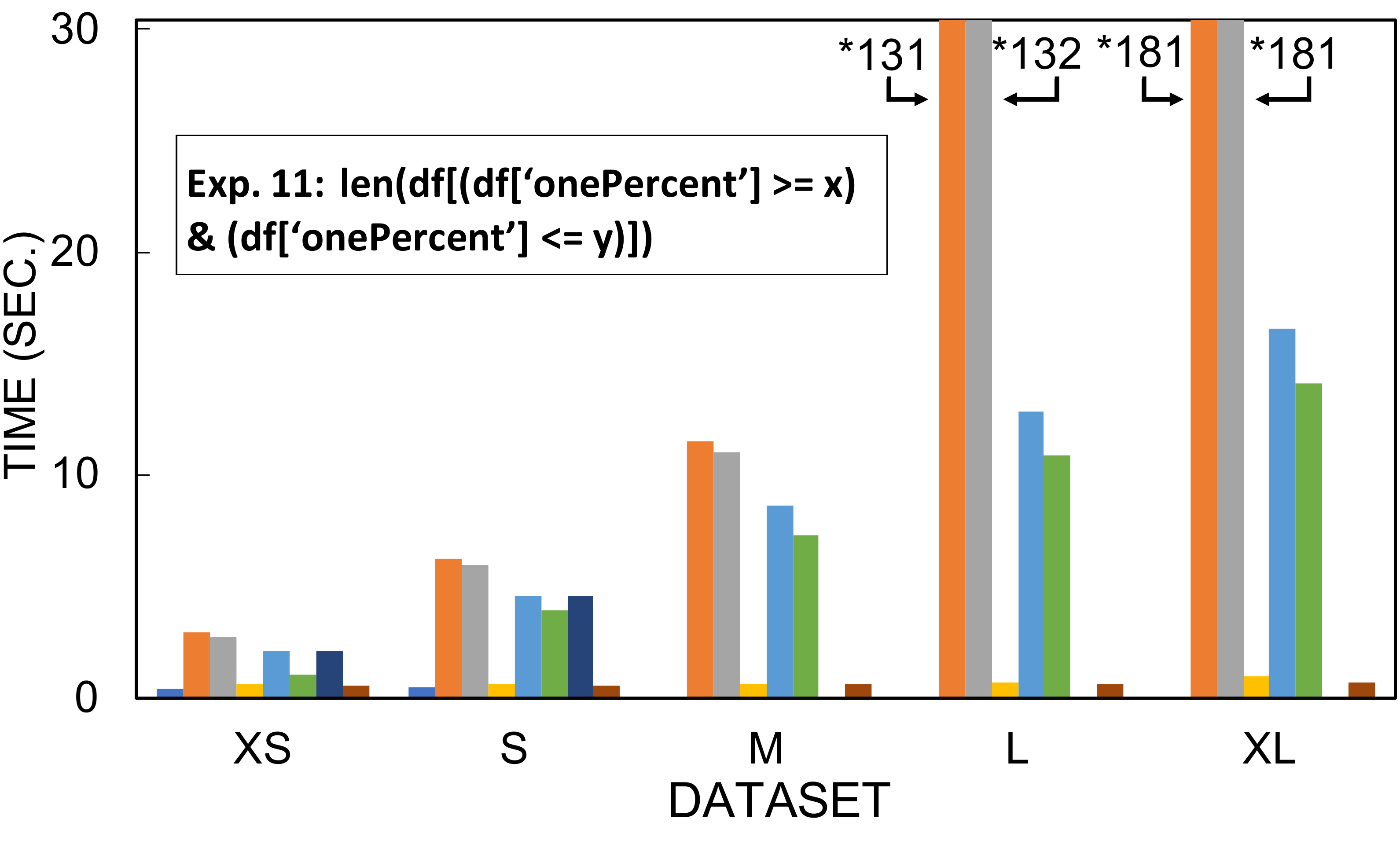}%
        \vspace{-1em}
        \caption{Expression 11 expression-only times}
        \label{fig:q11task}
    \end{subfigure}
    \begin{subfigure}[t]{0.42\textwidth}
        \includegraphics[trim=0.5cm 1.5 0 2,width=\textwidth,height=4cm]{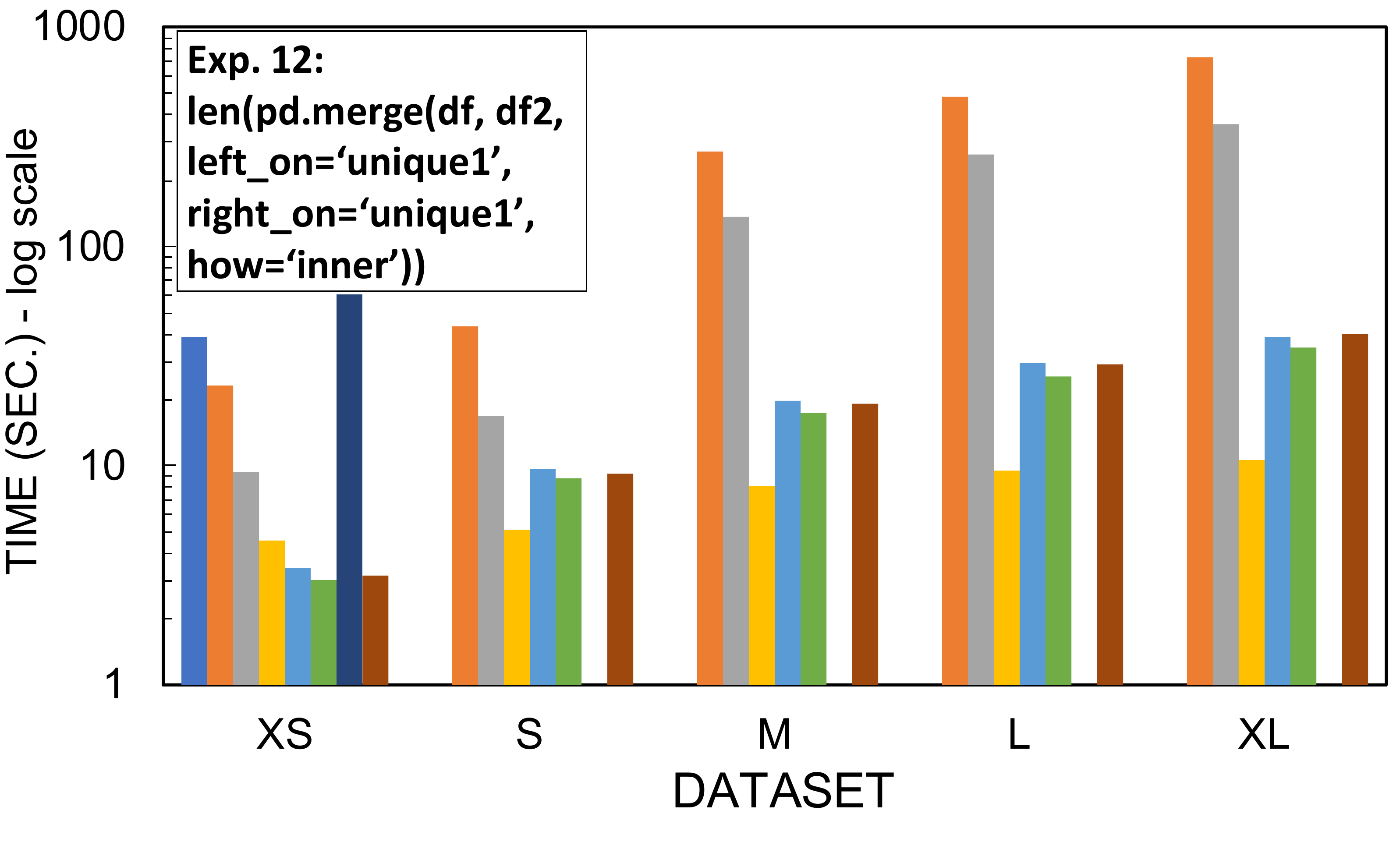}%
        \vspace{-1em}
        \caption{Expression 12 total times}
        \label{fig:q12total}
    \end{subfigure}
    \hspace{0.2cm}
    \begin{subfigure}[t]{0.42\textwidth}
        \includegraphics[trim=1cm 1.5 0 1.5,width=\textwidth,height=4cm]{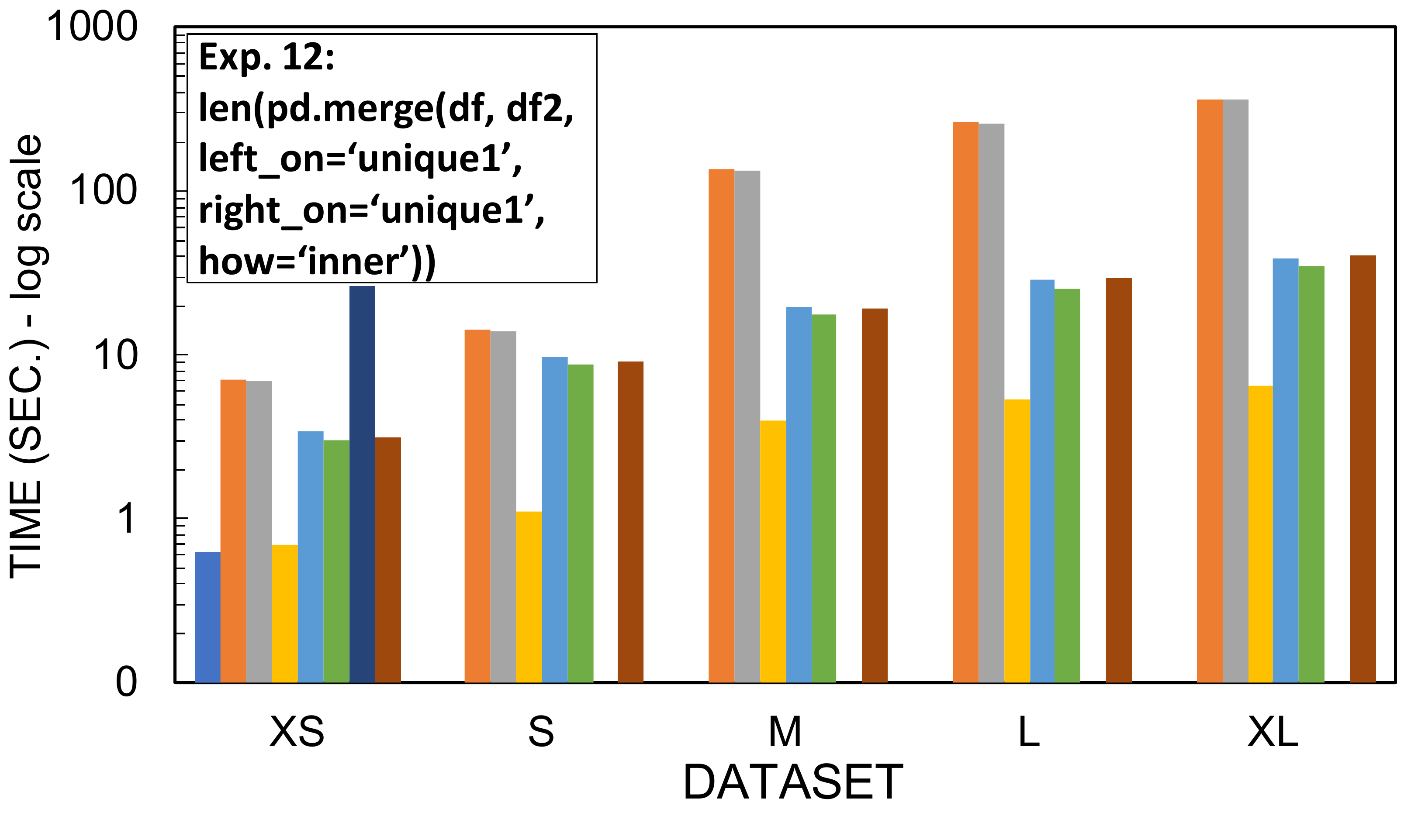}%
        \vspace{-1em}
        \caption{Expression 12 expression-only times}
        \label{fig:q12task}
    \end{subfigure}
    \caption{Single Node Evaluation: Expression 11-12 Results (* = value where the bar ends)}
    \vspace{-1em}
    \label{fig:single_node_11-12}
\end{figure*}

After the first evaluation round, we evaluated the systems' single-node data scalability by running each expression on all five different data sizes. As we can see from Figures~\ref{fig:single_node_2}, \ref{fig:single_node_6-10} and \ref{fig:single_node_11-12}, Pandas and Pandas on Ray were not able to complete the DataFrame creation process for the M-XL datasets (5-10 GB) due to insufficient memory. A possible workaround would be to load the data in smaller chunks; we did not consider applying this workaround because it would result in customizing the data chunk size and that would directly affect the performance evaluation. Pandas on Ray suffers from the same memory limitations as Pandas since it uses Pandas internally.

The results of our single node scalability evaluation are largely consistent with those from our first run of functionality checking. There are some interesting results in the cases of running Spark on L and XL datasets, which are 7.5 and 10 GB of JSON data (e.g., Figures \ref{fig:q1total}, \ref{fig:q1task}, \ref{fig:q3total}, \ref{fig:q3task}, \ref{fig:q4total},\ref{fig:q4task}, etc.). These results are much slower than the other datasets in terms of both the total and expression-only elapsed times. These results are explained by Spark's default settings and its memory management policy. By default, Spark reserves one GB less than the available memory (MAX\_MEMORY - 1) for its executor's memory. In our case this results in 7 GBs of memory being reserved for the executor tasks. When working with data that is larger than the available memory, Spark processes it in partitions and spills data to disk if it has insufficient memory. The L and XL datasets require Spark to spill to disk in order to complete the tasks, which results in long task execution times. In the Spark JSON case, providing a schema when creating a DataFrame from JSON files allows Spark to completely skips the schema inference step. This results in a lower total run time than when a schema is not provided. However, excluding the DataFrame creation time, whether or not the schema was provided, there was no significant performance difference between Spark JSON and Spark JSON Schema across all expressions.

In contrast to JSON, Spark's Parquet-based DataFrame performance results were consistent throughout all data sizes because the Parquet files are much smaller than the JSON files used to generate them. Since Parquet is supplied with a data schema and is a column-oriented format, it is especially suitable for column-based queries such as attribute projections. One factor to keep in mind is that even the Parquet-based DataFrame requires some DataFrame creation overhead. Figure~\ref{fig:q3total} displays the total elapsed time for expression 3, which asks for the count of records that satisfy column conditions. We can see that for the XS and S datasets, the Parquet-based DataFrame total time results were slower than AFrame. However, as the data size increases and the task processing time becomes more prominent, the Parquet-based DataFrame starts to have a better run time than AFrame. The Spark Parquet-based DataFrame starts to benefit when the operation time exceeds the DataFrame creation time. In turn, for the expressions that require access to whole records, such as expressions 5 and 10, as seen in Figures~\ref{fig:q5task} and \ref{fig:q10task}, Spark's JSON-based DataFrame performed significantly better than its Parquet-based (columnar) DataFrame. Even in the case that includes the DataFrame creation time, shown in Figures~\ref{fig:q5total} and \ref{fig:q10total}, Spark's JSON-based DataFrame with a pre-defined schema was faster than Parquet for all data sizes for expressions 5 and 10.

AFrame benefits from database optimizations like query planning and indexing. For expression 1, which asks for a total record count, AFrame with a primary key index performed the best for all data sizes. Similary, in expression 11 as shown in Figures~\ref{fig:q11total} and \ref{fig:q11task}, AFrame with an index on the range attribute was the fastest in the total time case and was competitive in the expression only case as well. 

AFrame also benefits from having indexes on the join attributes (Expression 12), as shown earlier in Figure~\ref{fig:6-10total}; also as the size of the dataset gets larger, the others suffer more from long DataFrame creation times because they have to scan an additional dataset for this expression. For this expression, both datasets are identical in content and size. As we can see from Figures~\ref{fig:q12total} and \ref{fig:q12task}, Pandas and Pandas on Ray even failed to complete DataFrame creation on dataset S as they had to load twice as much data. It is important to note that equality-based joins in AsterixDB default to use hybrid hash join algorithm (as it has good cost characteristics when joining large datasets). For AFrames without index (labeled AFrame and AFrame Schema), the join method used in expression 12 was a hash join algorithm while AFrame Index benefited from index nested-loop join. Hash joins are more efficient for large datasets, which is the reason why we started to see AFrame outperforming AFrame Index on dataset XL as probing a hash table by scanning the dataset in its entirety once could be faster than performing too many index lookups and traversing a b-tree index.


AFrame was faster than Spark's JSON-based DataFrames in most of the test cases in Figure~\ref{fig:single_node_1} and continued to be so as shown in Figures~\ref{fig:single_node_2}, \ref{fig:single_node_6-10}, and \ref{fig:single_node_11-12} for both expression-only and total times evaluations. AFrame without indexes was slower than Spark Parquet in most of the column-based expression-only times. However, for whole-row-based expressions, such as expression 10 (Figures~\ref{fig:q10total} and \ref{fig:q10task}), AFrame without indexes performed better than Spark Parquet and were the best for both the expression-only and total run time evaluations. 

%% file: multinode_speedup.tex
\begin{figure*}[ht!]
     \centering
    \begin{subfigure}[t]{0.45\textwidth}
        \includegraphics[trim=0 1.5 0 1.5,width=\textwidth,height=0.7cm]{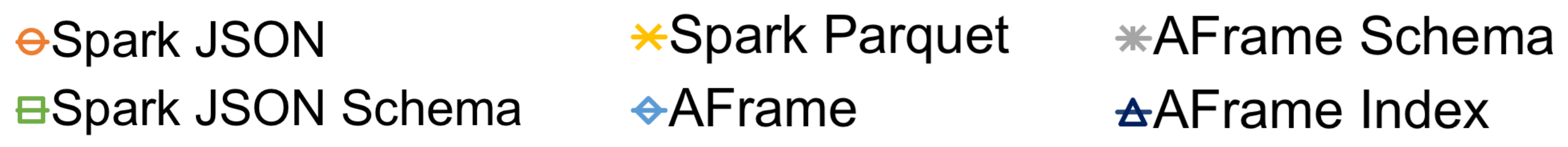}
    \end{subfigure}
    \hspace{15cm}
    \begin{subfigure}[t]{0.24\textwidth}
        \includegraphics[trim=0 1.5 0 1.5,width=\textwidth,height=3.5cm]{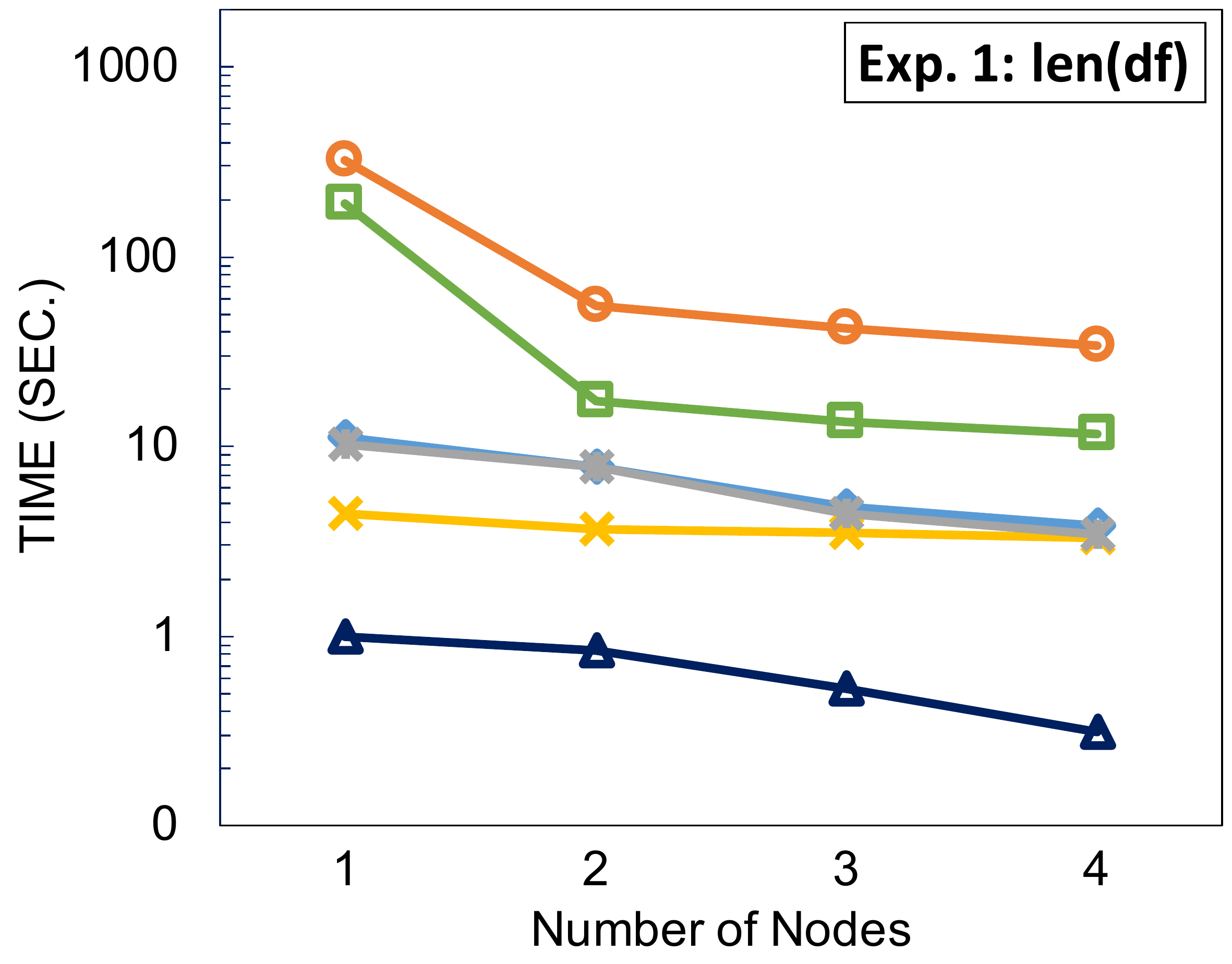}
        \caption{Expression 1: total times}
        \label{fig:q1_speedup}
    \end{subfigure}
    \begin{subfigure}[t]{0.24\textwidth}
        \includegraphics[trim=0 1.5 0 1.5,width=\textwidth,height=3.5cm]{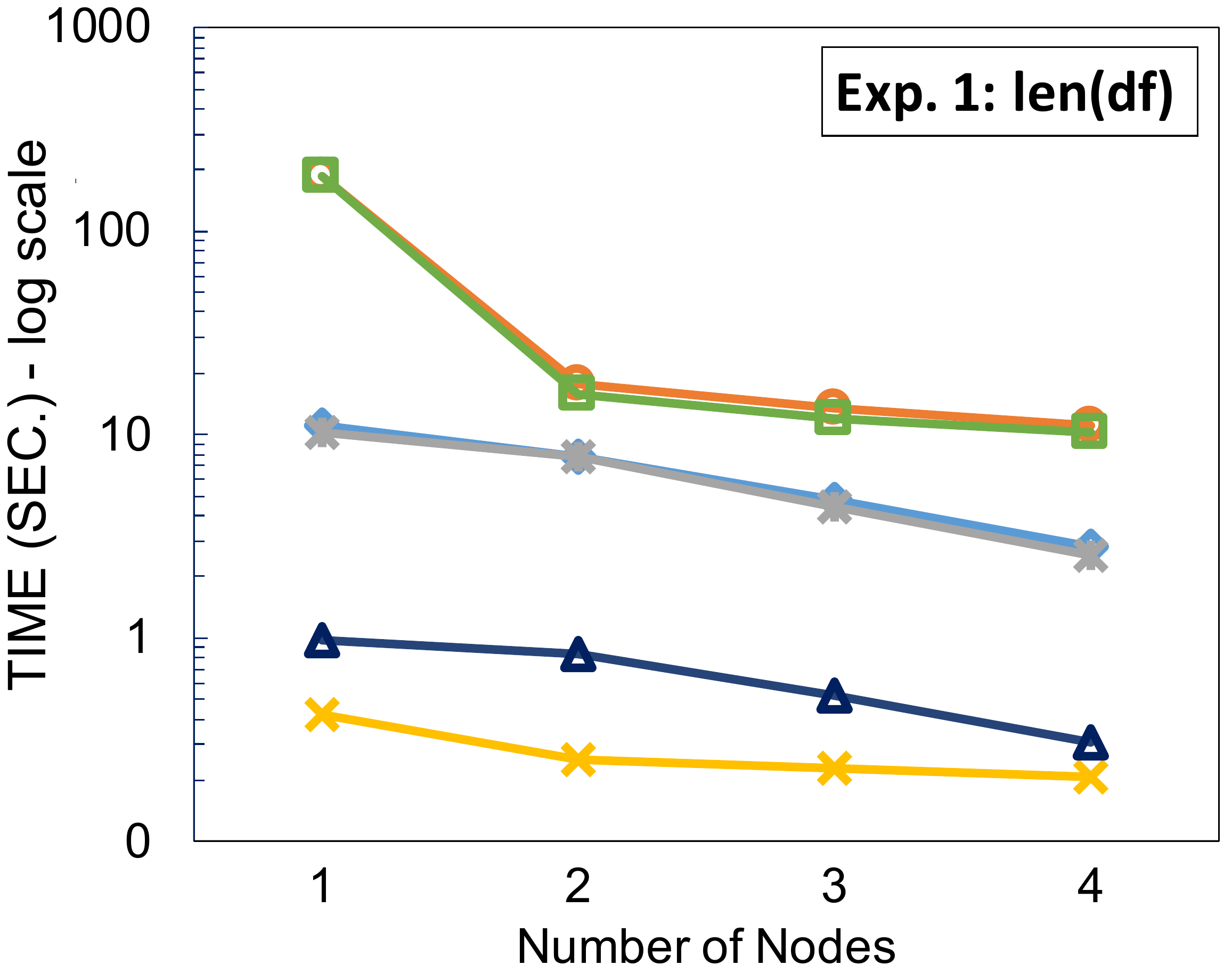}%
        \caption{Expression 1: expression-only times}
        \label{fig:q1_speedup_wo}
    \end{subfigure}
    \hfill
    \begin{subfigure}[t]{0.24\textwidth}
        \includegraphics[trim=0 1.5 0 1.5,width=\textwidth,height=3.5cm]{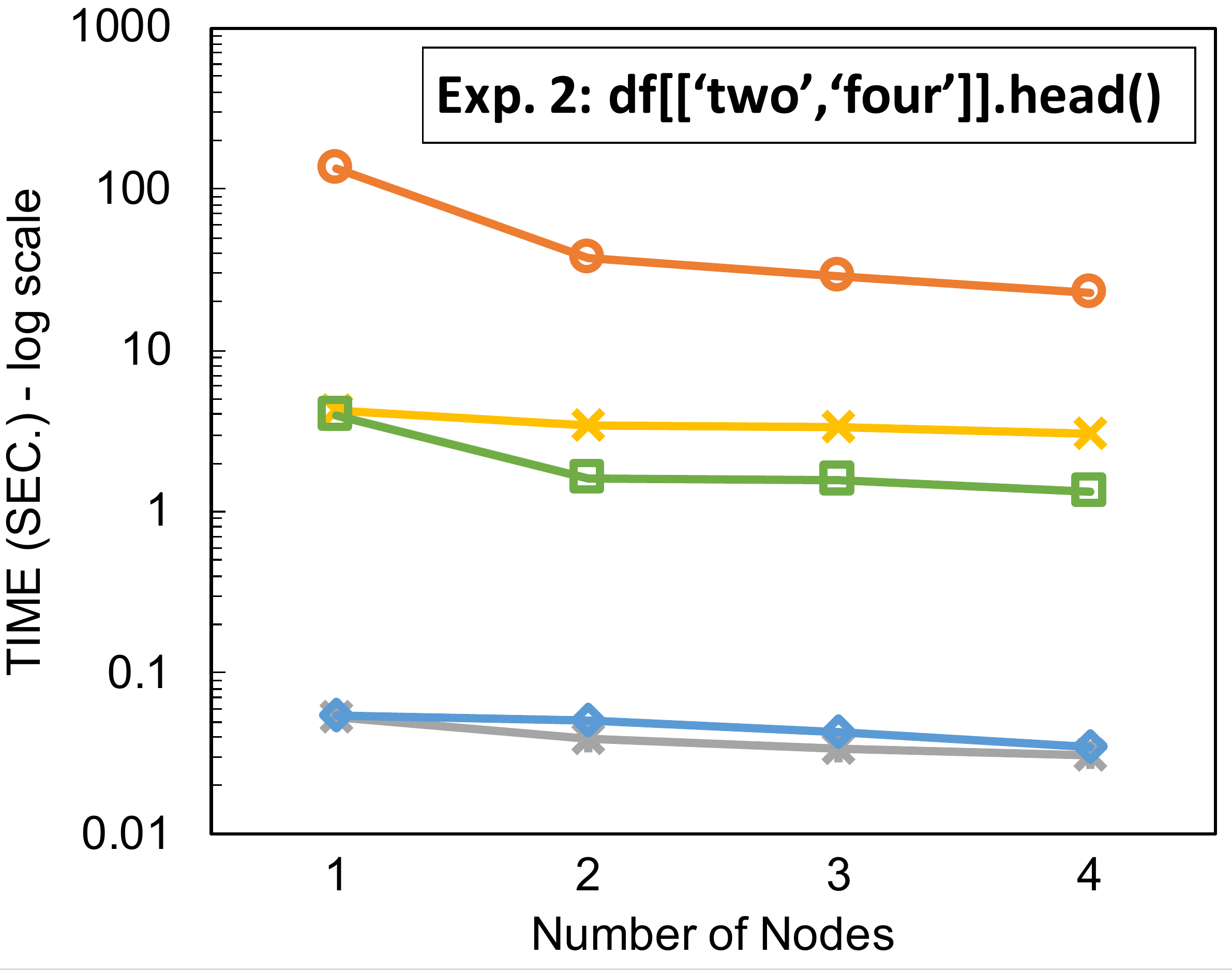}
        \caption{Expression 2: total times}
        \label{fig:q2_speedup}
    \end{subfigure}
    \begin{subfigure}[t]{0.24\textwidth}
        \includegraphics[trim=0 1.5 0 1.5,width=\textwidth,height=3.5cm]{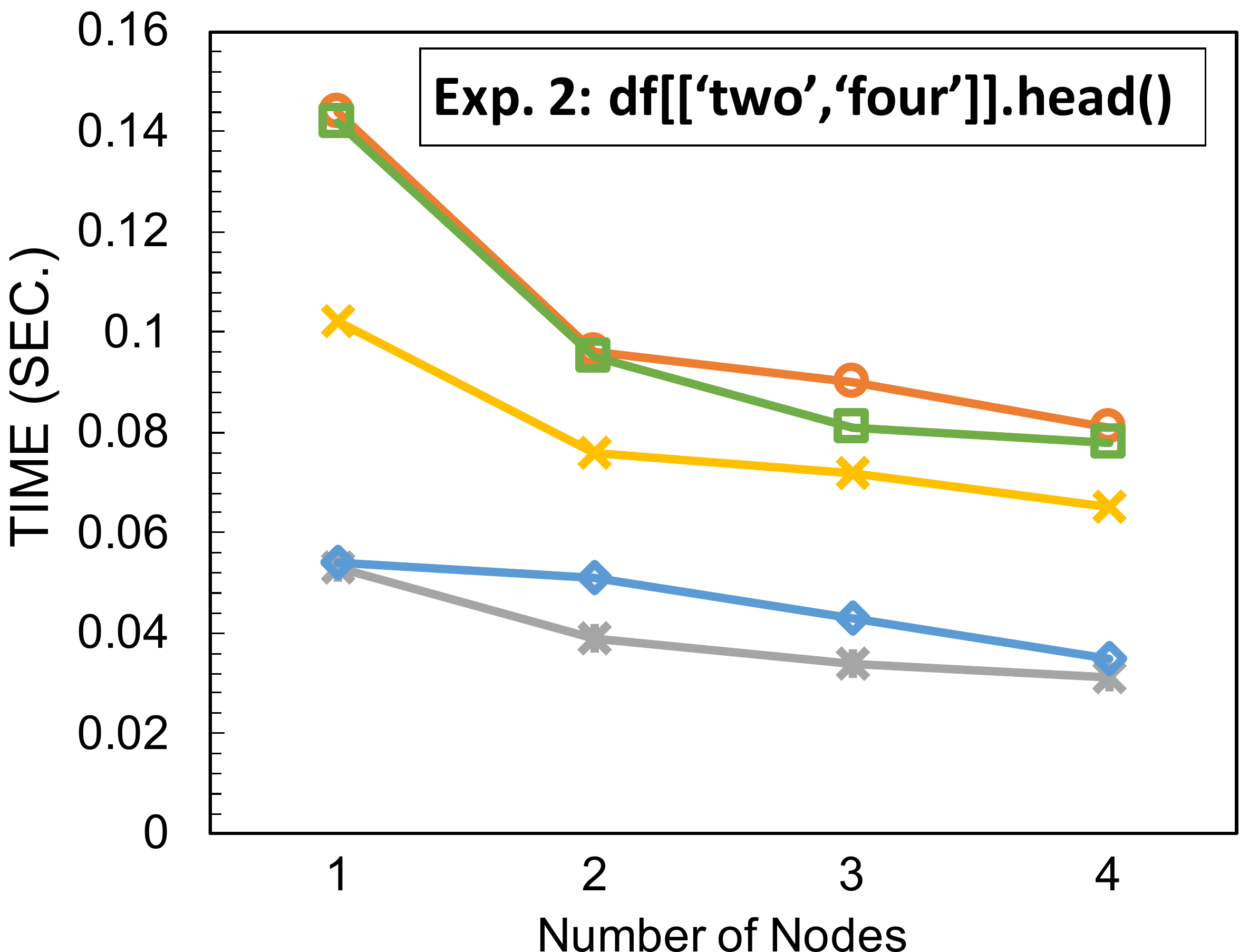}%
        \caption{Expression 2: expression-only times}
        \label{fig:q2_speedup_wo}
    \end{subfigure}
    \begin{subfigure}[t]{0.24\textwidth}
        \includegraphics[trim=1.5 1.5 0.5cm 1.5,width=\textwidth,height=3.5cm]{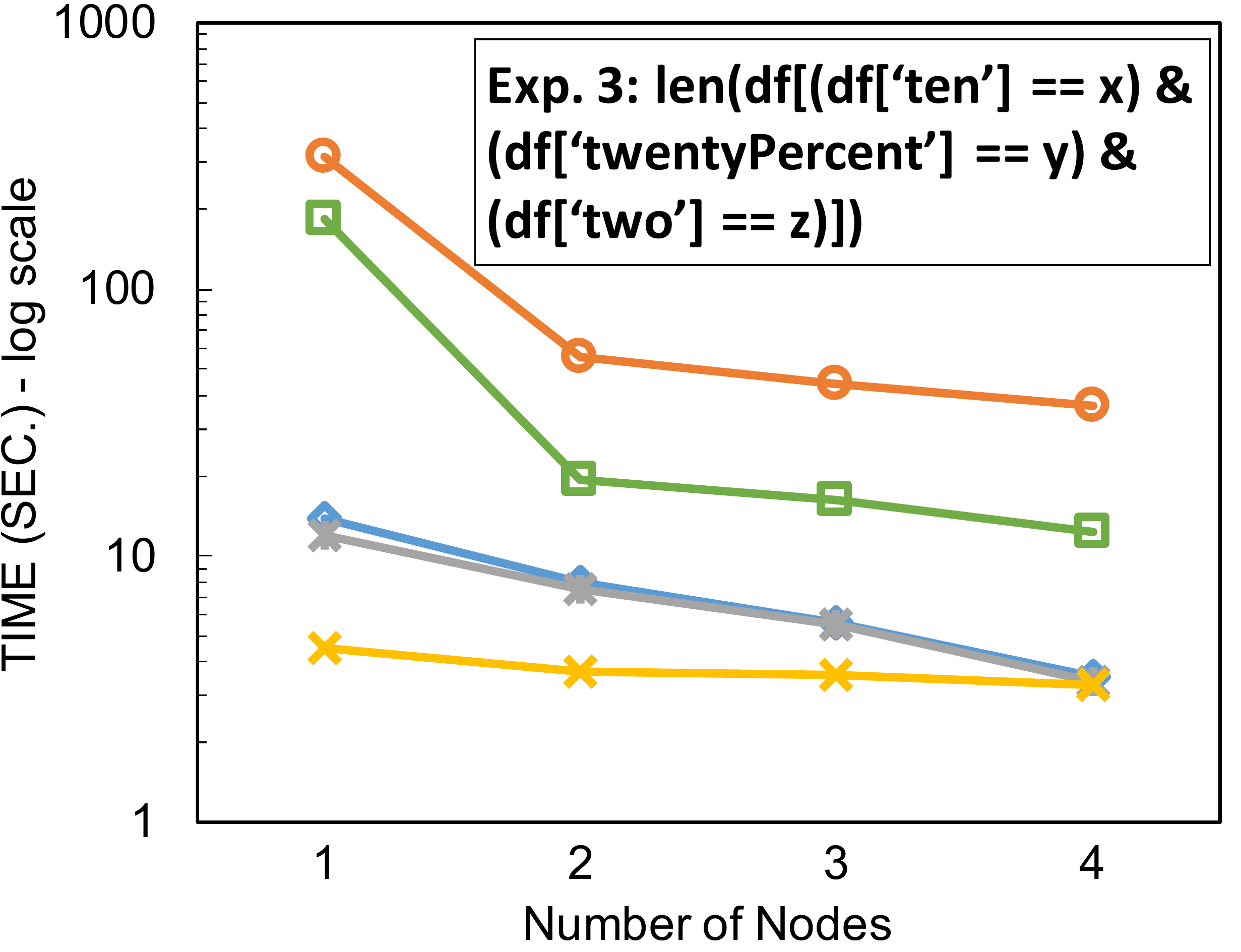}
        \caption{Expression 3: total times}
        \label{fig:q3_speedup}
    \end{subfigure}
    \begin{subfigure}[t]{0.24\textwidth}
        \includegraphics[trim=0 1.5 0 1.5,width=\textwidth,height=3.5cm]{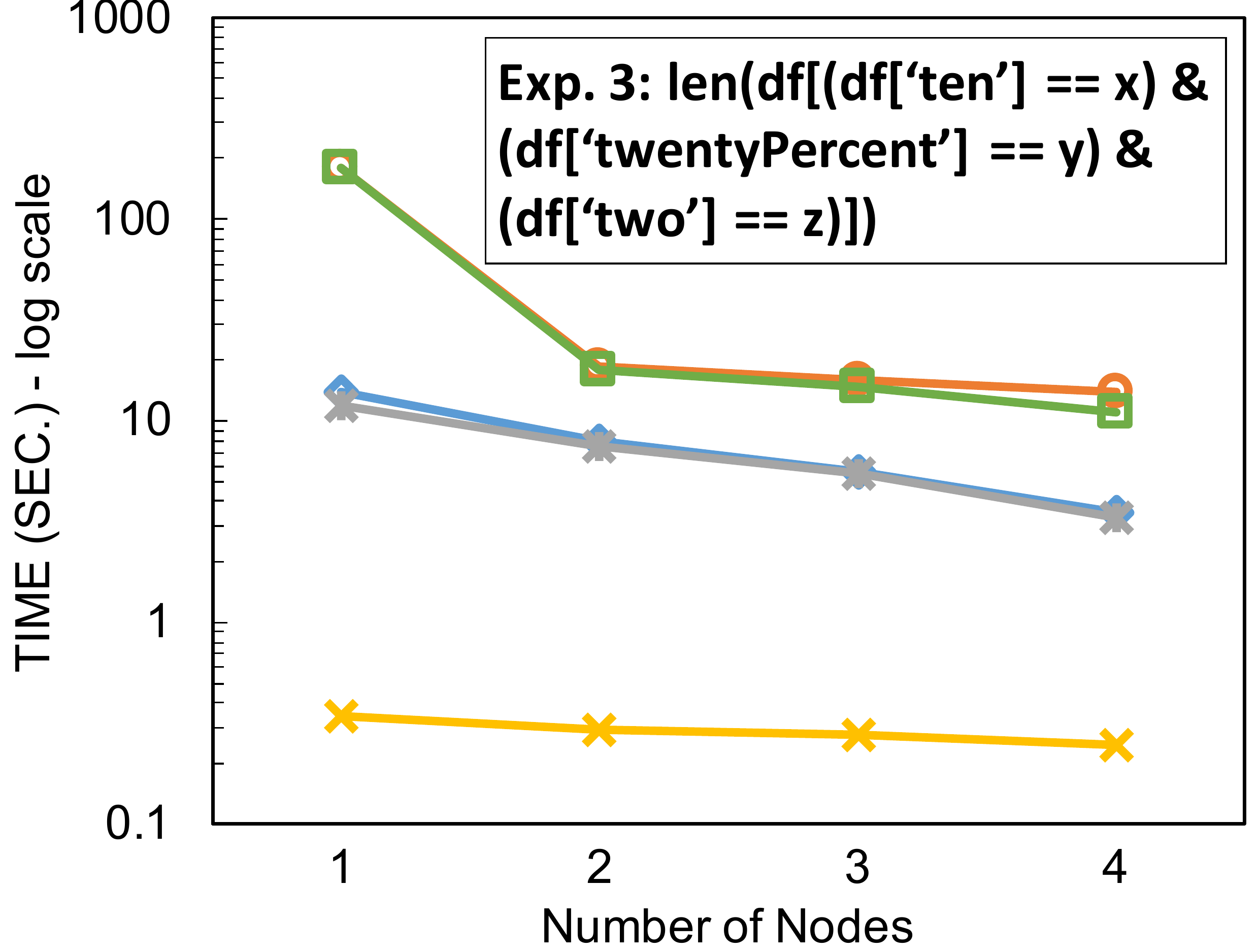}%
        \caption{Expression 3: expression-only times}
        \label{fig:q3_speedup_wo}
    \end{subfigure}
    \hfill
    \begin{subfigure}[t]{0.24\textwidth}
        \includegraphics[trim=0 1.5 0 1.5,width=\textwidth,height=3.5cm]{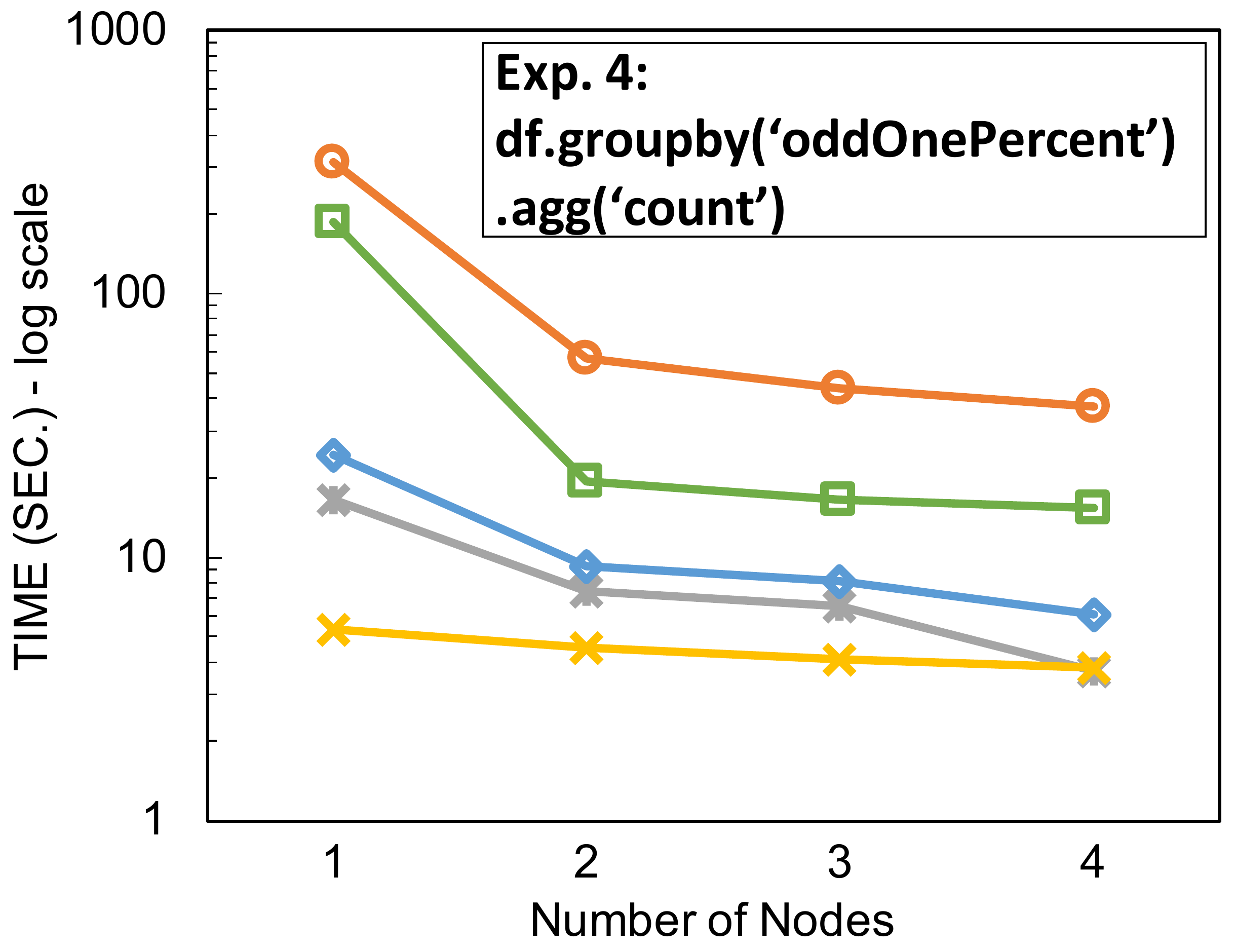}
        \caption{Expression 4: total times}
        \label{fig:q4_speedup}
    \end{subfigure}
    \begin{subfigure}[t]{0.24\textwidth}
        \includegraphics[trim=0 1.5 0 1.5,width=\textwidth,height=3.5cm]{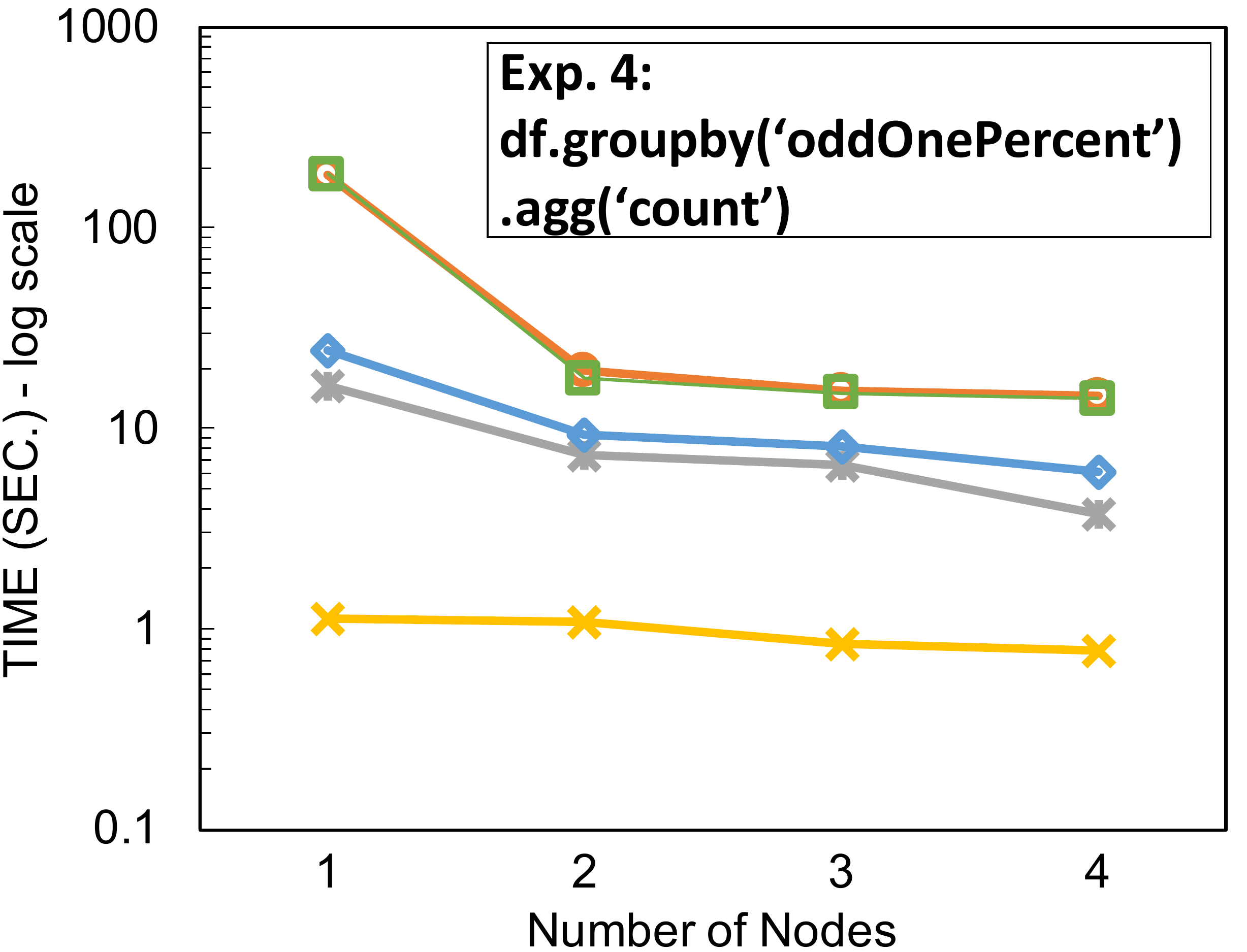}%
        \caption{Expression 4: expression-only times}
        \label{fig:q4_speedup_wo}
    \end{subfigure}
    \begin{subfigure}[t]{0.24\textwidth}
        \includegraphics[trim=0 1.5 0 1.5,width=\textwidth,height=3.5cm]{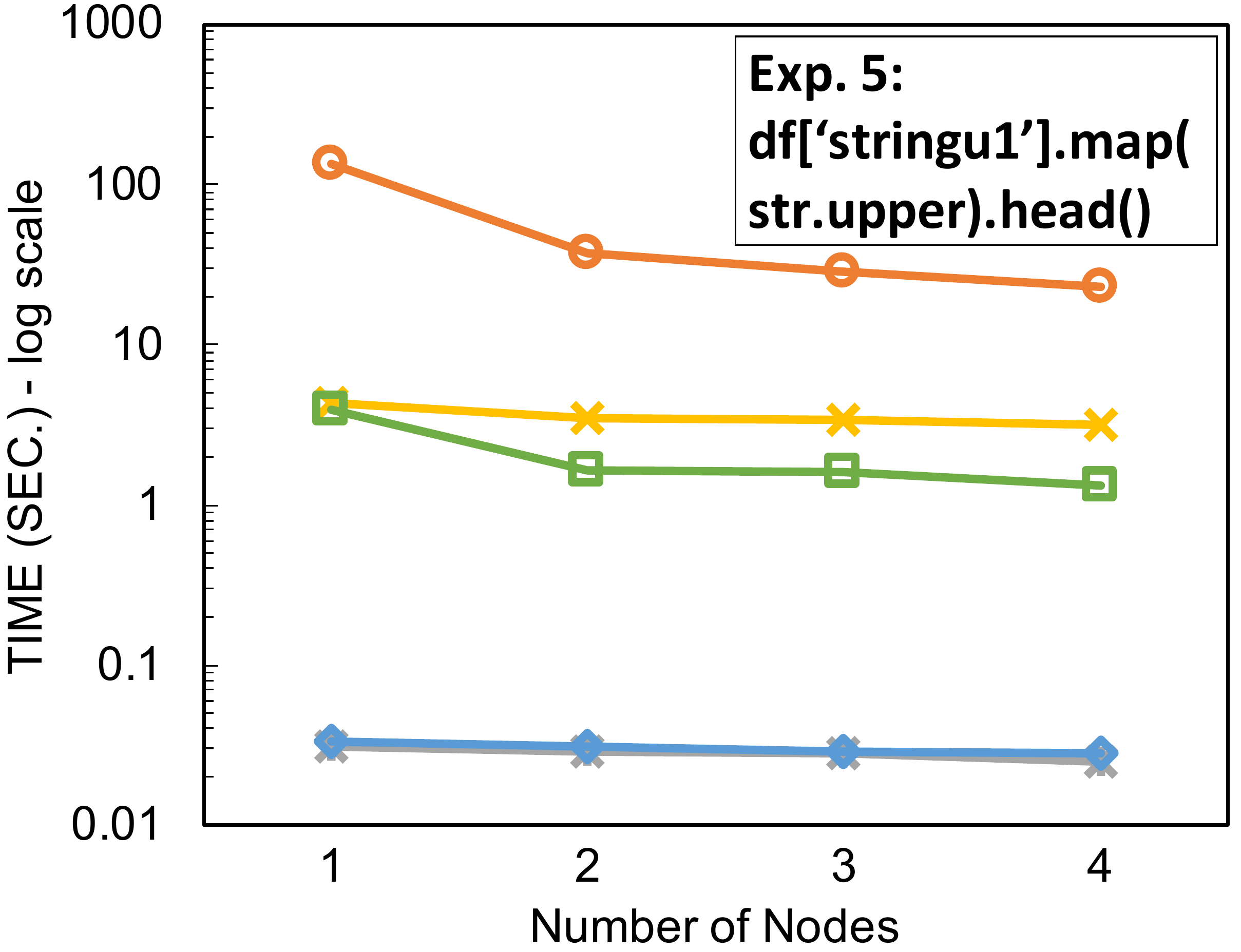}
        \caption{Expression 5: total times}
        \label{fig:q5_speedup}
    \end{subfigure}
    \begin{subfigure}[t]{0.24\textwidth}
        \includegraphics[trim=0 1.5 0 1.5,width=\textwidth,height=3.5cm]{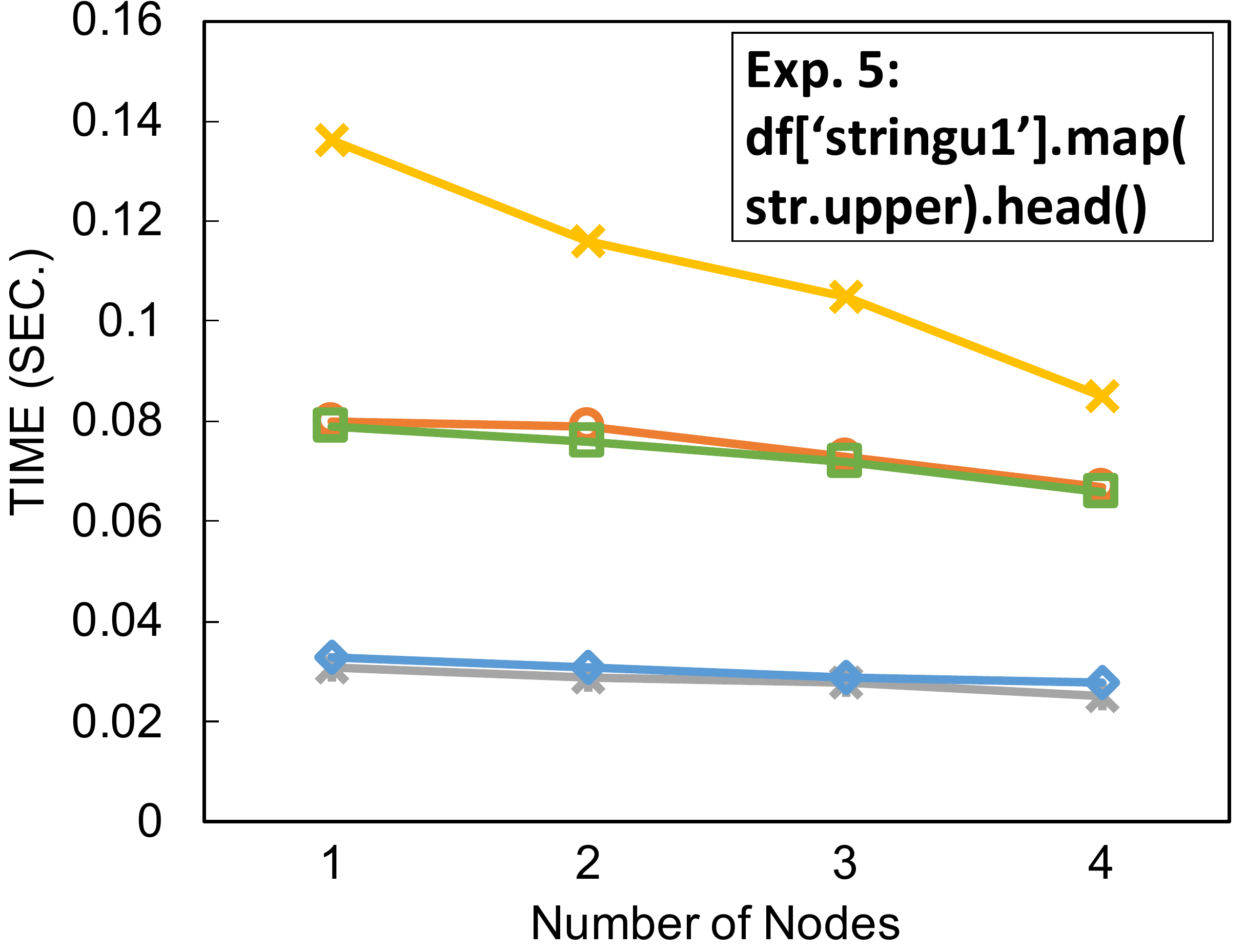}%
        \caption{Expression 5: expression-only times}
        \label{fig:q5_speedup_wo}
    \end{subfigure}
    \hfill
    \begin{subfigure}[t]{0.24\textwidth}
        \includegraphics[trim=0 1.5 0 1.5,width=\textwidth,height=3.5cm]{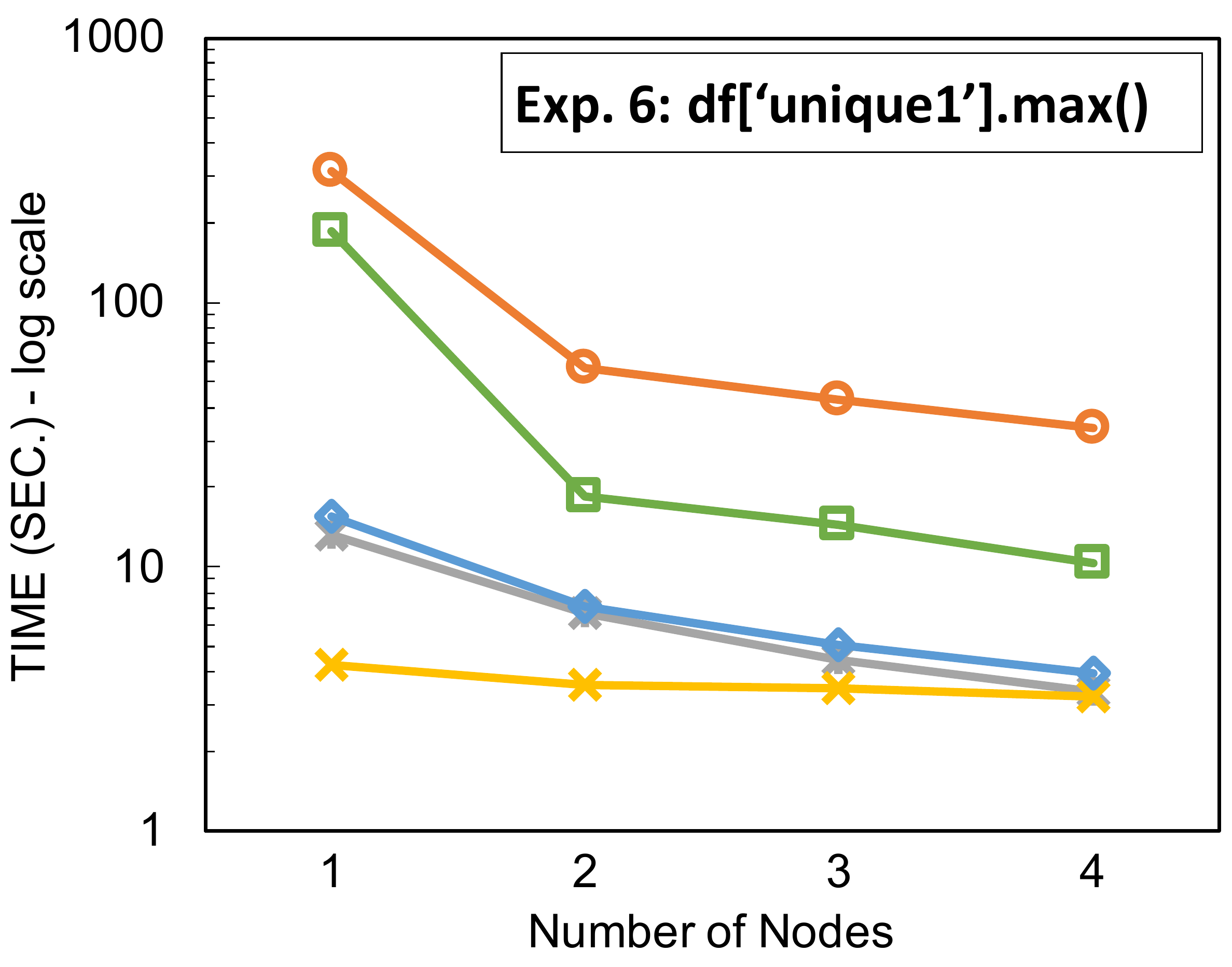}
        \caption{Expression 6: total times}
        \label{fig:q6_speedup}
    \end{subfigure}
    \begin{subfigure}[t]{0.24\textwidth}
        \includegraphics[trim=0 1.5 0 1.5,width=\textwidth,height=3.5cm]{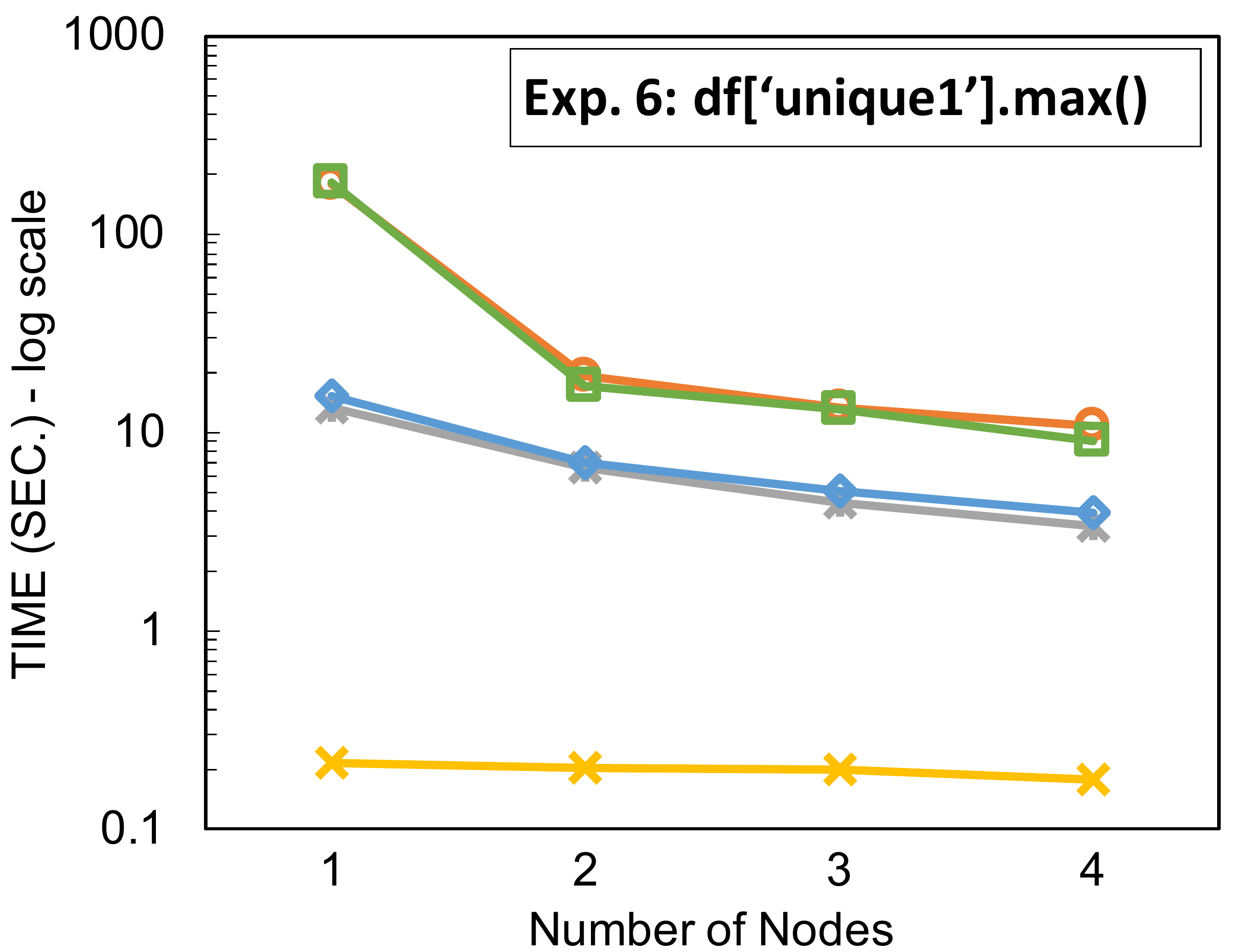}%
        \caption{Expression 6: expression-only times}
        \label{fig:q6_speedup_wo}
    \end{subfigure}
    \begin{subfigure}[t]{0.24\textwidth}
        \includegraphics[trim=0 1.5 0 1.5,width=\textwidth,height=3.5cm]{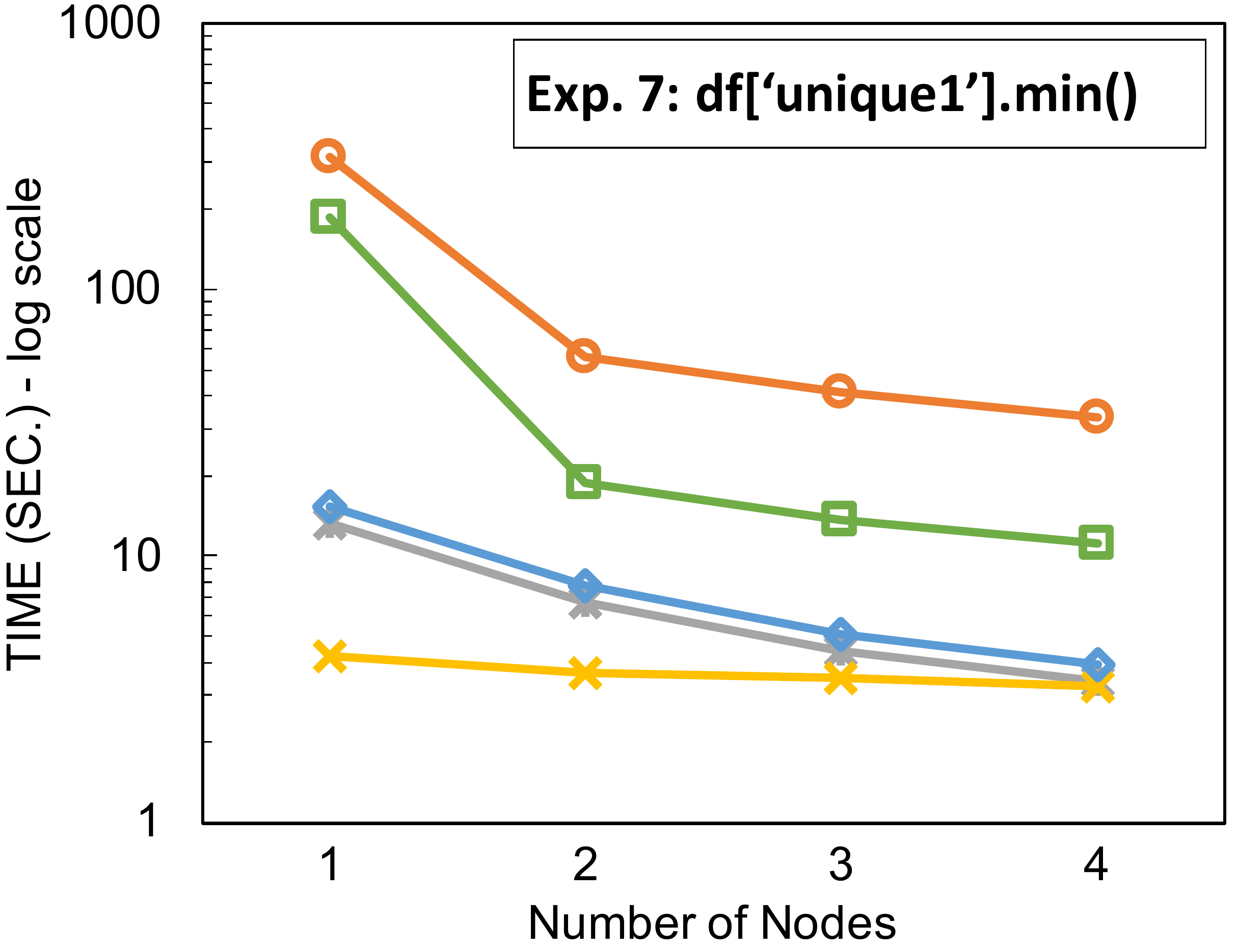}
        \caption{Expression 7: total times}
        \label{fig:q7_speedup}
    \end{subfigure}
    \begin{subfigure}[t]{0.24\textwidth}
        \includegraphics[trim=0 1.5 0 1.5,width=\textwidth,height=3.5cm]{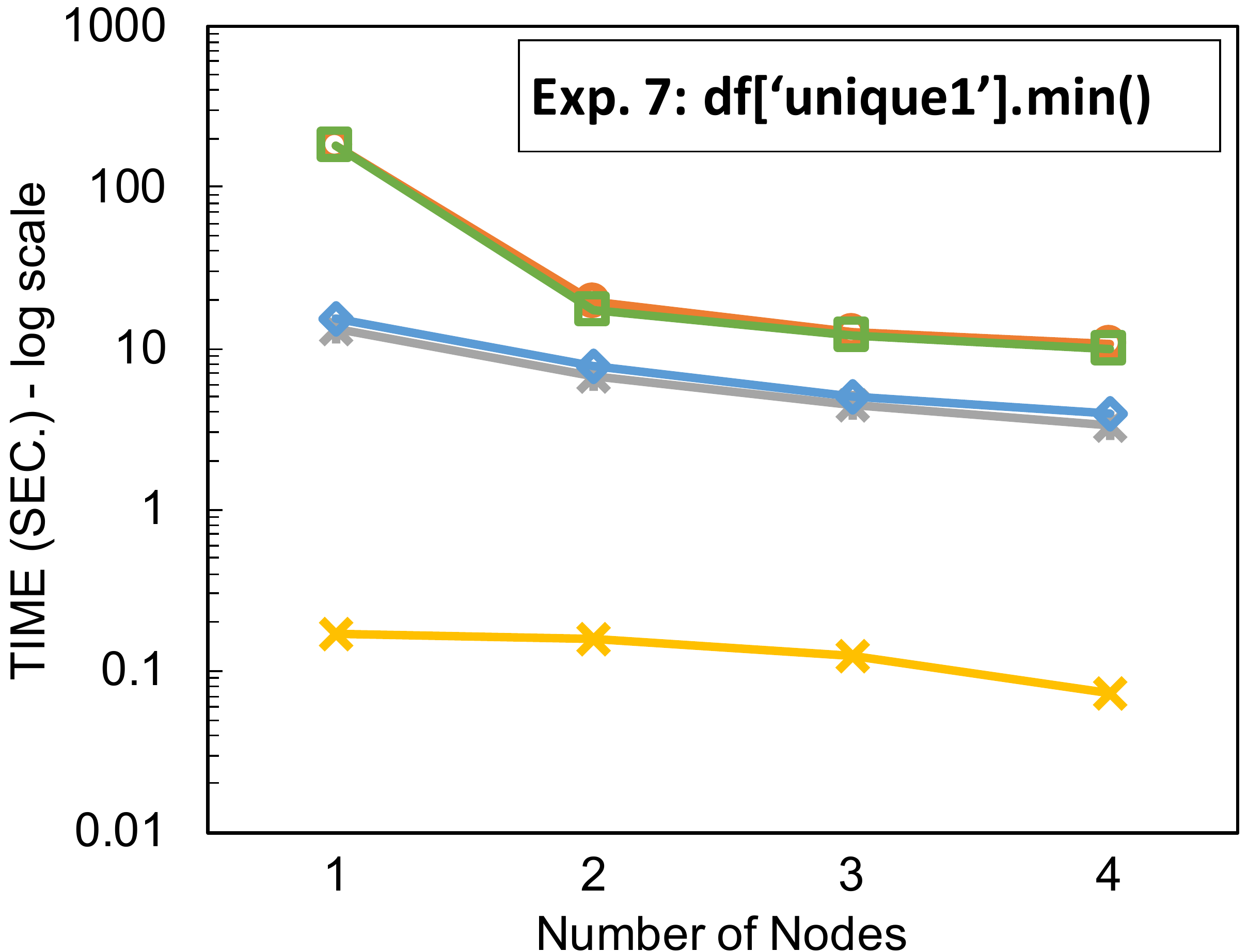}%
        \caption{Expression 7: expression-only times}
        \label{fig:q7_speedup_wo}
    \end{subfigure}
    \hfill
    \begin{subfigure}[t]{0.24\textwidth}
        \includegraphics[trim=0 1.5 0 1.5,width=\textwidth,height=3.5cm]{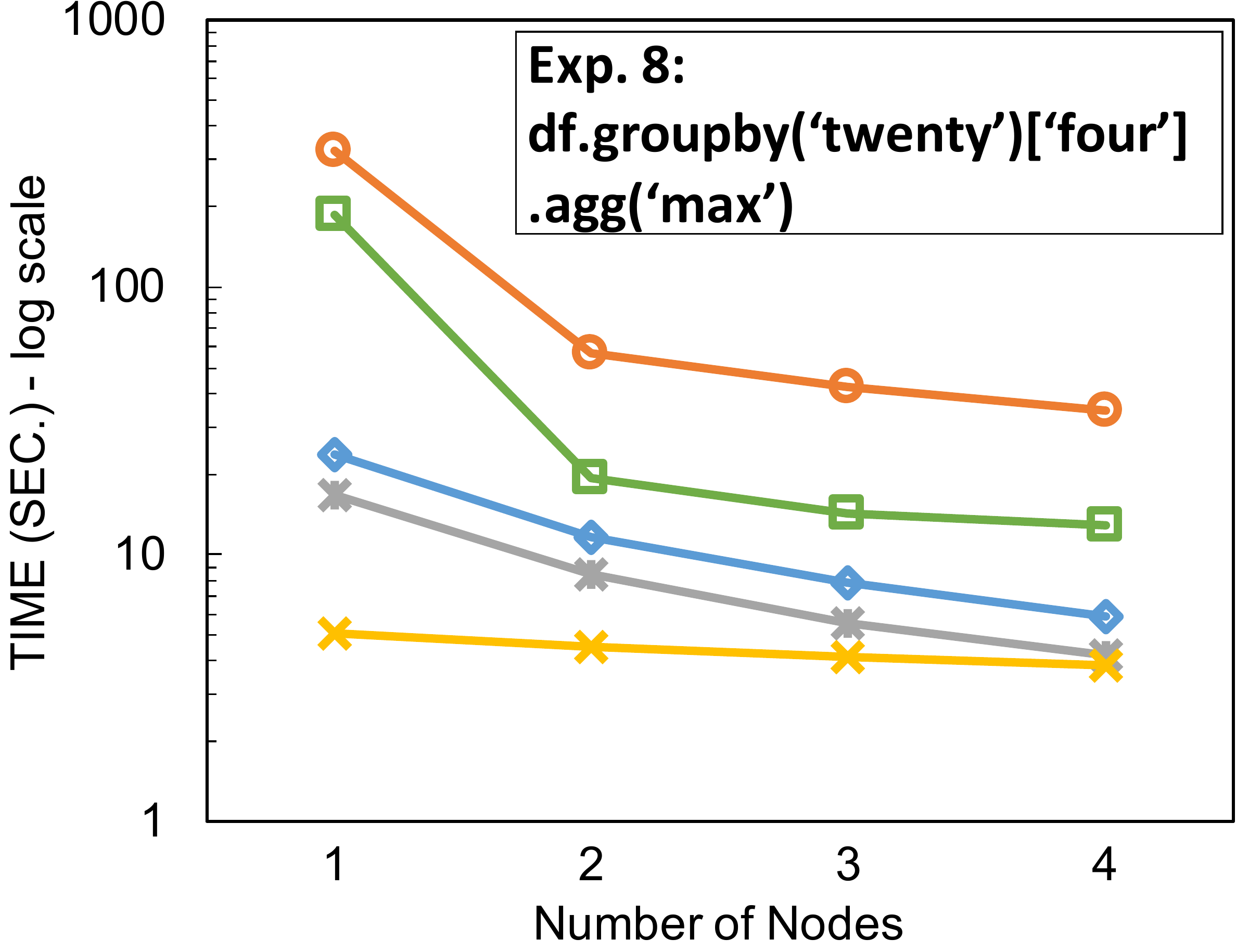}
        \caption{Expression 8: total times}
        \label{fig:q8_speedup}
    \end{subfigure}
    \begin{subfigure}[t]{0.24\textwidth}
        \includegraphics[trim=0 1.5 0 1.5,width=\textwidth,height=3.5cm]{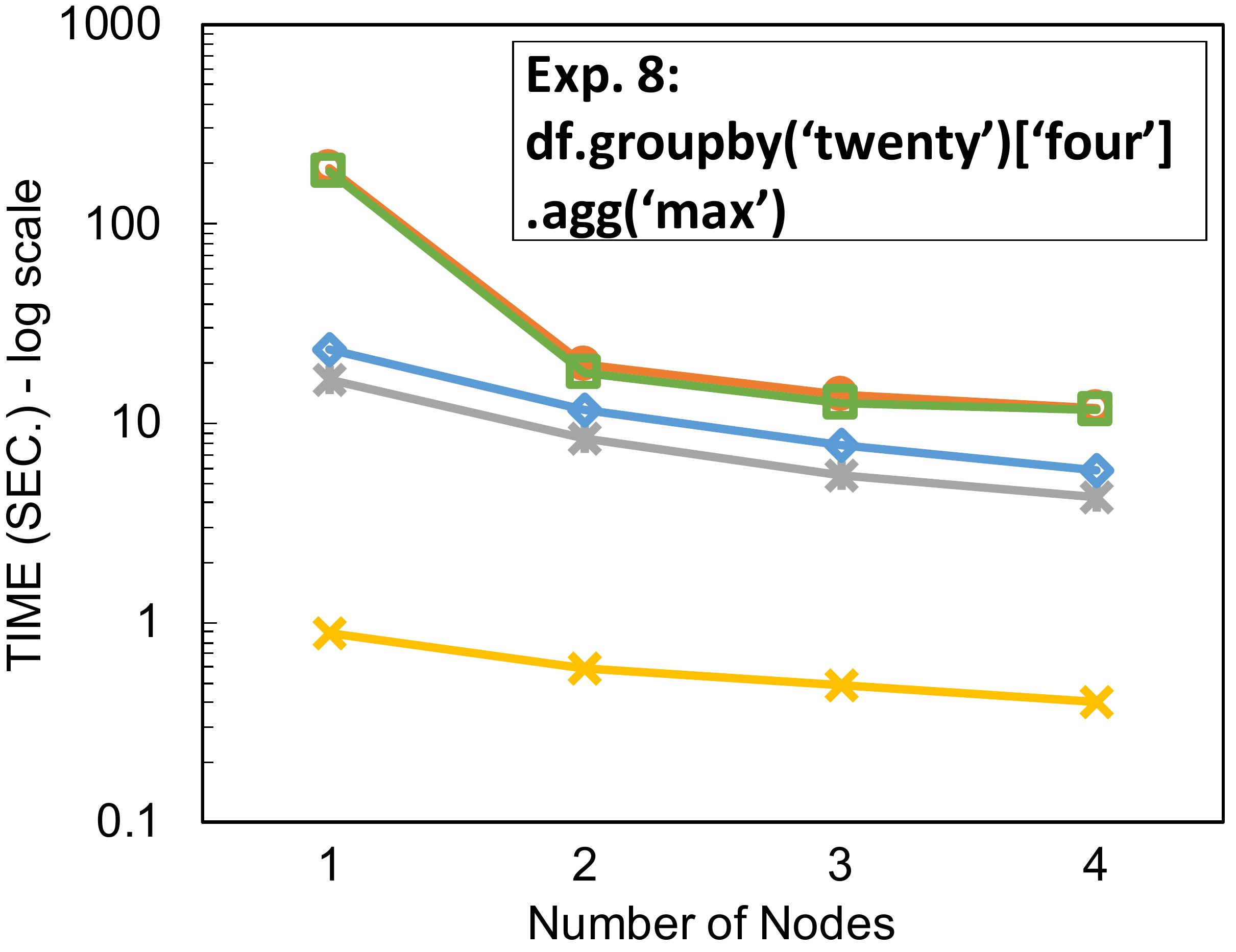}%
        \caption{Expression 8: expression-only times}
        \label{fig:q8_speedup_wo}
    \end{subfigure}
    \begin{subfigure}[t]{0.24\textwidth}
        \includegraphics[trim=0 1.5 0 1.5,width=\textwidth,height=3.5cm]{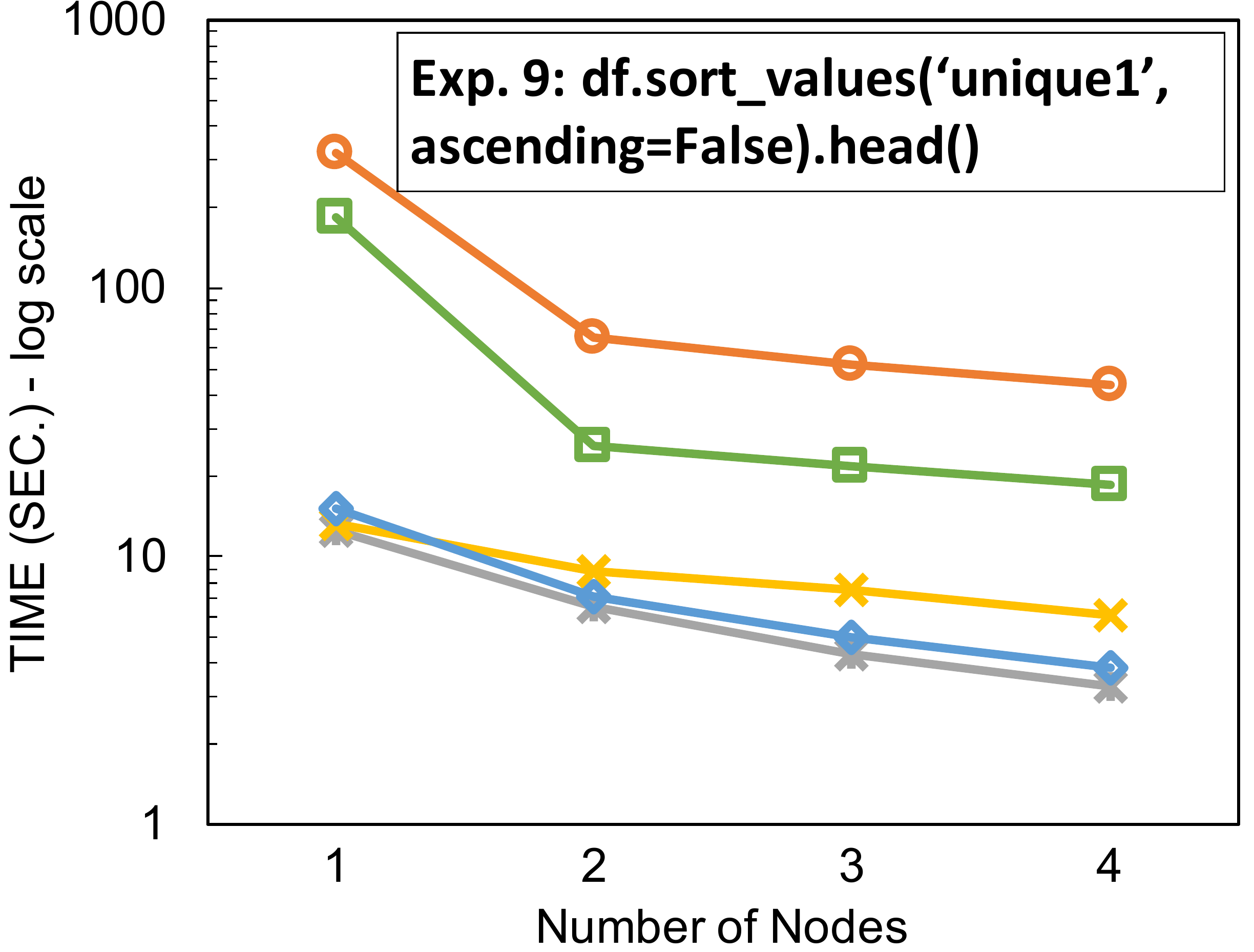}
        \caption{Expression 9: total times}
        \label{fig:q9_speedup}
    \end{subfigure}
    \begin{subfigure}[t]{0.24\textwidth}
        \includegraphics[trim=0 1.5 0 1.5,width=\textwidth,height=3.5cm]{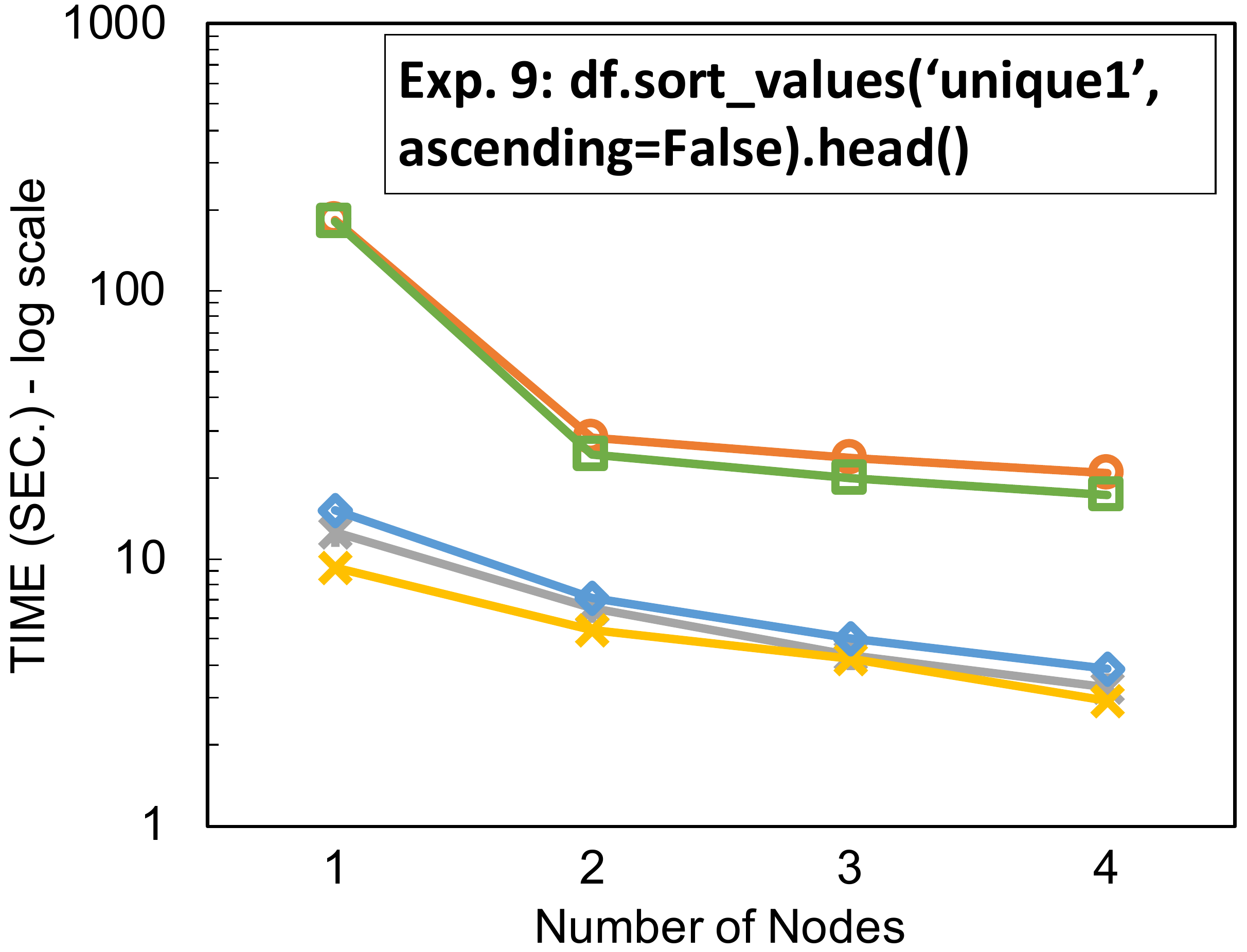}%
        \caption{Expression 9: expression-only times}
        \label{fig:q9_speedup_wo}
    \end{subfigure}
    \hfill
    \begin{subfigure}[t]{0.24\textwidth}
        \includegraphics[trim=0 1.5 0 1.5,width=\textwidth,height=3.5cm]{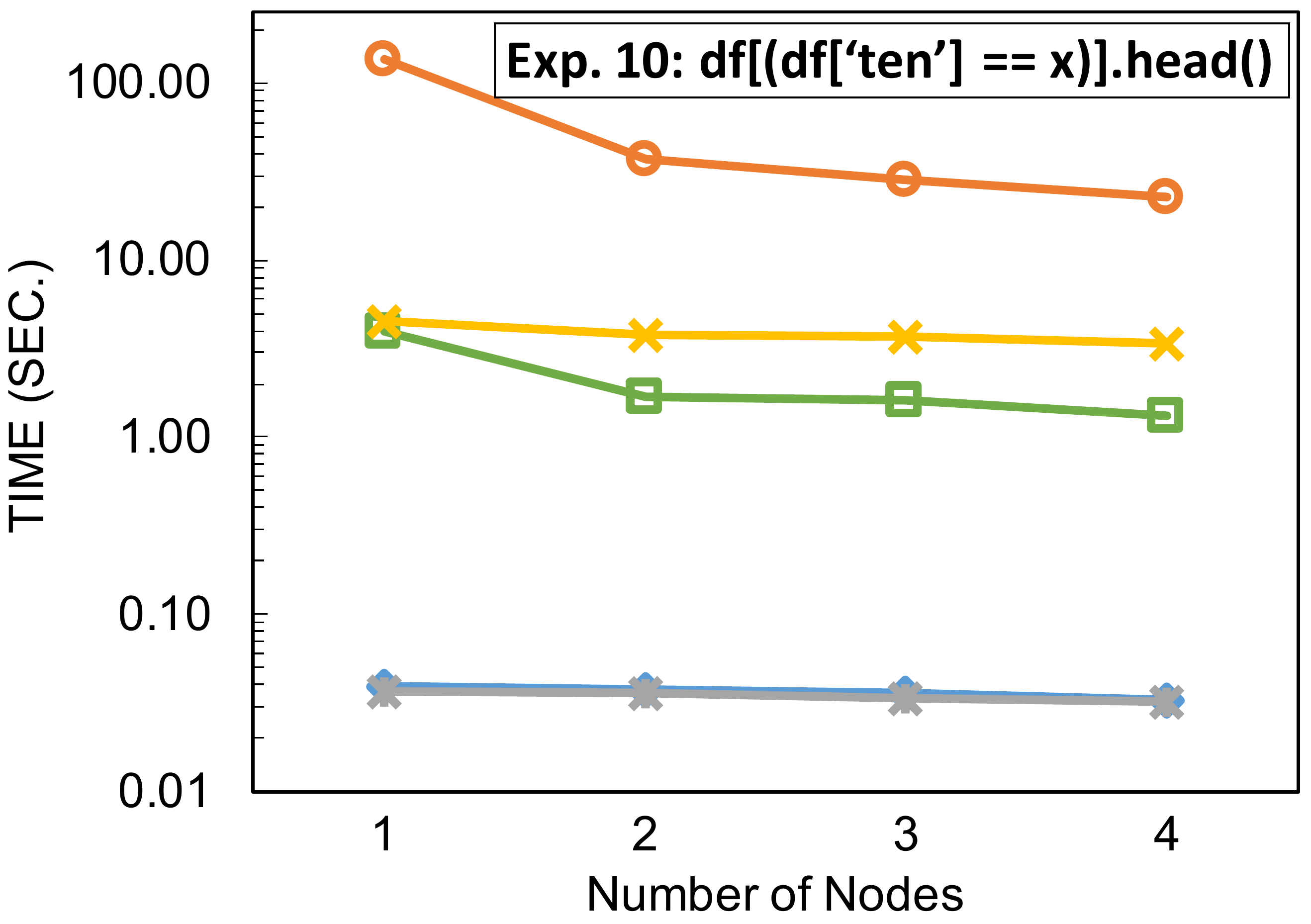}
        \caption{Expression 10: total times}
        \label{fig:q10_speedup}
    \end{subfigure}
    \begin{subfigure}[t]{0.24\textwidth}
        \includegraphics[trim=0 1.5 0 1.5,width=\textwidth,height=3.5cm]{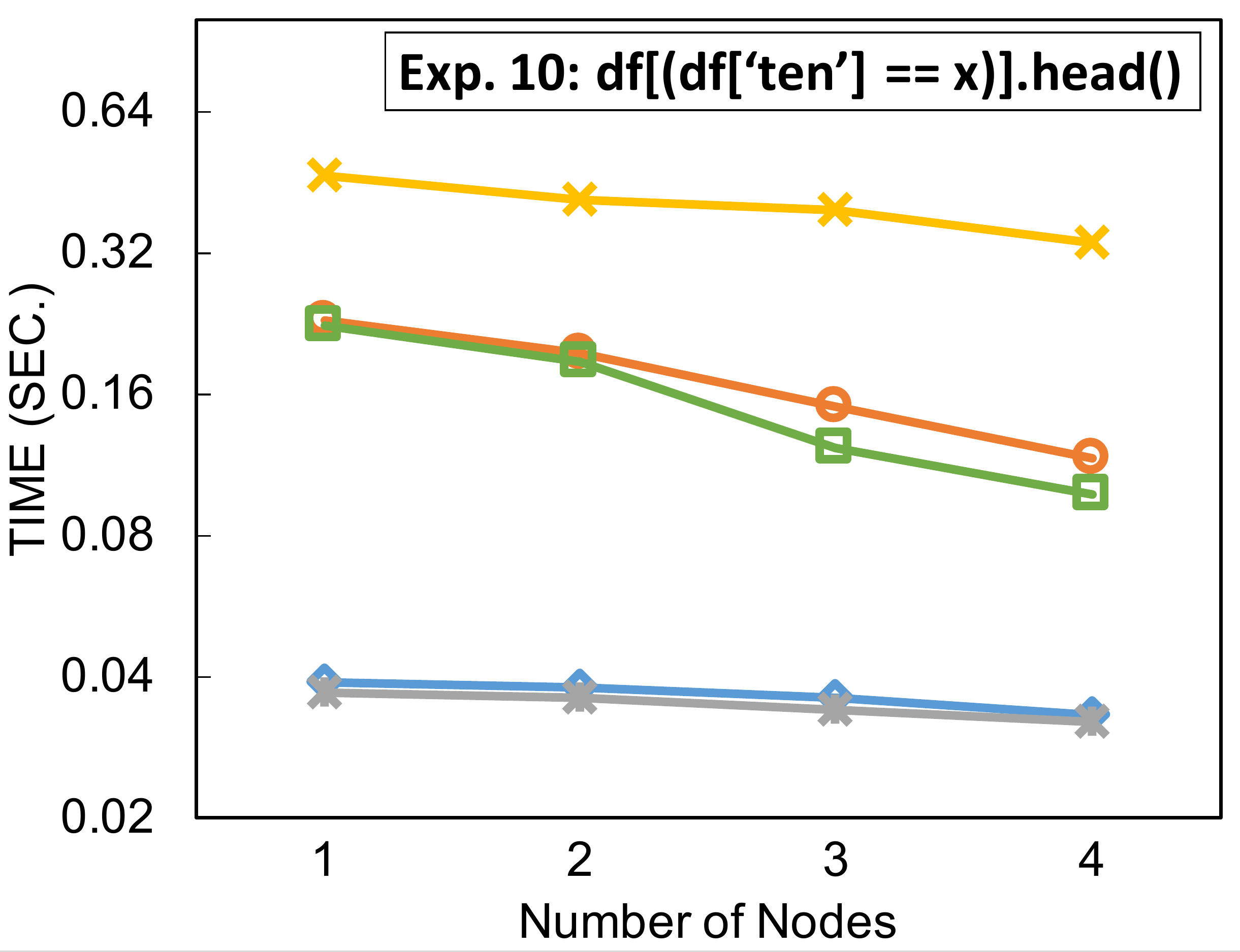}%
        \caption{Expression 10: expression-only times}
        \label{fig:q10_speedup_wo}
    \end{subfigure}

    \caption{Multi-Node Speedup Evaluation Results}

    \label{fig:speedup_results}
    \vspace{-1.5em}
\end{figure*}

\begin{figure*}[!ht]
     \centering
    \begin{subfigure}[t]{0.45\textwidth}
        \includegraphics[trim=0 1.5 0 1.5,width=\textwidth,height=0.7cm]{figures/scaleup_legend.pdf}
    \end{subfigure}
    \hspace{15cm}
    \begin{subfigure}[t]{0.24\textwidth}
        \includegraphics[trim=0 1.5 0 1.5,width=\textwidth,height=3.5cm]{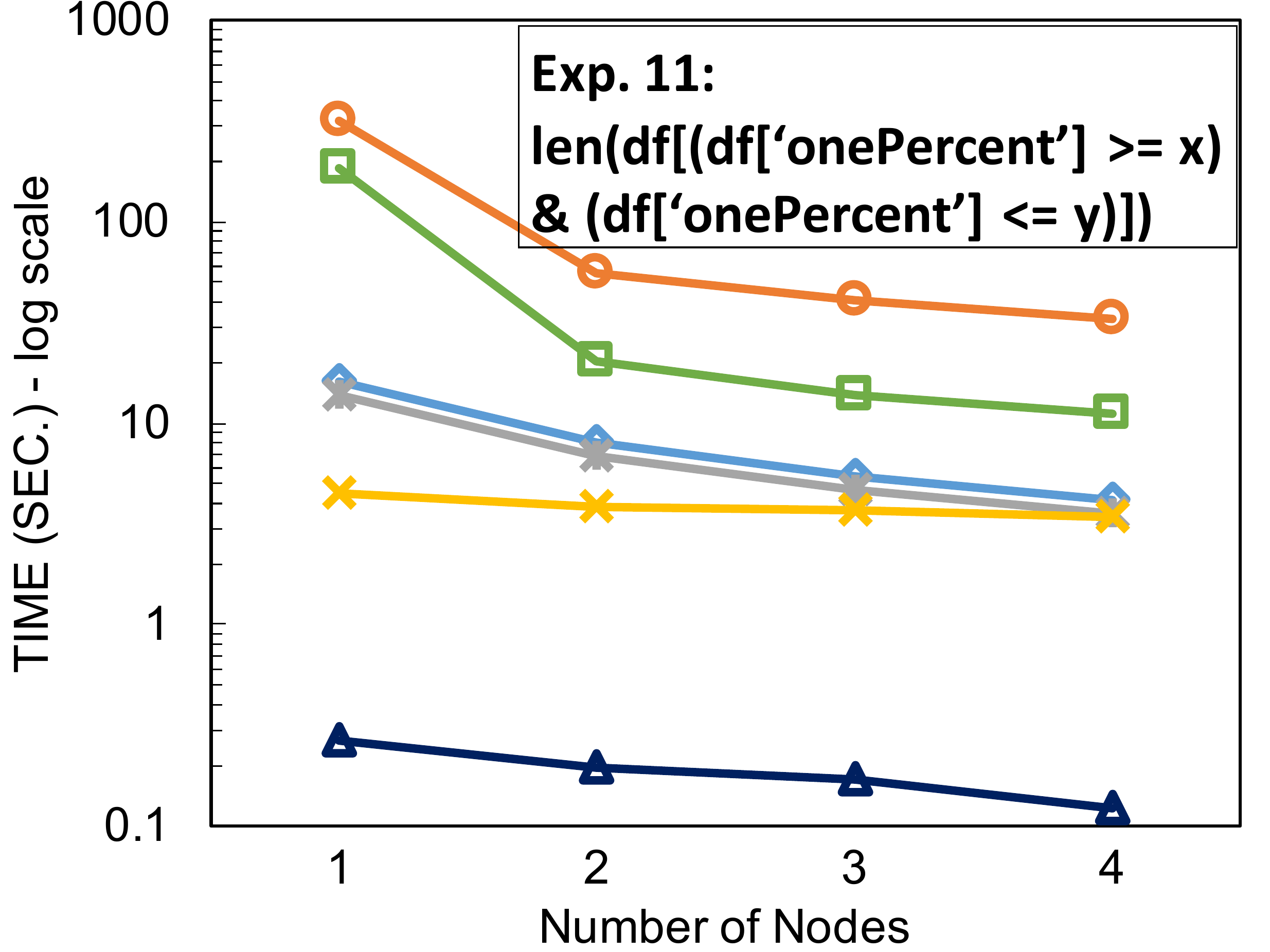}
        \caption{Expression 11: total times}
        \label{fig:q11_speedup}
    \end{subfigure}
    \begin{subfigure}[t]{0.24\textwidth}
        \includegraphics[trim=0 1.5 0 1.5,width=\textwidth,height=3.5cm]{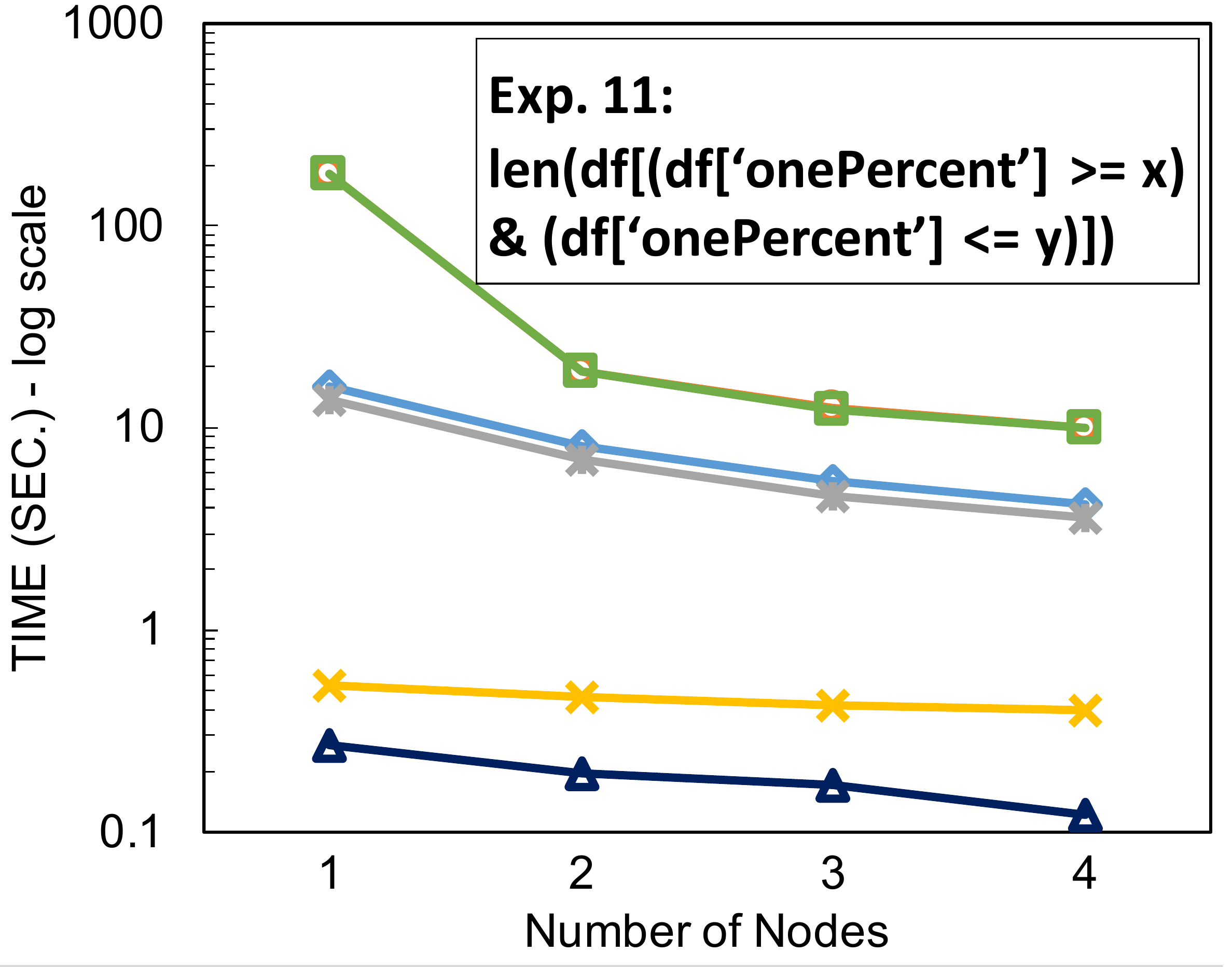}%
        \caption{Expression 11: expression-only times}
        \label{fig:q11_speedup_wo}
    \end{subfigure}
    \hfill
    \begin{subfigure}[t]{0.24\textwidth}
        \includegraphics[trim=0 1.5 0 1.5,width=\textwidth,height=3.5cm]{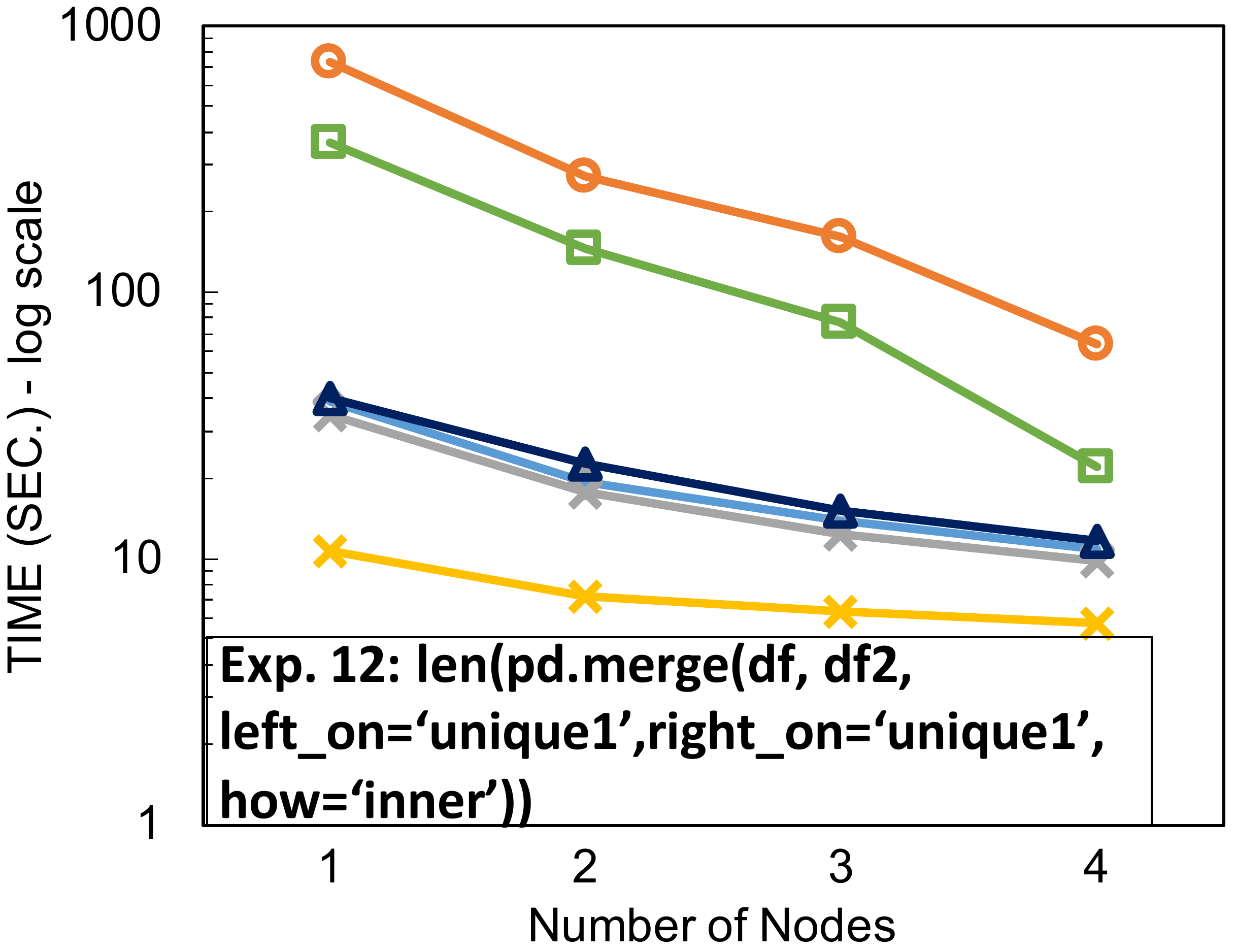}
        \caption{Expression 12: total times}
        \label{fig:q12_speedup}
    \end{subfigure}
    \begin{subfigure}[t]{0.24\textwidth}
        \includegraphics[trim=0.5cm 1.5 0.5cm 1.5,width=\textwidth,height=3.5cm]{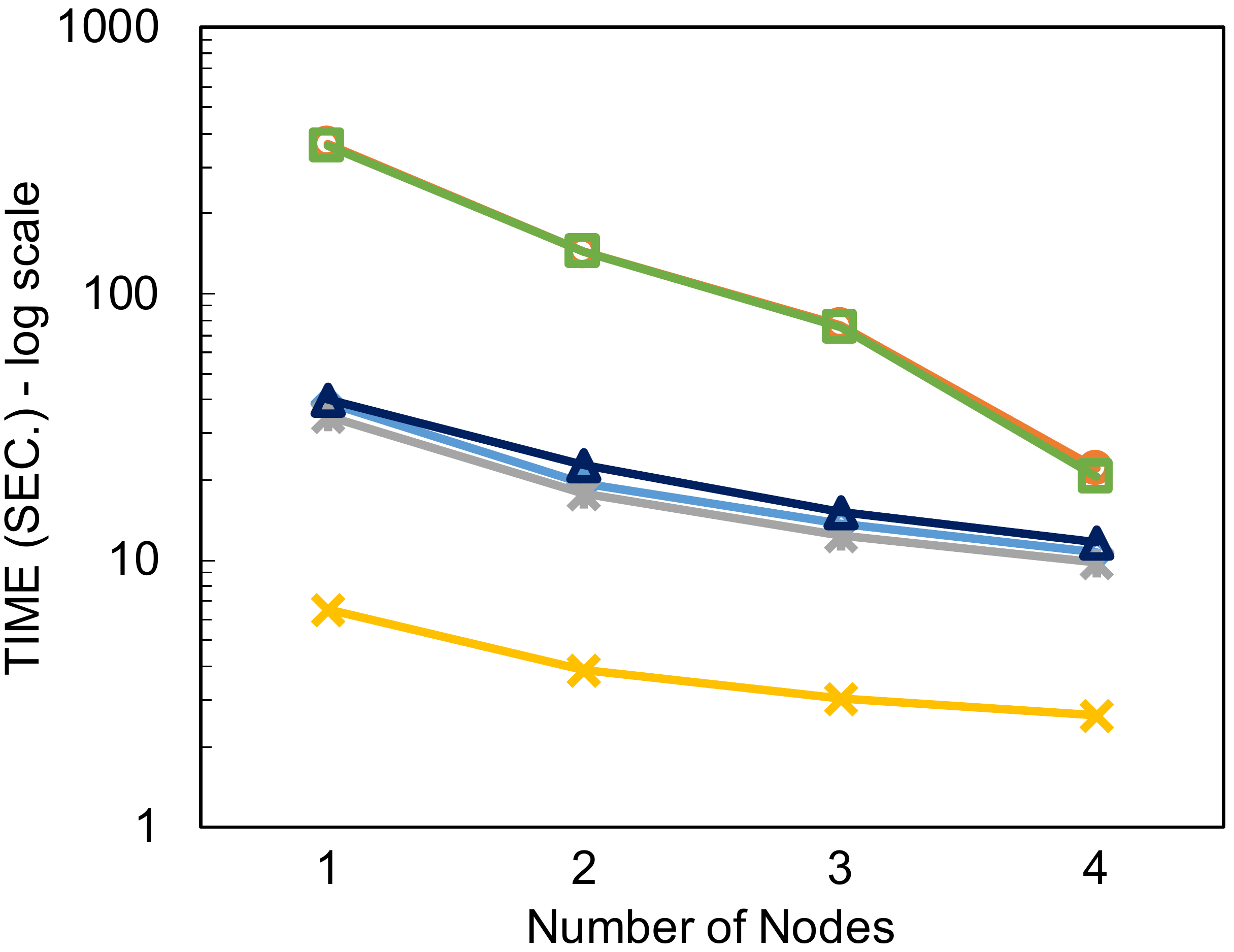}%
        \caption{Expression 12: expression-only times}
        \label{fig:q12_speedup_wo}
    \end{subfigure}
    \caption{Multi-Node Speedup Evaluation Results (continued)}
    \label{fig:speedup_results_11-12}
    \vspace{-1.5em}
\end{figure*}
\subsection{Multi-Node Results}
For the distributed environment evaluation, as mentioned earlier, we have only evaluated Spark and AFrame. We evaluated Spark on the same three DataFrame creation sources: JSON, JSON with schema, and Parquet. Likewise, we evaluated AFrame on its same three datasets, which are datasets with an open datatype, with a schema, and with an index.

For the multi-node evaluation, we evaluated the systems' performance in a distributed environment. As we observed in the single node evaluation, Spark spills to disk for both the L and XL datasets (7.5 and 10 GB), which significantly affected its performance. In order to observe the effect of clusters processing data that is larger than the available aggregate memory, we chose to start our multi-node evaluation with the 10-GB dataset. Here we evaluated both systems according to both the speedup and scaleup metrics. 

The multi-node evaluation was performed on ec2 machines with the same specifications as the single node evaluation. 


\subsubsection{Speedup Results}

In addition to the plots in Figures~\ref{fig:speedup_results} and \ref{fig:speedup_results_11-12}, the raw averaged run times for multi-node speedup results are included in Table~\ref{tab:speedup} in the appendix section. The results for both Spark and AFrame are consistent with their single-node results in terms of their performance rankings. Both systems processed the tasks faster when increasing the number of processors while maintaining the same data size. Spark's performance improved drastically when the distributed data begin to fit in memory in the case of JSON DataFrames. \Cref{fig:speedup_results,fig:speedup_results_11-12} show that increasing the number of processing nodes reduces Spark JSON-based DataFrame's run time by an order of magnitude in the case of going from a single node to a 2-node cluster. This is especially more visible in the total run time case. However, once the data fits in memory, increasing the number of nodes no longer results in such a drastic change (as we can see from the flatter lines for both of Spark's JSON-based DataFrames going from 2 nodes to 4 nodes).

For expression 1 (Figures~\ref{fig:q1_speedup} and \ref{fig:q1_speedup_wo}), AFrame with an index and Spark's Parquet-based DataFrame performed the best. AFrame operating on a dataset with a primary key index was faster than Spark in the total time case, and Spark's Parquet-based DataFrame was best in terms of the expression-only time. 

Similar to the single node results, the Parquet-based DataFrame was the slowest in expression-only evaluation when access to the entire data record is required, as seen for expression 10 in Figure~\ref{fig:q10_speedup_wo} across different numbers of nodes. AFrame with and without schema are the fastest in both expression-only (\ref{fig:q10_speedup_wo}) and total time (Figures~\ref{fig:q10_speedup}) for this expression. However, for expression 11 (Figures~\ref{fig:q11_speedup} and \ref{fig:q11_speedup_wo}), AFrame with an index on the range attribute was the fastest in both expression-only and total time evaluations because Spark Parquet incurred certain DataFrame creation overheads while AFrame translated this expression into a query that was executed as an index-only query on AsterixDB.

For expression 12 (Figures~\ref{fig:q12_speedup} and \ref{fig:q12_speedup_wo}), the similar behavior from the single node evaluation is also visible here. AFrame was faster than Spark JSON-based DataFrames on all cluster sizes but it was slower than Spark Parquet. AFrame and AFrame Schema display results of hash join on the XL dataset (10 GB JSON data) with increasing number of processing nodes while AFrame Index displays results of a broadcast-based index nested-loop join. 



%% file: multinode_scaleup.tex
\begin{figure*}[!ht]
     \centering
    \begin{subfigure}[t]{0.45\textwidth}
        \includegraphics[trim=1.5 1.5 1cm 1.5,width=\textwidth,height=0.7cm]{figures/scaleup_legend.pdf}
    \end{subfigure}
    \hspace{15cm}
    \begin{subfigure}[t]{0.24\textwidth}
        \includegraphics[trim=0 1.5 0 1.5,width=\textwidth,height=3.5cm]{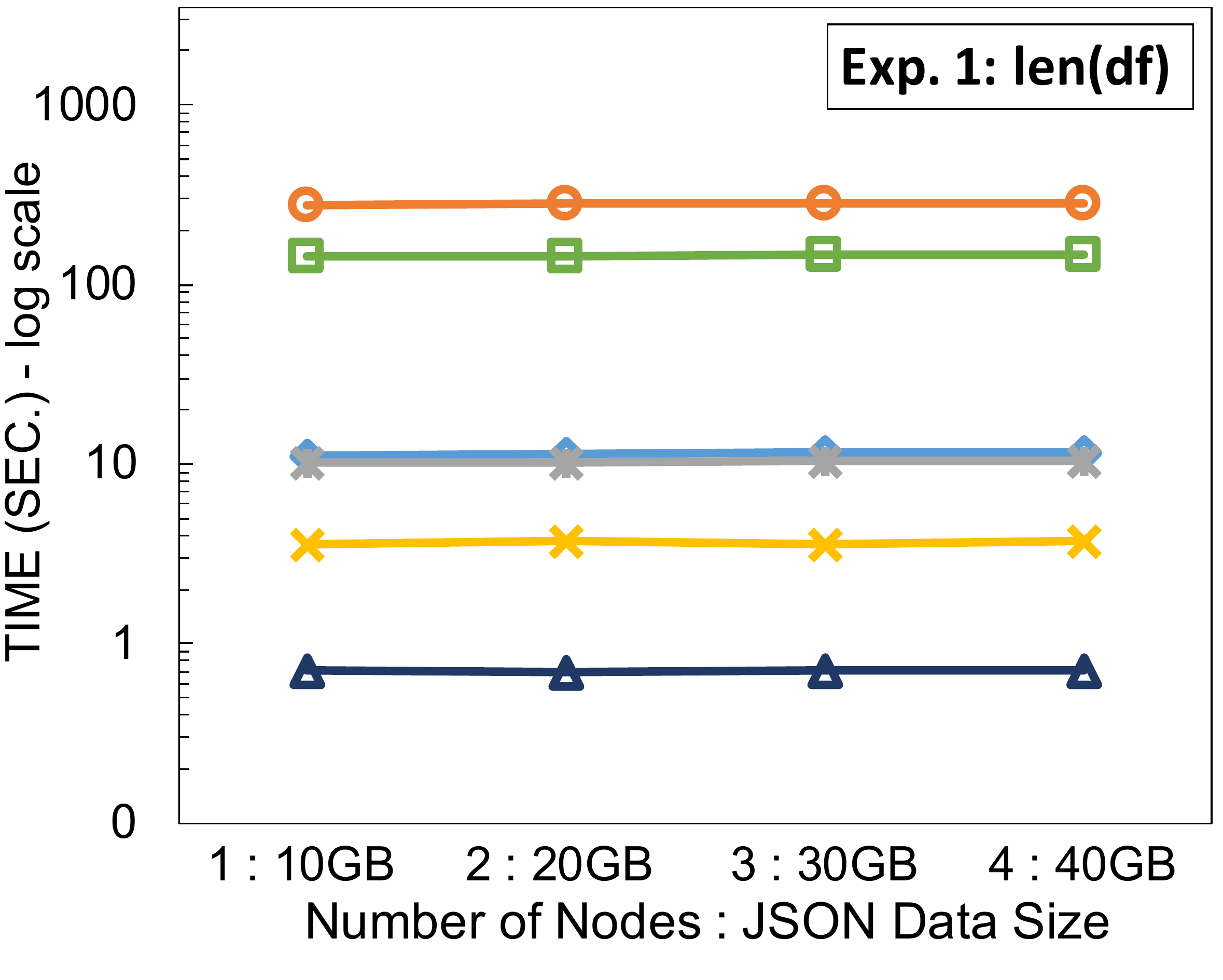}
        \caption{Expression 1: total times}
        \label{fig:q1_scaleup}
    \end{subfigure}
    \begin{subfigure}[t]{0.24\textwidth}
        \includegraphics[trim=0 1.5 0 1.5,width=\textwidth,height=3.5cm]{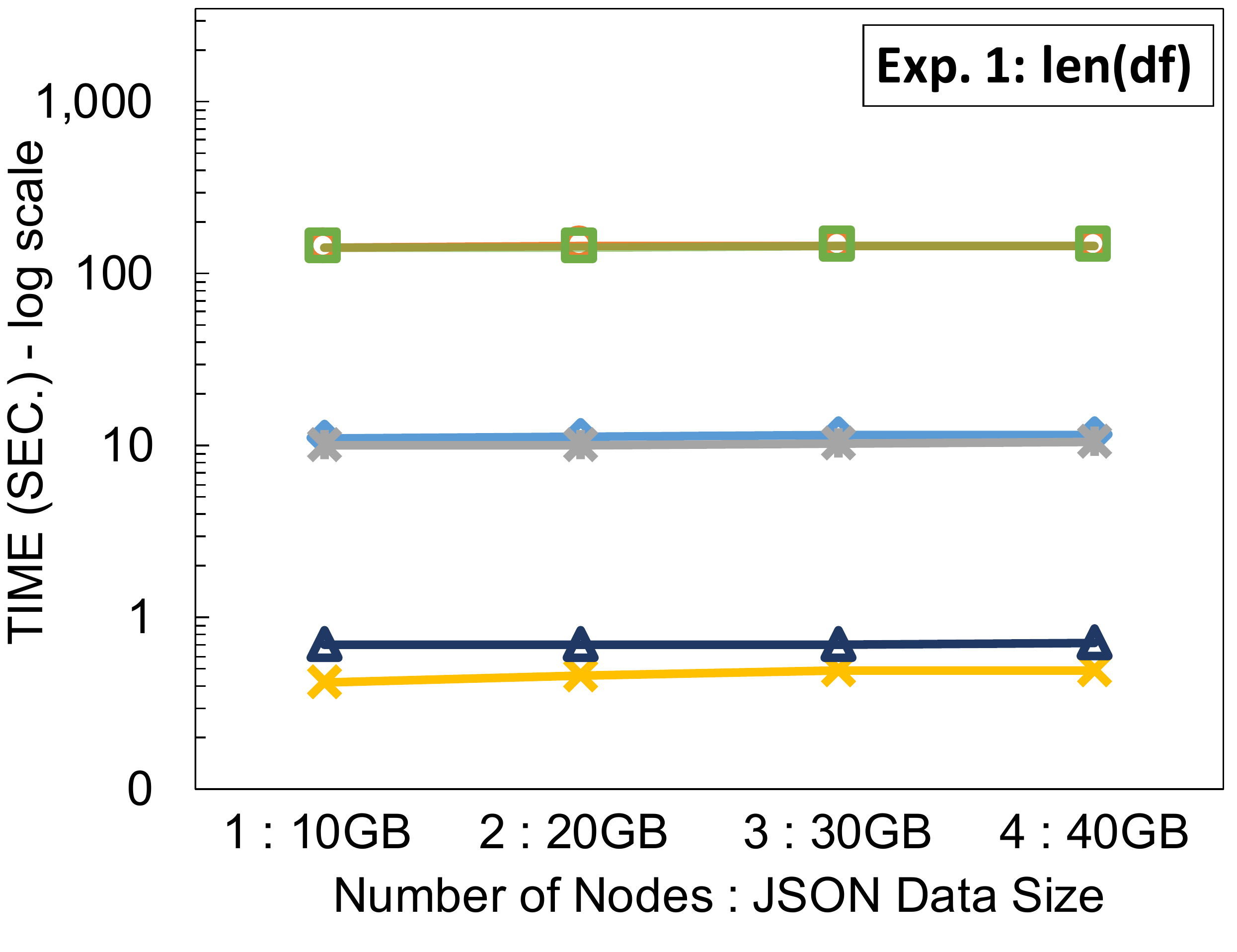}%
        \caption{Expression 1: expression-only times}
        \label{fig:q1_scaleup_wo}
    \end{subfigure}
    \hfill
    \begin{subfigure}[t]{0.24\textwidth}
        \includegraphics[trim=0 1.5 0 1.5,width=\textwidth,height=3.5cm]{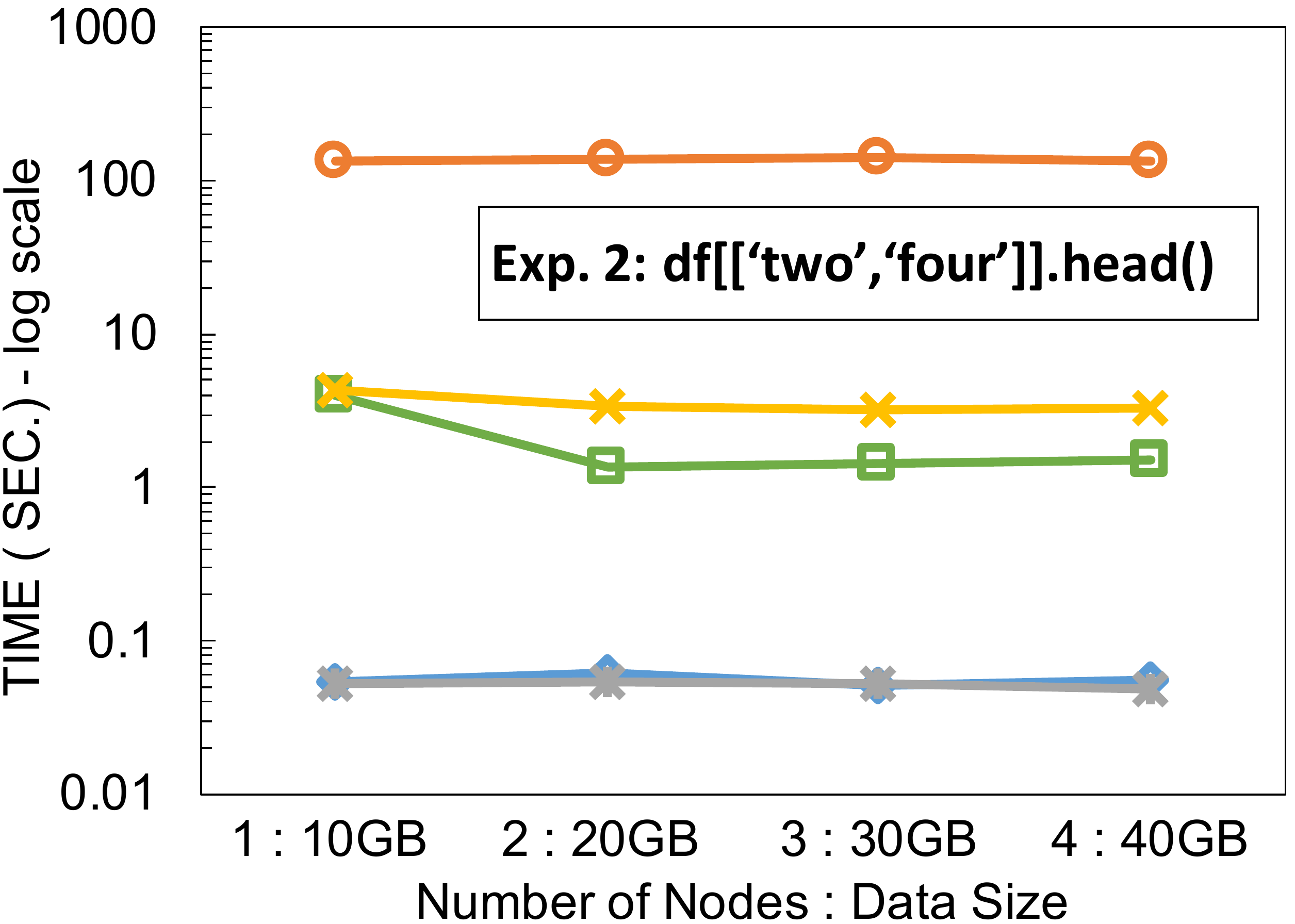}
        \caption{Expression 2: total times}
        \label{fig:q2_scaleup}
    \end{subfigure}
    \begin{subfigure}[t]{0.24\textwidth}
        \includegraphics[trim=0 1.5 0 1.5,width=\textwidth,height=3.5cm]{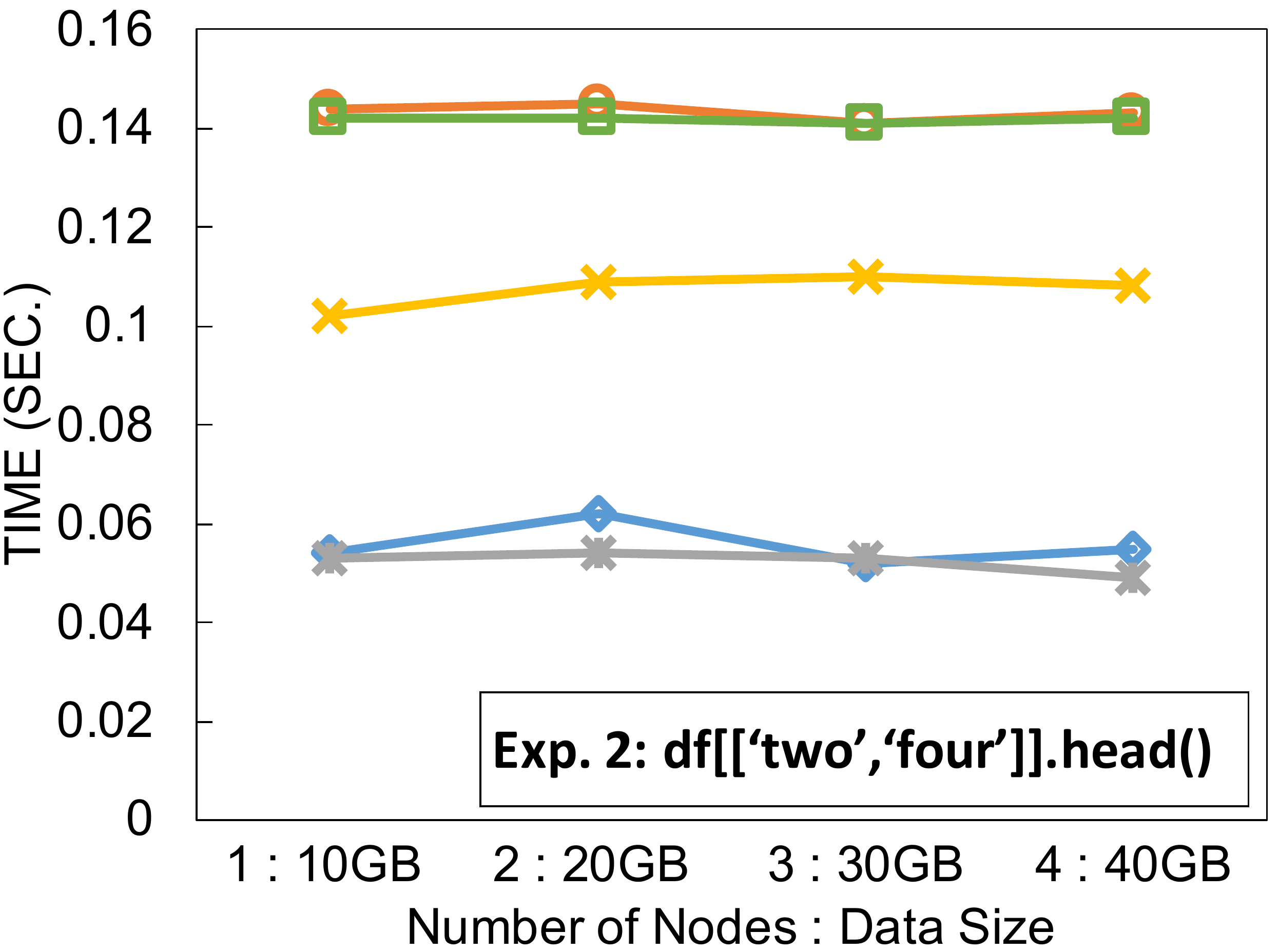}%
        \caption{Expression 2: expression-only times}
        \label{fig:q2_scaleup_wo}
    \end{subfigure}
    \begin{subfigure}[t]{0.24\textwidth}
        \includegraphics[trim=0 1.5 0 1.5,width=\textwidth,height=3.5cm]{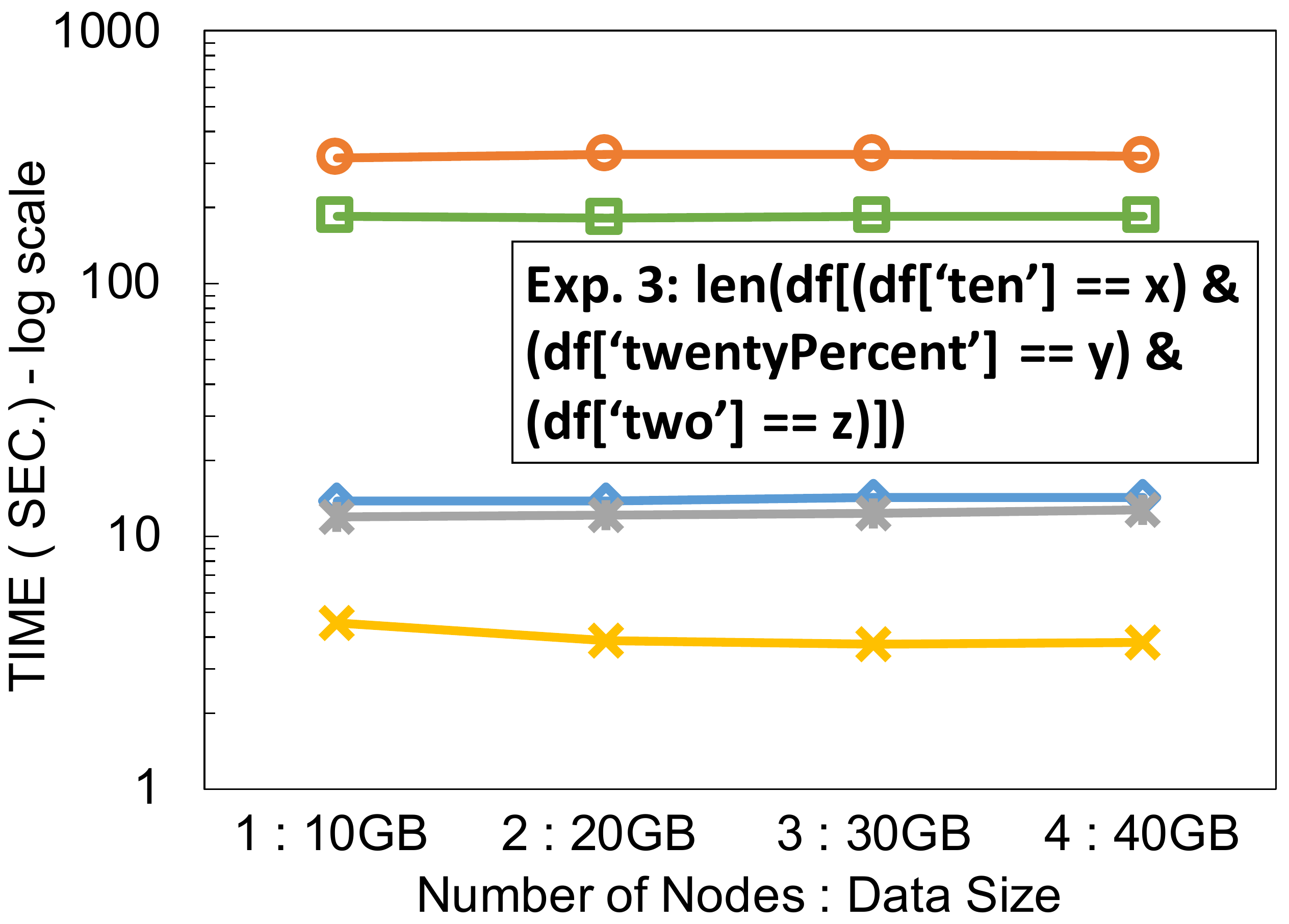}
        \caption{Expression 3: total times}
        \label{fig:q3_scaleup}
    \end{subfigure}
    \begin{subfigure}[t]{0.24\textwidth}
        \includegraphics[trim=0 1.5 0 1.5,width=\textwidth,height=3.5cm]{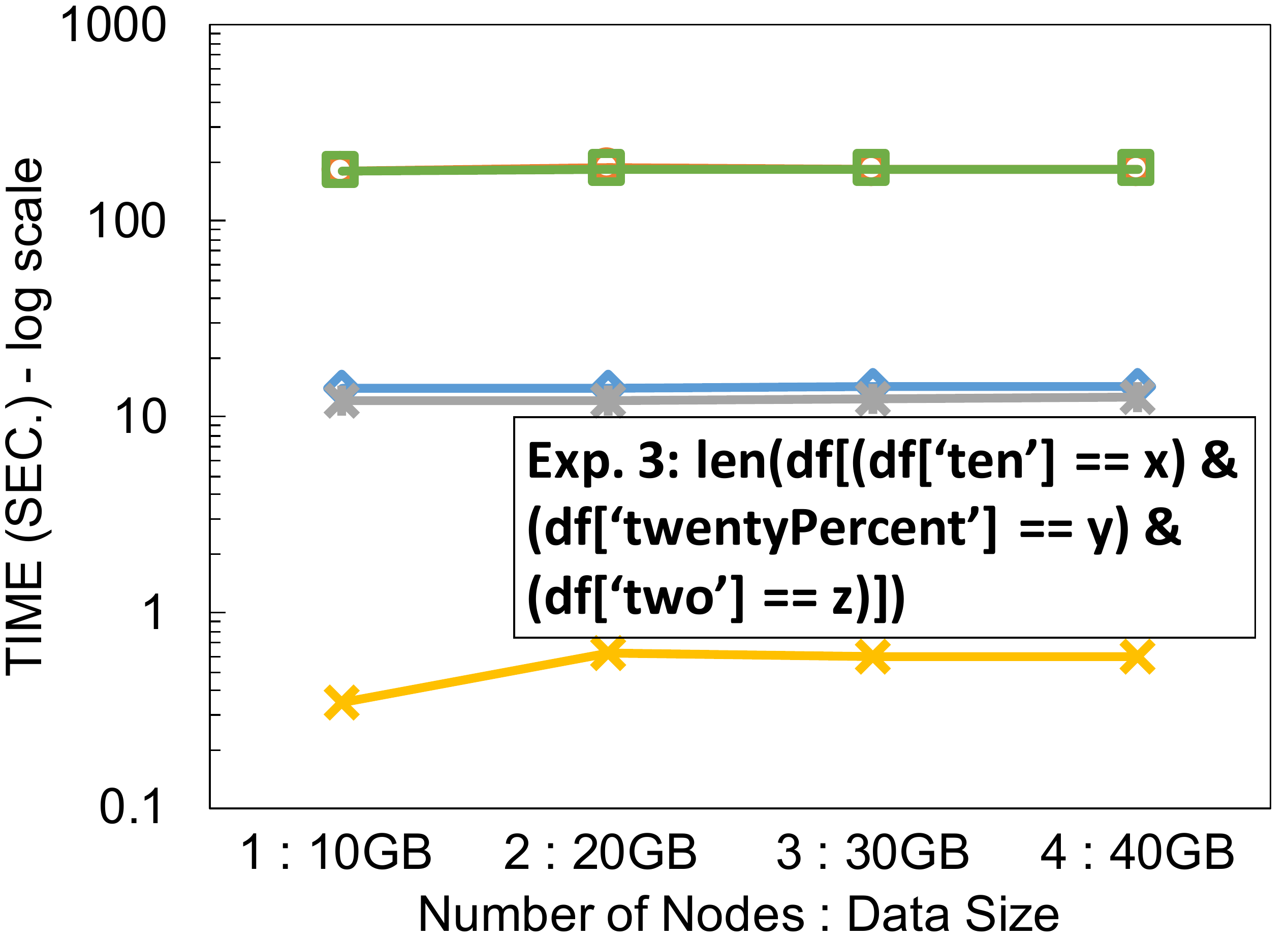}%
        \caption{Expression 3: expression-only times}
        \label{fig:q3_scaleup_wo}
    \end{subfigure}
    \hfill
    \begin{subfigure}[t]{0.24\textwidth}
        \includegraphics[trim=0 1.5 0 1.5,width=\textwidth,height=3.5cm]{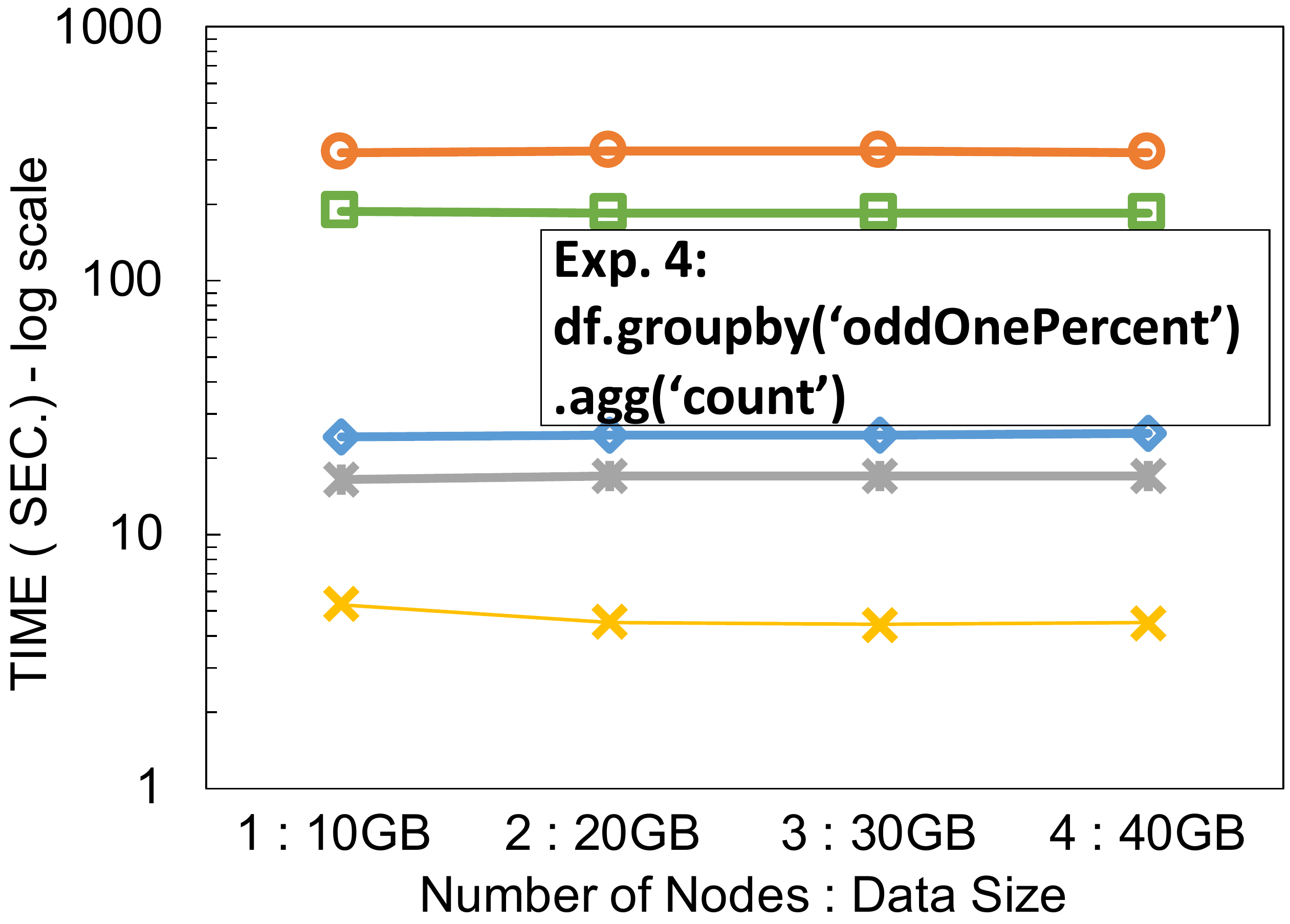}
        \caption{Expression 4: total times}
        \label{fig:q4_scaleup}
    \end{subfigure}
    \begin{subfigure}[t]{0.24\textwidth}
        \includegraphics[trim=0 1.5 0 1.5,width=\textwidth,height=3.5cm]{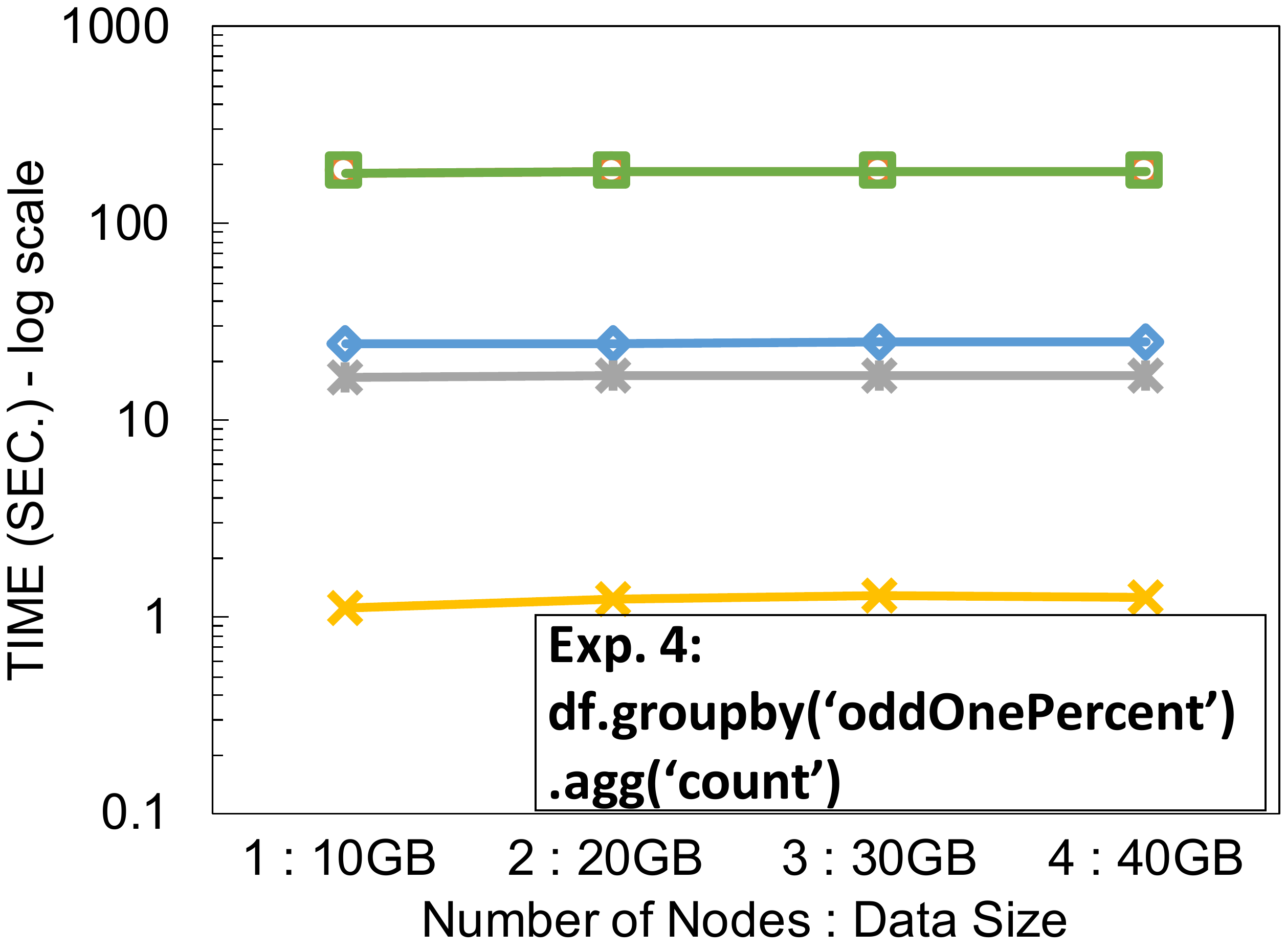}%
        \caption{Expression 4: expression-only times}
        \label{fig:q4_scaleup_wo}
    \end{subfigure}
    \begin{subfigure}[t]{0.24\textwidth}
        \includegraphics[trim=0 1.5 0 1.5,width=\textwidth,height=3.5cm]{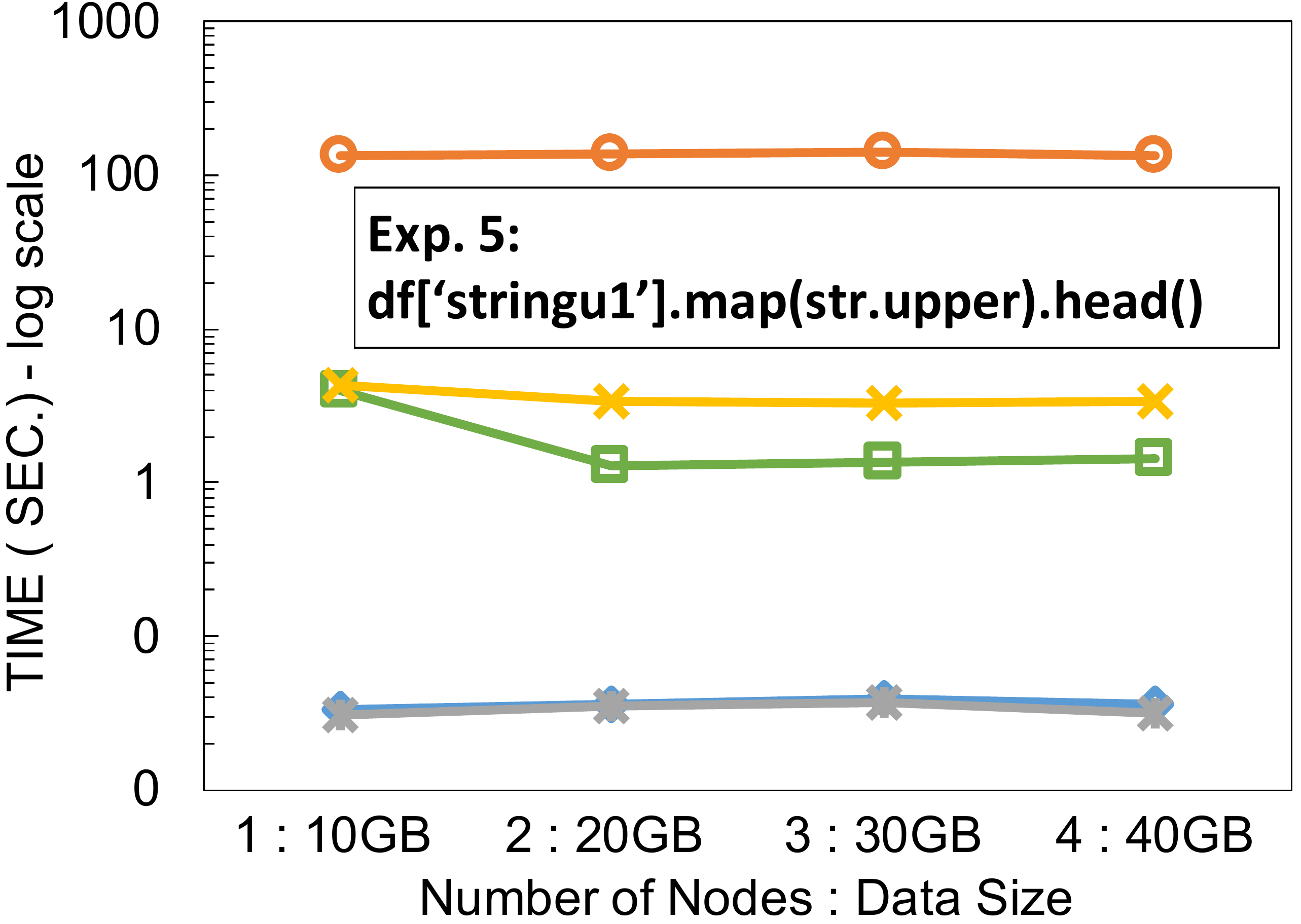}
        \caption{Expression 5: total times}
        \label{fig:q5_scaleup}
    \end{subfigure}
    \begin{subfigure}[t]{0.24\textwidth}
        \includegraphics[trim=0 1.5 0 1.5,width=\textwidth,height=3.5cm]{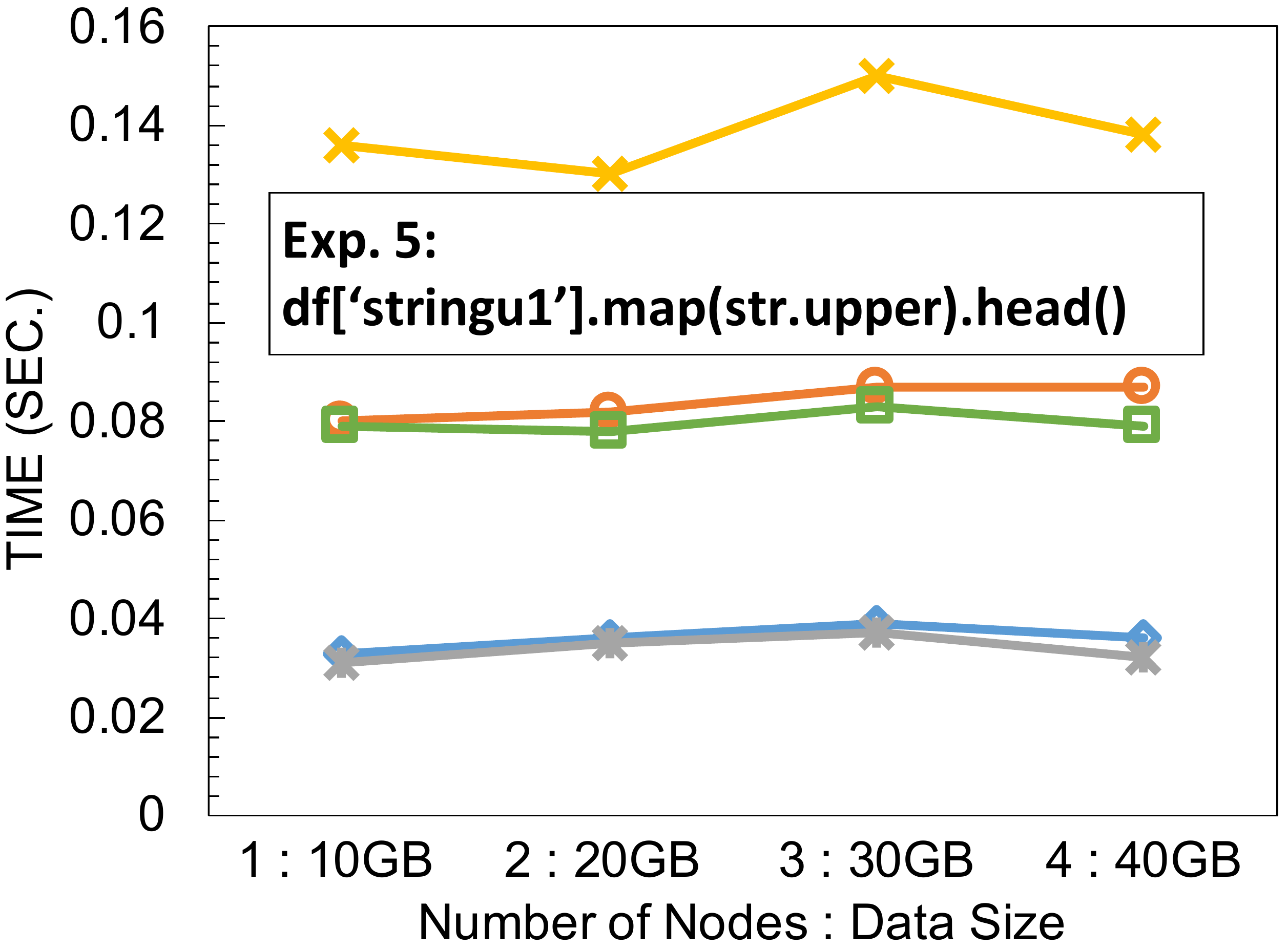}%
        \caption{Expression 5: expression-only times}
        \label{fig:q5_scaleup_wo}
    \end{subfigure}
    \hfill
    \begin{subfigure}[t]{0.24\textwidth}
        \includegraphics[trim=0 1.5 0 1.5,width=\textwidth,height=3.5cm]{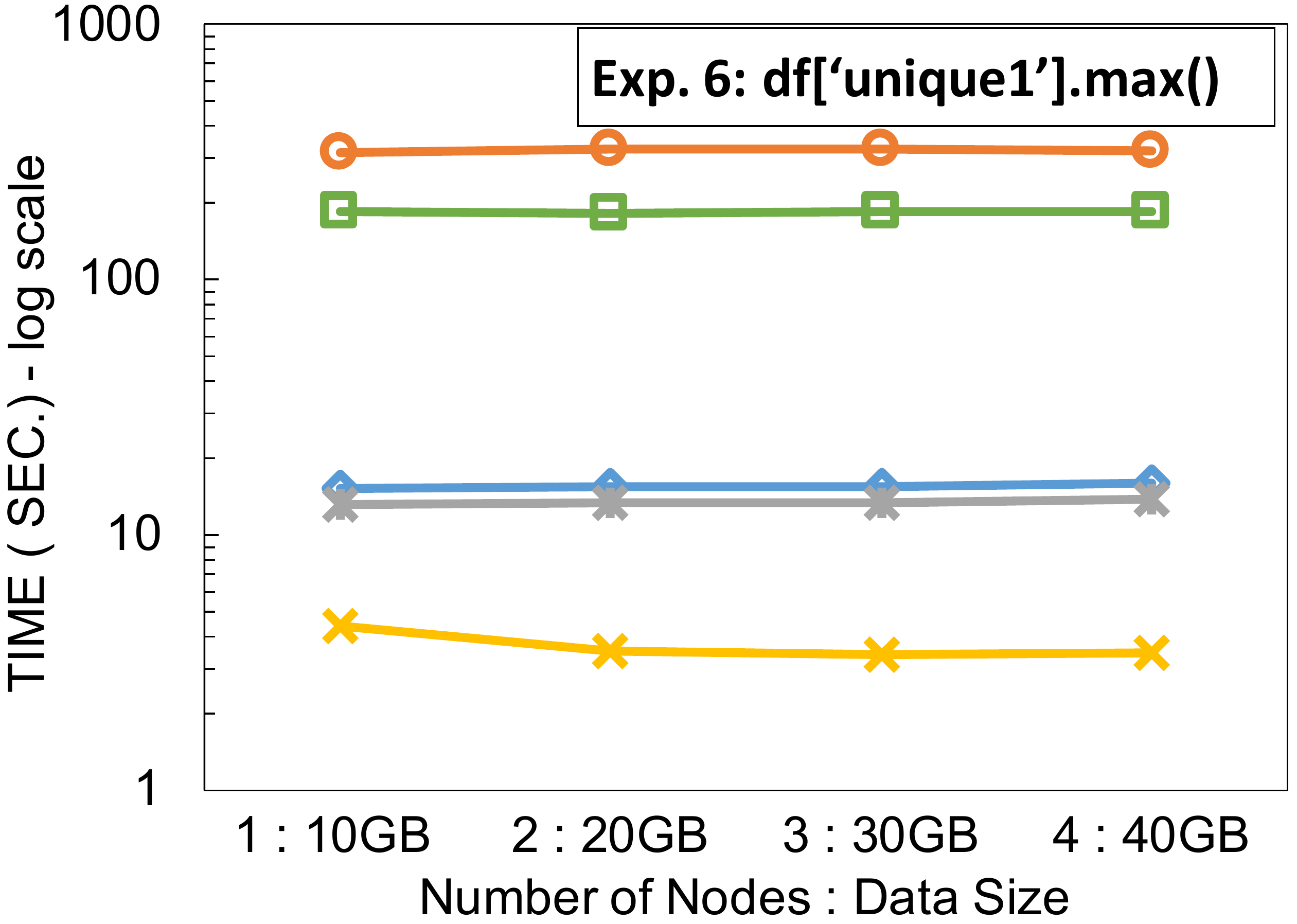}
        \caption{Expression 6: total times}
        \label{fig:q6_scaleup}
    \end{subfigure}
    \begin{subfigure}[t]{0.24\textwidth}
        \includegraphics[trim=0 1.5 0 1.5,width=\textwidth,height=3.5cm]{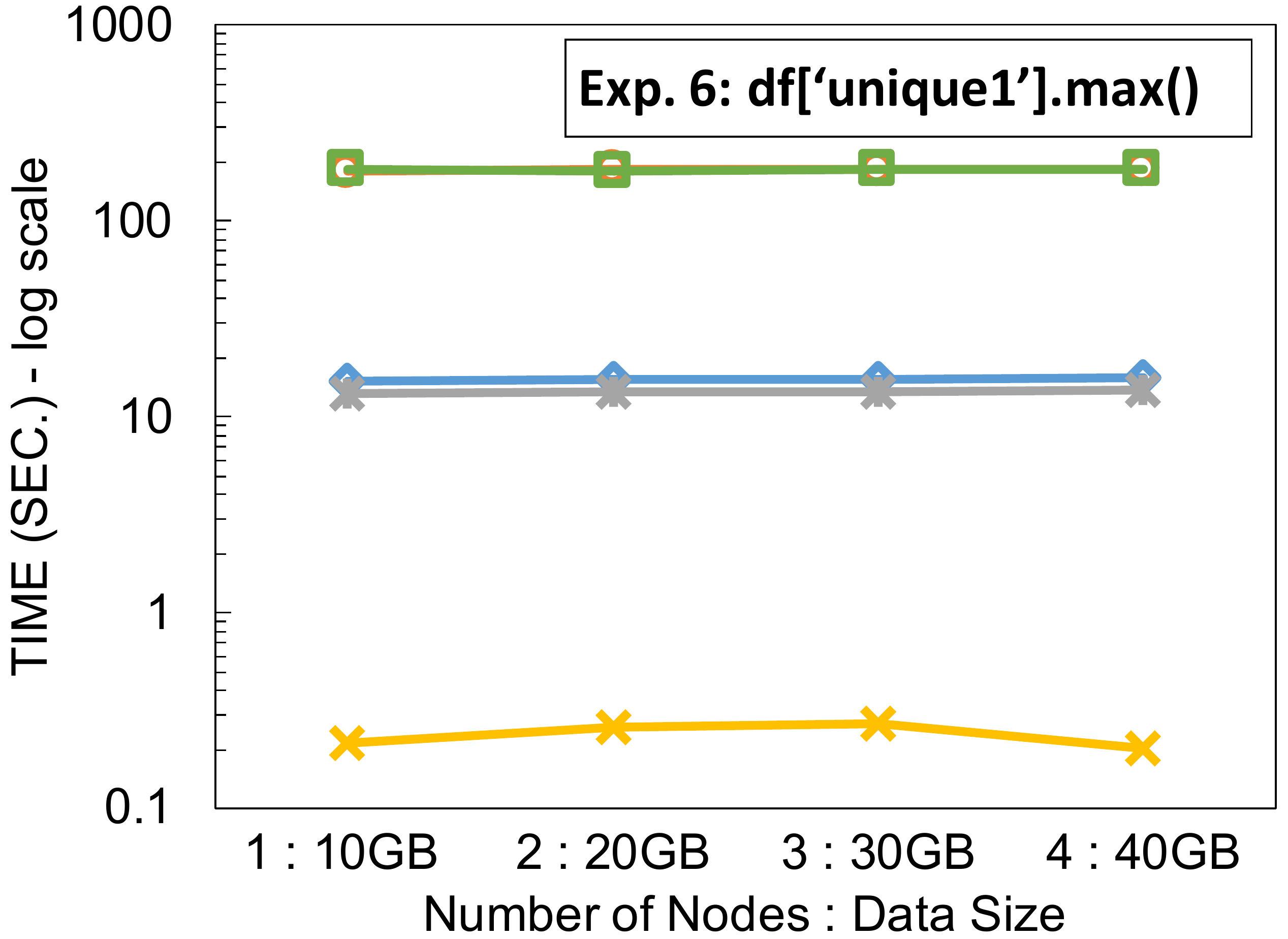}%
        \caption{Expression 6: expression-only times}
        \label{fig:q6_scaleup_wo}
    \end{subfigure}
    \begin{subfigure}[t]{0.24\textwidth}
        \includegraphics[trim=0 1.5 0 1.5,width=\textwidth,height=3.5cm]{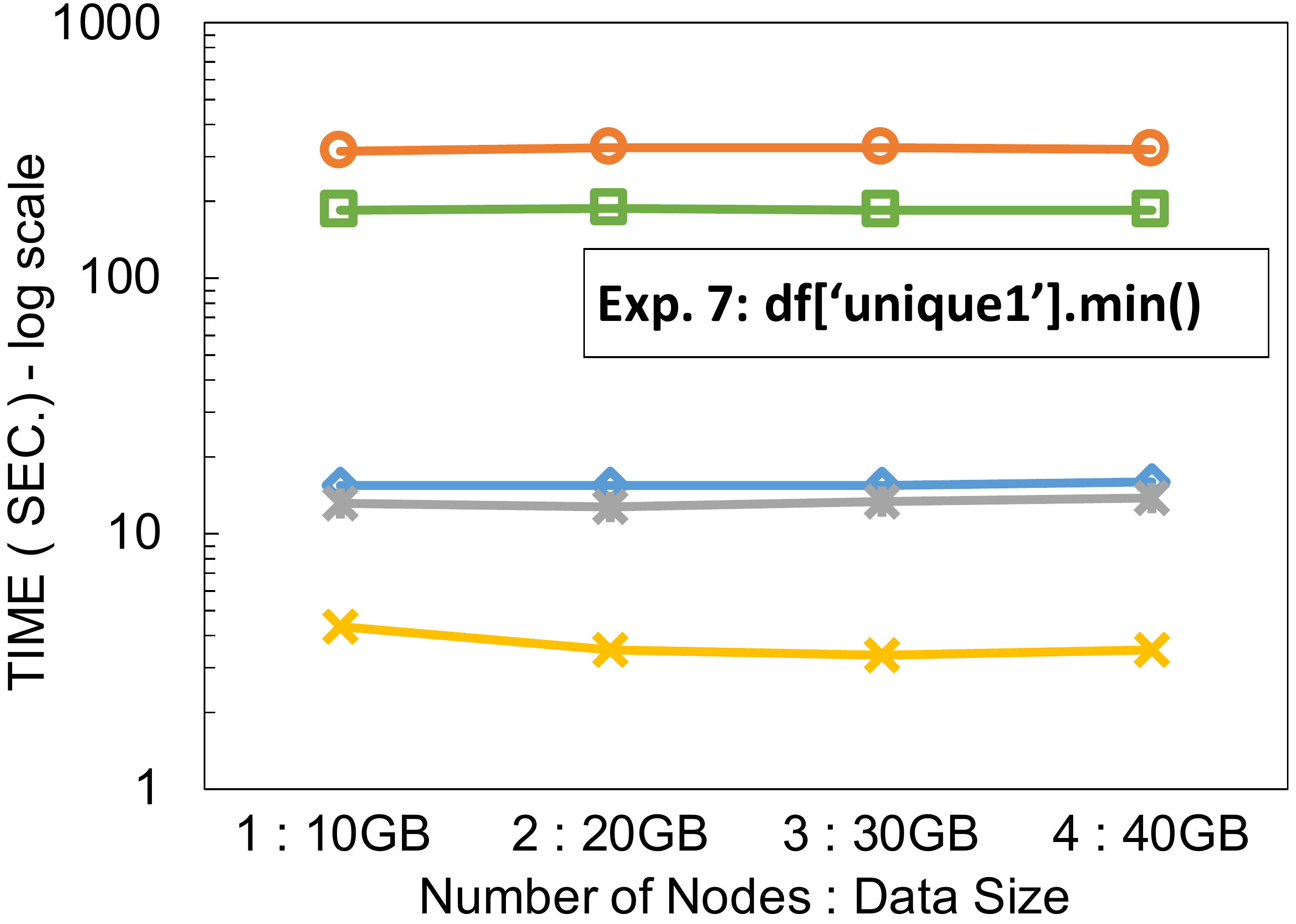}
        \caption{Expression 7: total times}
        \label{fig:q7_scaleup}
    \end{subfigure}
    \begin{subfigure}[t]{0.24\textwidth}
        \includegraphics[trim=0 1.5 0 1.5,width=\textwidth,height=3.5cm]{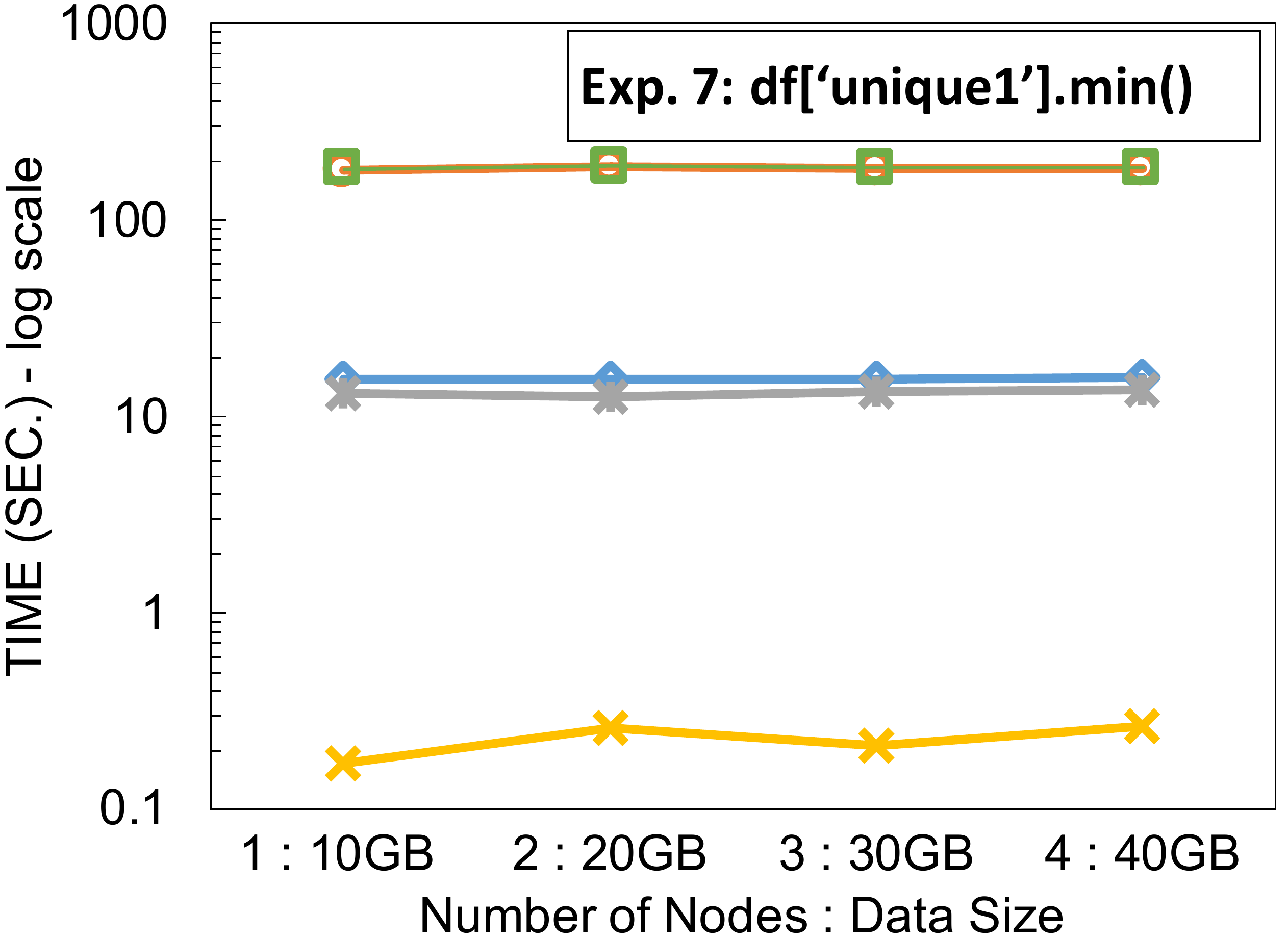}%
        \caption{Expression 7: expression-only times}
        \label{fig:q7_scaleup_wo}
    \end{subfigure}
    \hfill
    \begin{subfigure}[t]{0.24\textwidth}
        \includegraphics[trim=0 1.5 0 1.5,width=\textwidth,height=3.5cm]{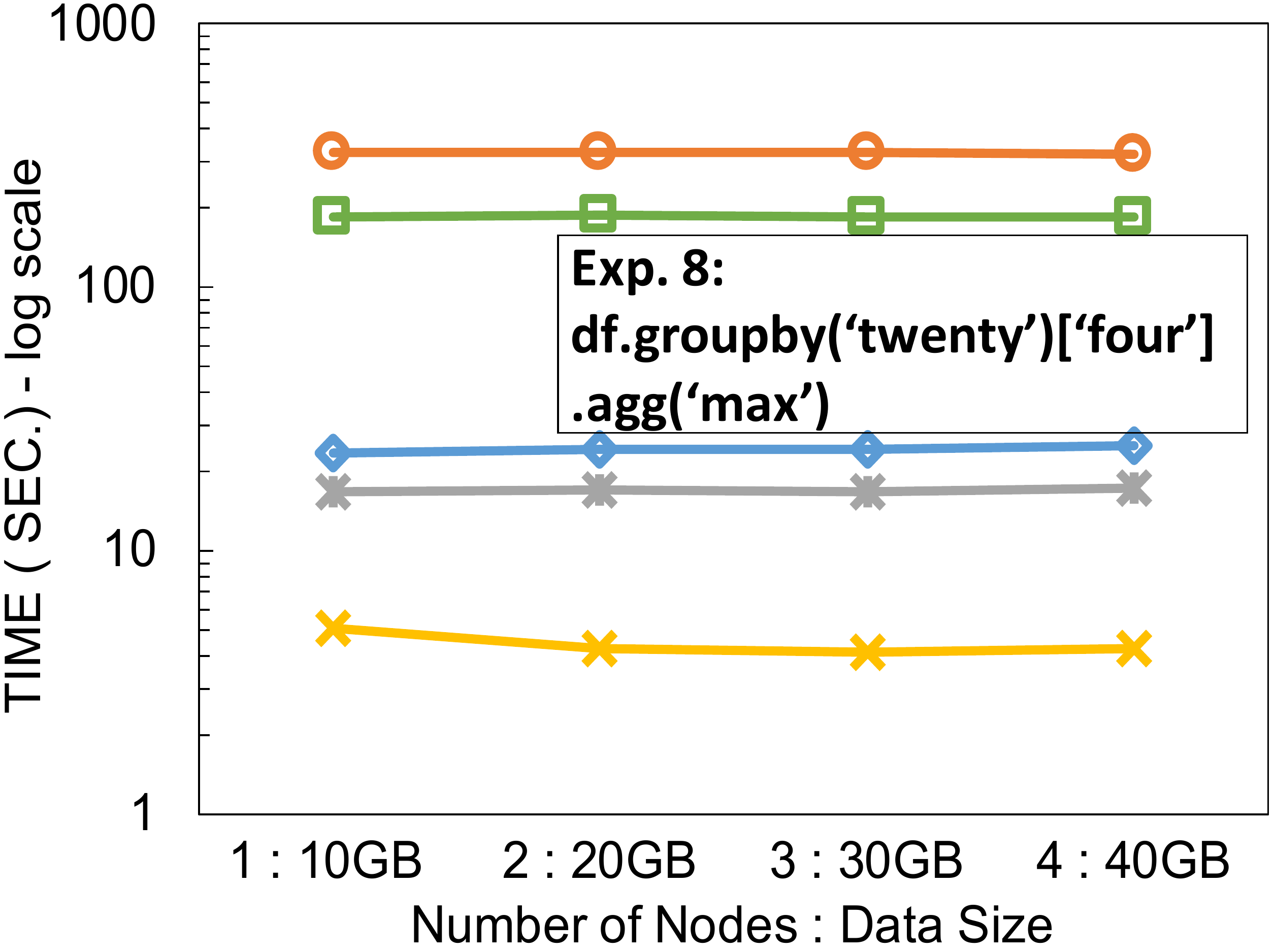}
        \caption{Expression 8: total times}
        \label{fig:q8_scaleup}
    \end{subfigure}
    \begin{subfigure}[t]{0.24\textwidth}
        \includegraphics[trim=0 1.5 0 1.5,width=\textwidth,height=3.5cm]{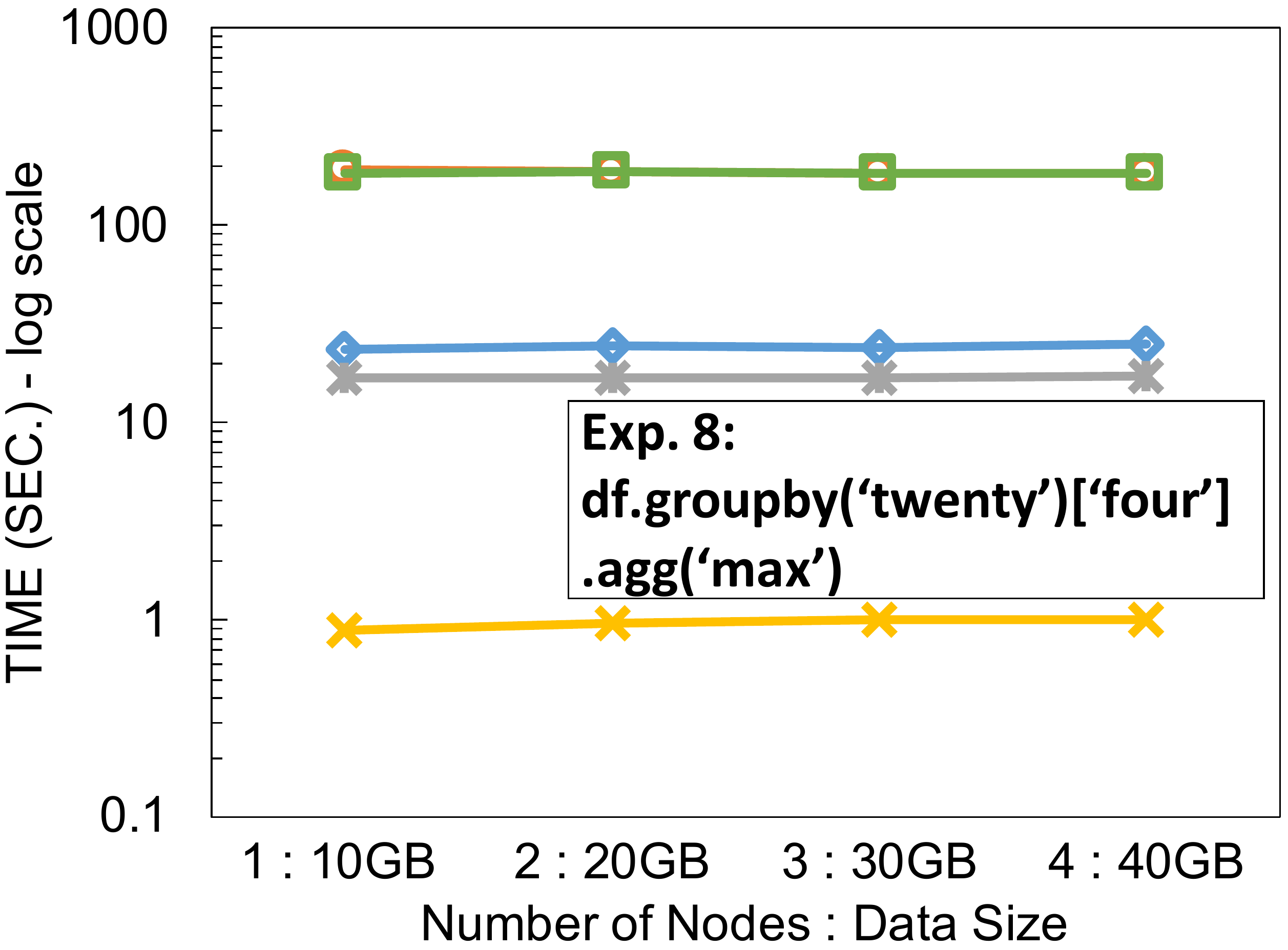}%
        \caption{Expression 8: expression-only times}
        \label{fig:q8_scaleup_wo}
    \end{subfigure}
    \begin{subfigure}[t]{0.24\textwidth}
        \includegraphics[trim=0 1.5 0 1.5,width=\textwidth,height=3.5cm]{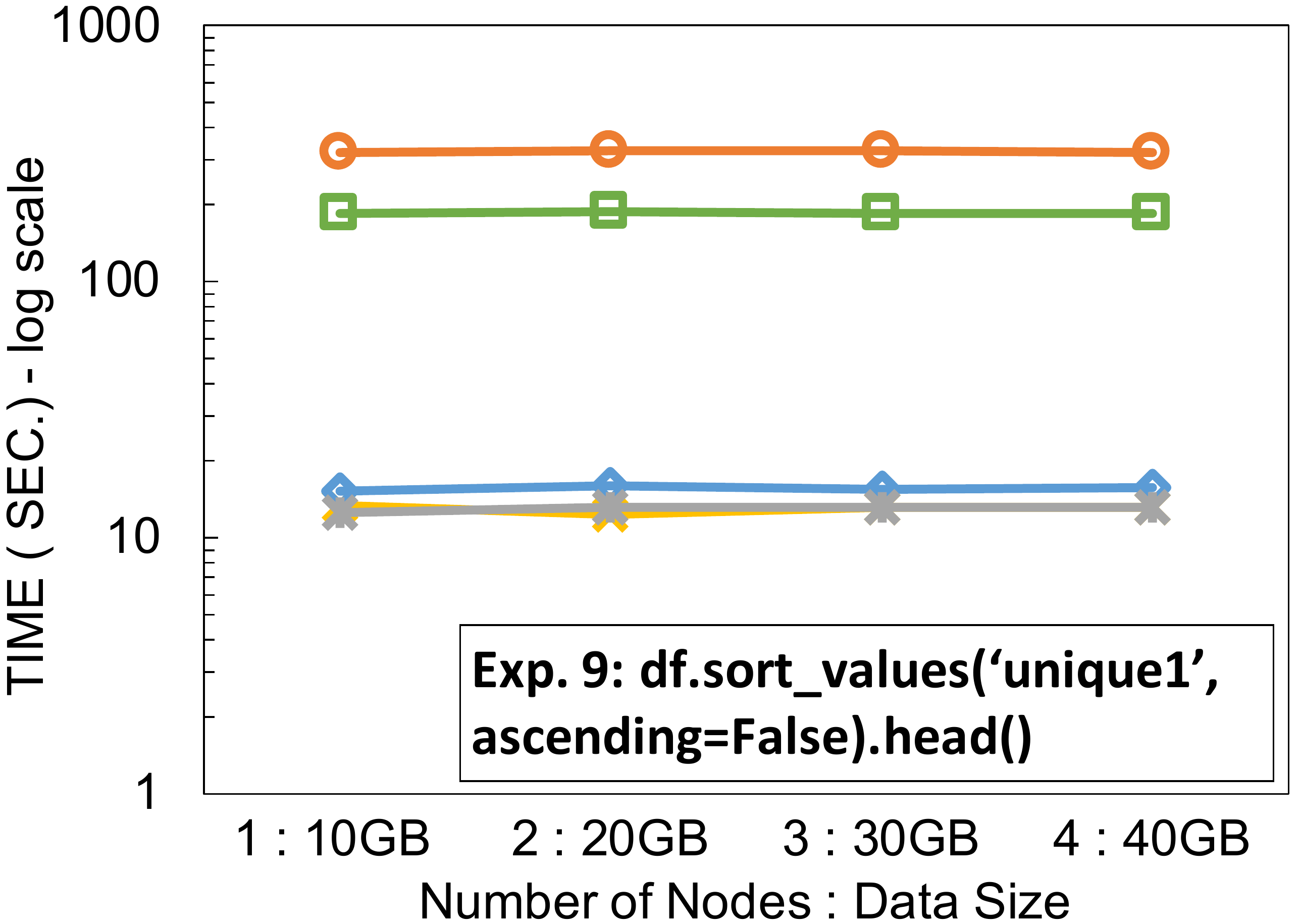}
        \caption{Expression 9: total times}
        \label{fig:q9_scaleup}
    \end{subfigure}
    \begin{subfigure}[t]{0.24\textwidth}
        \includegraphics[trim=0 1.5 0 1.5,width=\textwidth,height=3.5cm]{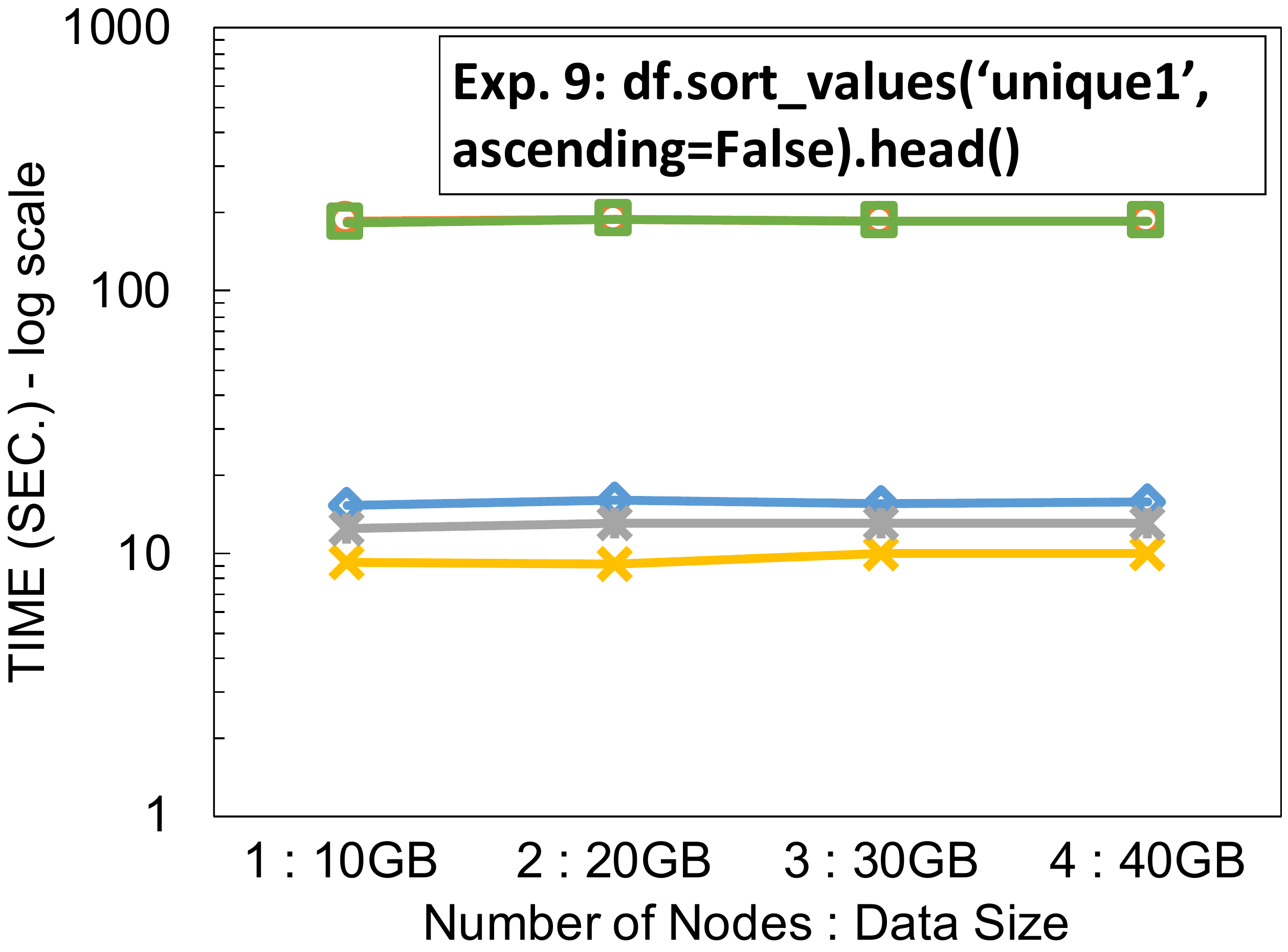}%
        \caption{Expression 9: expression-only times}
        \label{fig:q9_scaleup_wo}
    \end{subfigure}
    \hfill
    \begin{subfigure}[t]{0.24\textwidth}
        \includegraphics[trim=0 1.5 0 1.5,width=\textwidth,height=3.5cm]{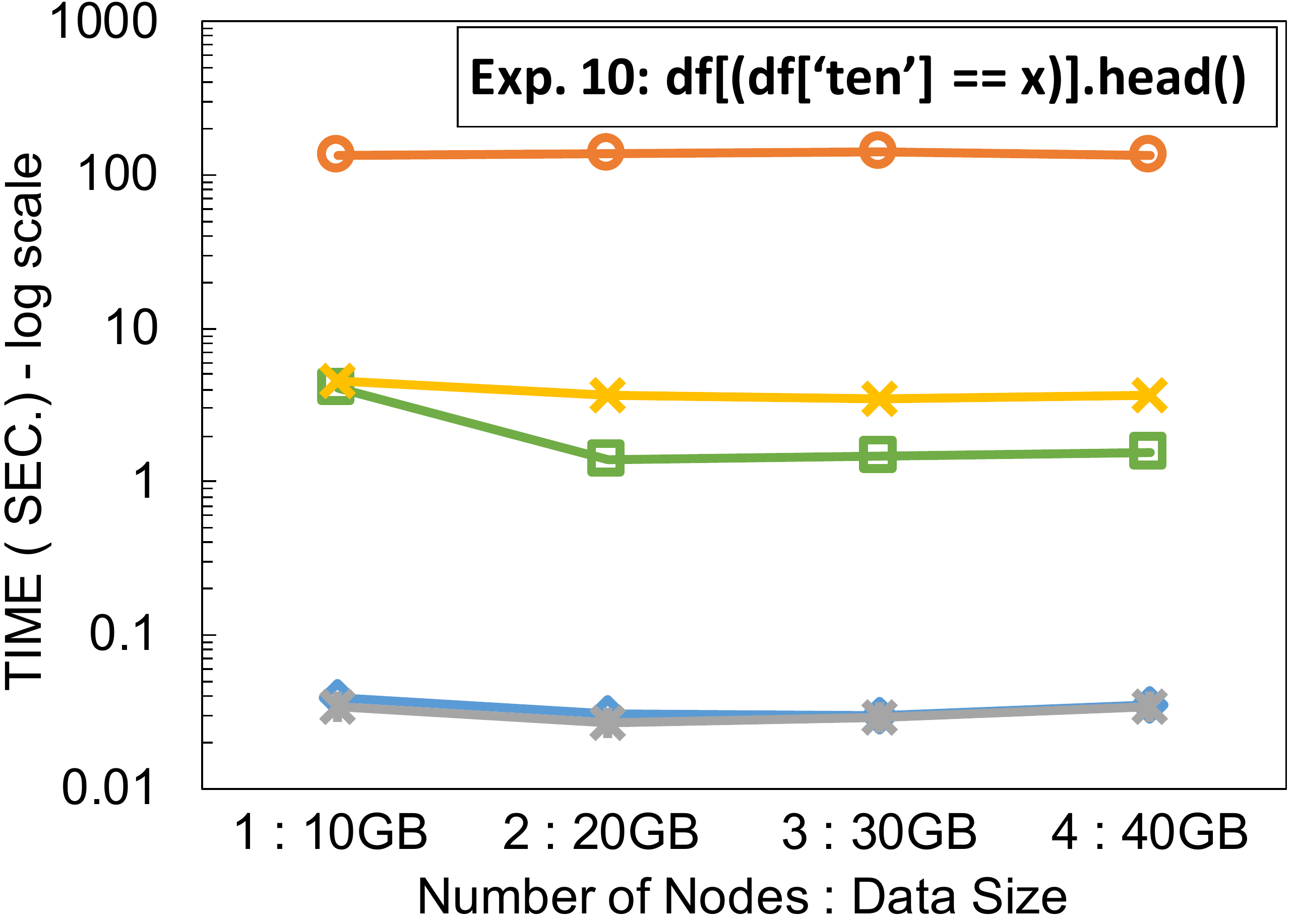}
        \caption{Expression 10: total times}
        \label{fig:q10_scaleup}
    \end{subfigure}
    \begin{subfigure}[t]{0.24\textwidth}
        \includegraphics[trim=0 1.5 0 1.5,width=\textwidth,height=3.5cm]{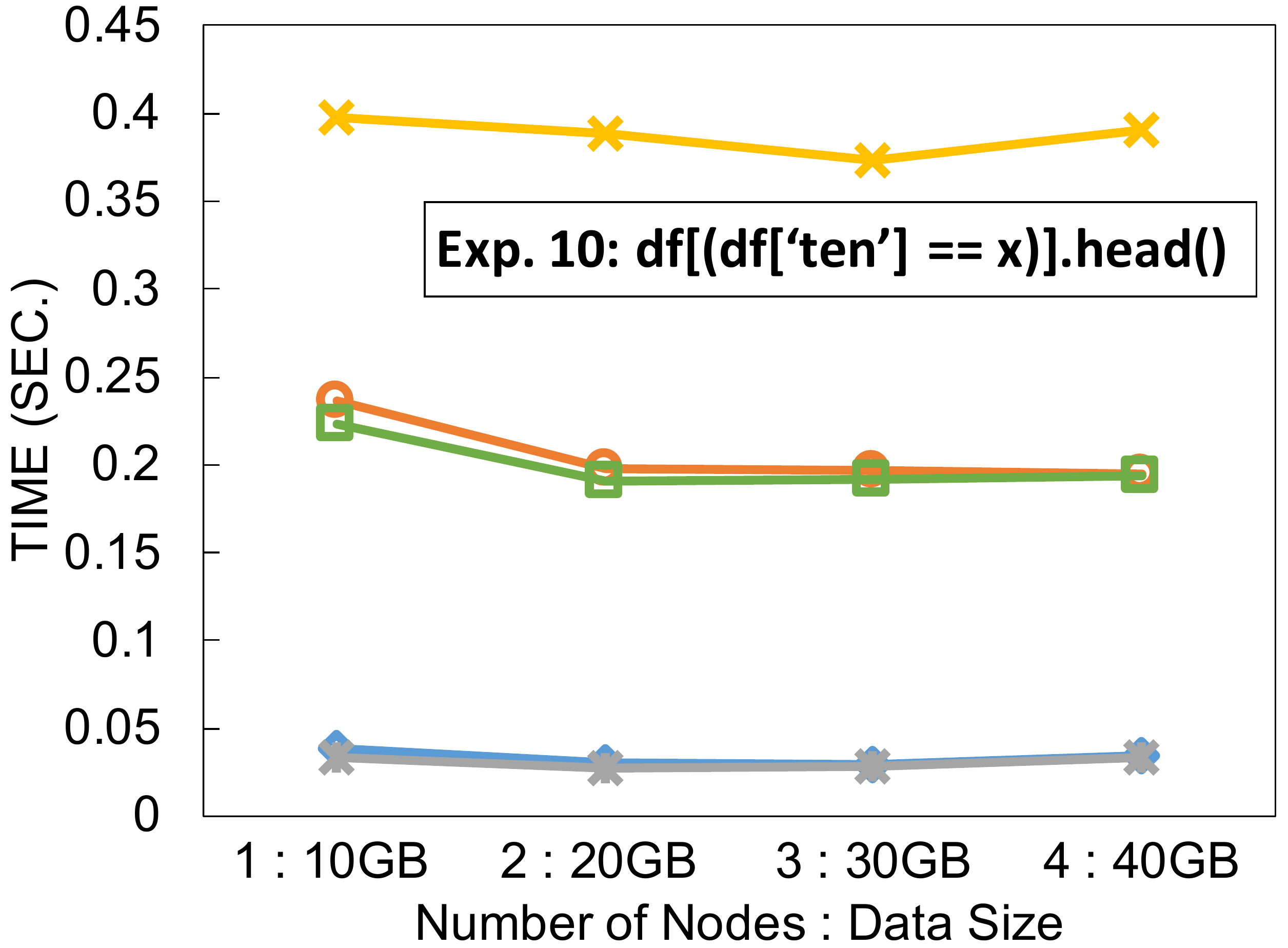}%
        \caption{Expression 10: expression-only times}
        \label{fig:q10_scaleup_wo}
    \end{subfigure}
    \caption{Multi-Node Scaleup Evaluation Results}

    \label{fig:scaleup_results}
    \vspace{-1em}
\end{figure*}

\begin{figure*}[!ht]
     \centering
    \begin{subfigure}[t]{0.45\textwidth}
        \includegraphics[trim=1.5 1.5 1cm 1.5,width=\textwidth,height=0.7cm]{figures/scaleup_legend.pdf}
    \end{subfigure}
    \hspace{15cm}
    \begin{subfigure}[t]{0.24\textwidth}
        \includegraphics[trim=0 1.5 0 1.5,width=\textwidth,height=3.5cm]{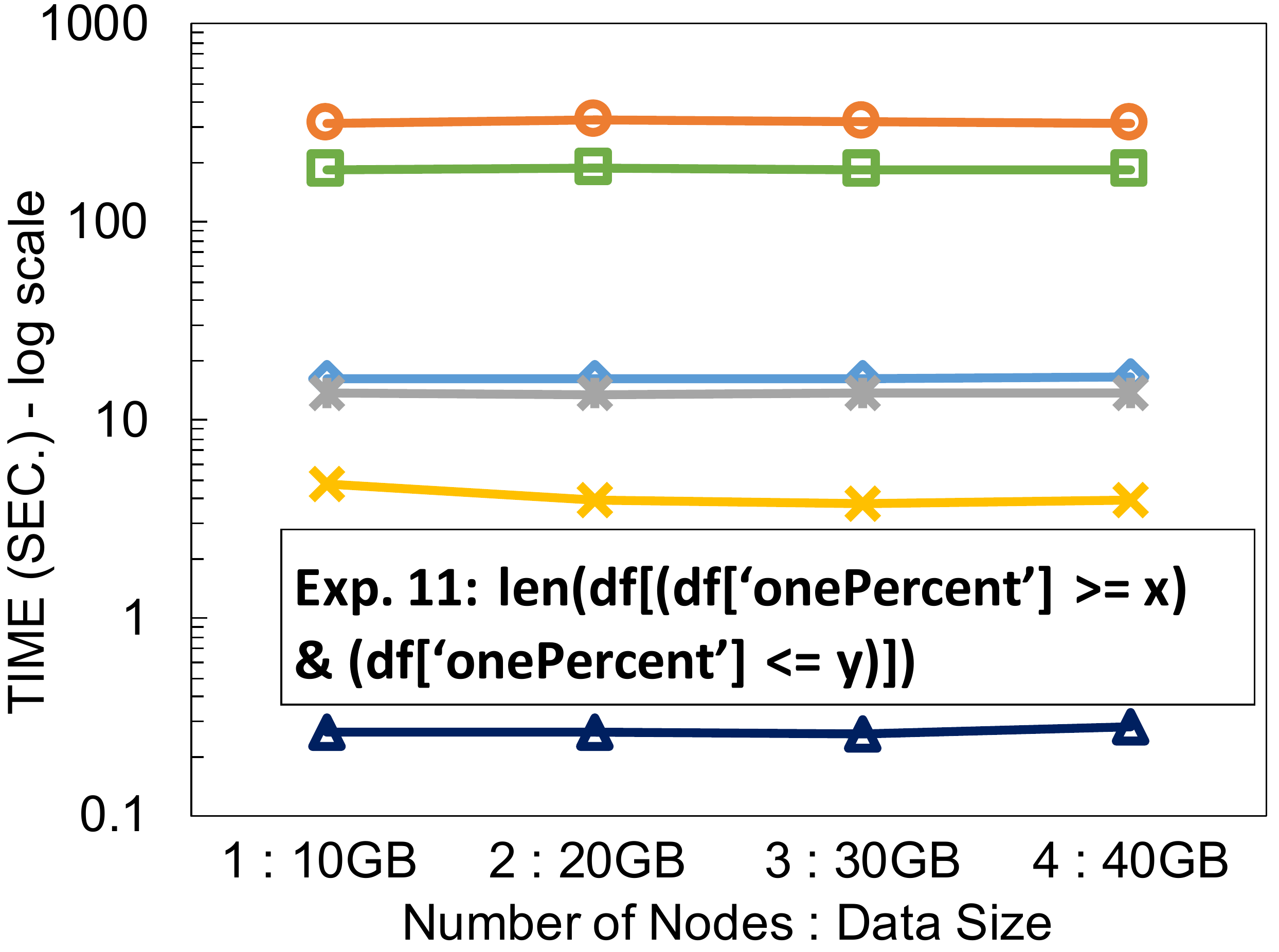}
        \caption{Expression 11: total times}
        \label{fig:q11_scaleup}
    \end{subfigure}
    \begin{subfigure}[t]{0.24\textwidth}
        \includegraphics[trim=0 1.5 0 1.5,width=\textwidth,height=3.5cm]{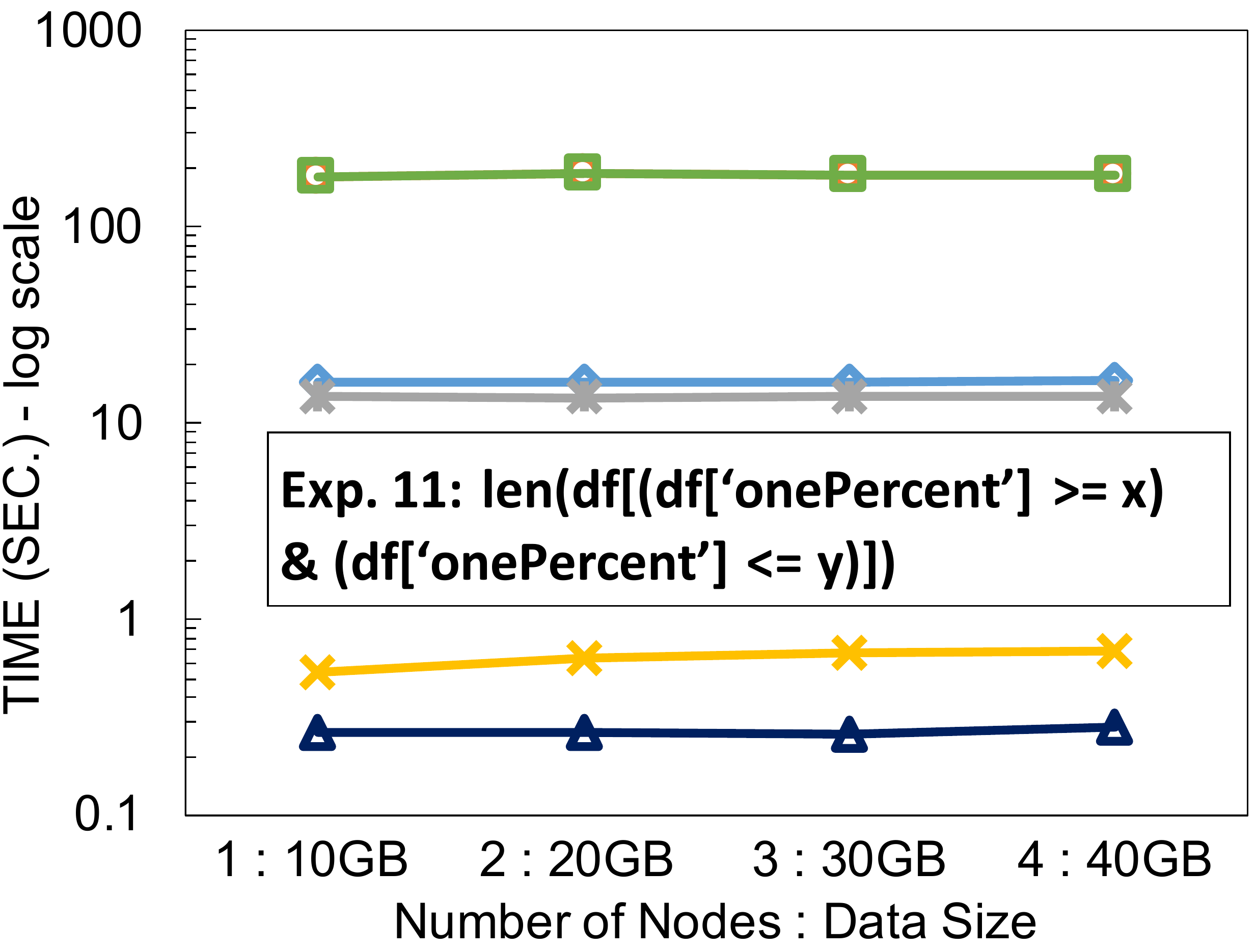}%
        \caption{Expression 11: expression-only times}
        \label{fig:q11_scaleup_wo}
    \end{subfigure}
    \hfill
    \begin{subfigure}[t]{0.24\textwidth}
        \includegraphics[trim=0 1.5 0 1.5,width=\textwidth,height=3.5cm]{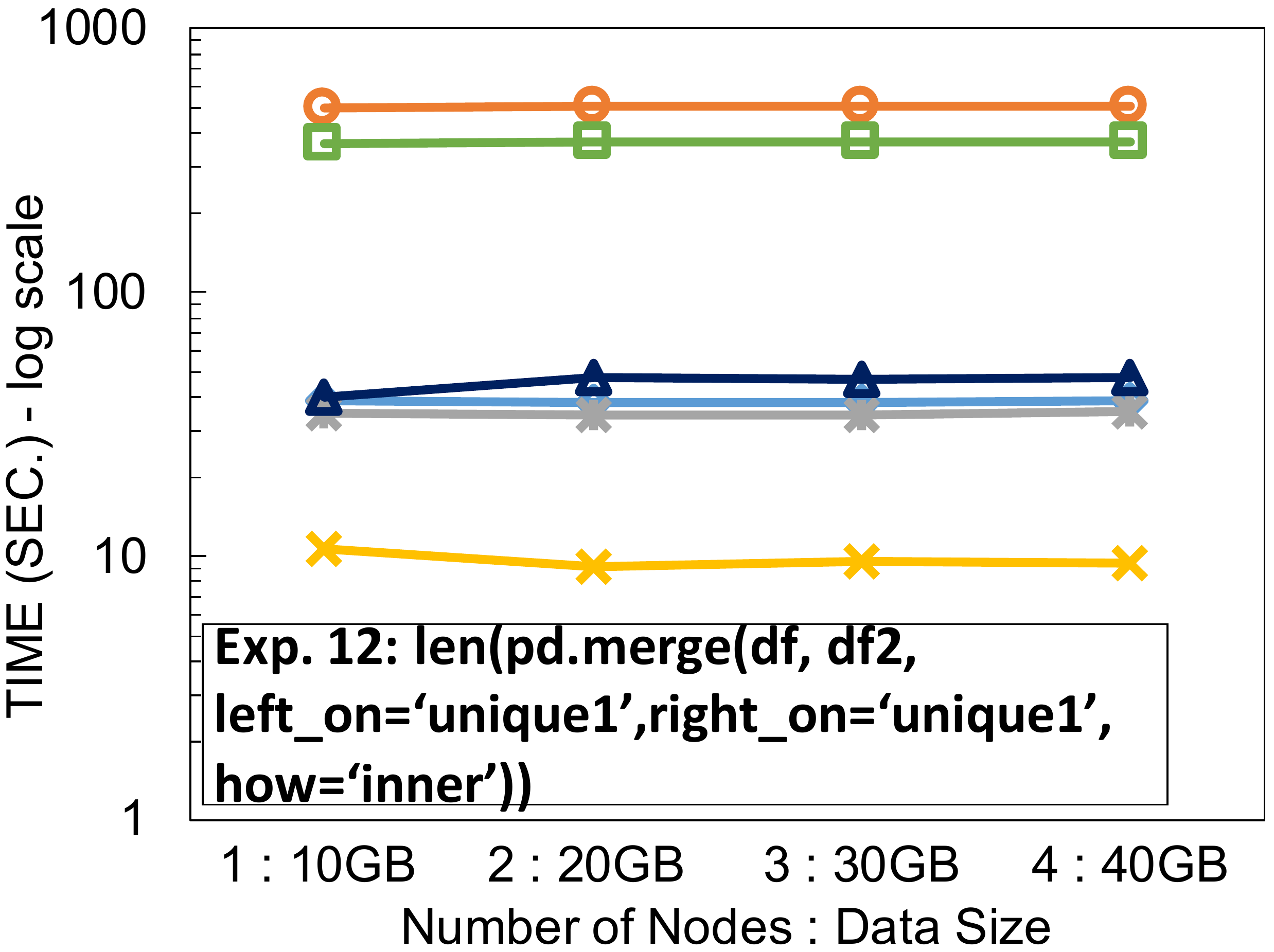}
        \caption{Expression 12: total times}
        \label{fig:q12_scaleup}
    \end{subfigure}
    \begin{subfigure}[t]{0.24\textwidth}
        \includegraphics[trim=0 1.5 0 1.5,width=\textwidth,height=3.5cm]{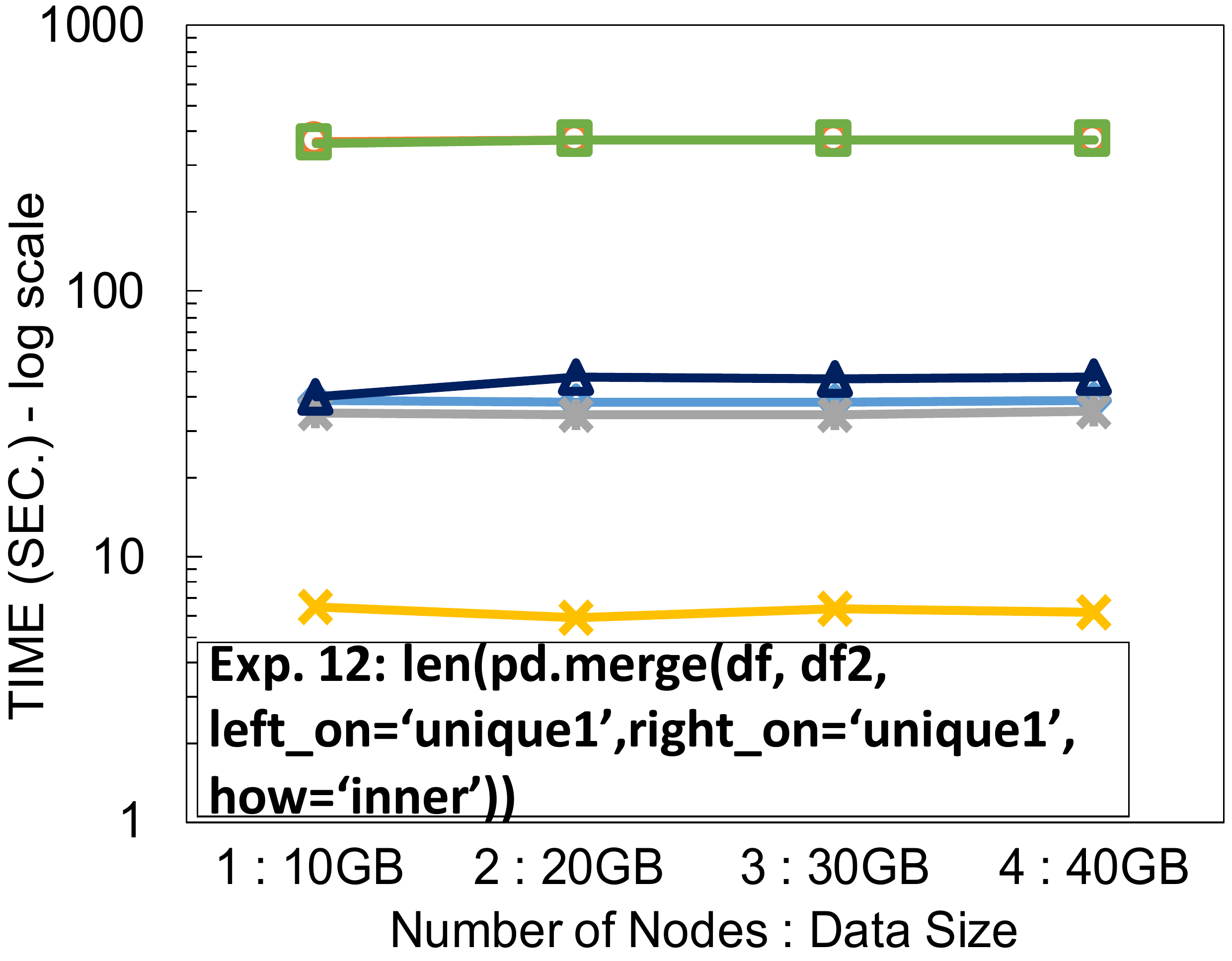}%
        \caption{Expression 12: expression-only times}
        \label{fig:q12_scaleup_wo}
    \end{subfigure}
    \caption{Multi-Node Scaleup Evaluation Results (continued)}

    \label{fig:scaleup_results_11-12}
    \vspace{-1em}
\end{figure*}

\subsubsection{Scaleup Results}
In addition to the plots in Figures~\ref{fig:scaleup_results} and \ref{fig:scaleup_results_11-12}, the raw averaged run times for multi-node scaleup results are included in Table~\ref{tab:scaleup} in the appendix section. In Figures \ref{fig:scaleup_results} and \ref{fig:scaleup_results_11-12}, scaleup results for Spark and AFrame are presented. No single system performed the best across all tasks. Spark's Parquet-based DataFrame was the fastest for column-based expressions (e.g., expressions 6 and 7) and was consistently competitive, but it also incurred an overhead for DataFrame creation. However, for row-based expressions (e.g., expressions 5 and 10), AFrame continued to follow the same trend from the single node case with the XL dataset, outperforming Spark Parquet. 

As we saw in Figure~\ref{fig:scaleup_results}, in the total time evaluations, by providing the JSON-based DataFrames with a schema, the total time is reduced by an order of magnitude, especially when only a subset of data is required. Expressions 2 (Figure \ref{fig:q2_scaleup}), 5 (Figure \ref{fig:q5_scaleup}), and 10 (Figure \ref{fig:q10_scaleup}) only sample a few records from a large dataset, which causes the schema inference time to otherwise dominate the actual expression execution time.

%% file: discussion.tex
\subsection{Discussion}

Pandas performed competitively on all tasks for a single node when the data fits in memory. However, its weaknesses lie in resource utilization and scalability. The memory requirement for Pandas is large and it can only take advantage of a single processing core. In addition, Pandas' eager evaluation strategy has disadvantages when expressions involve potentially repetitive tasks. Exploratory operations that only view a small subset of the data took longer on Pandas than on frameworks that utilize parallel processing and/or lazy evaluation.

Pandas on Ray did an excellent job in functionally covering Pandas operations. It reroutes operations to the default Pandas when its parallel work distribution has not been enabled for an operation. While treating Pandas DataFrame as a black box does not solve the problem of its memory requirement, it utilizes parallel processing for loading and processing data in order to speed up the computation. Evaluating the system as-is reveals that there can be significant overhead associated with work distribution for Pandas on Ray. This is a known issue which is mentioned in the project's own benchmarking results~\cite{pandasonray}, where the authors provide an explanation of the issues and give insights as to when the benefit of its work distribution will be significant. Its experimental out-of-core support will be worth looking into once it is enabled and distributed installation instructions are provided.
 
Spark DataFrame provides similar syntax to that of Pandas' with the ability to operate on data that exceeds the per-node memory limit; it provides a friendly interface to the Apache Spark distributed compute engine. While Spark can operate on large datasets, its performance drastically degrades when having to work with insufficient cluster memory as its strength lies in in-memory computation. As a result, on large datasets, its JSON-based DataFrame was an order of magnitude slower than AFrame. On the other hand, its Parquet-based DataFrame performed quite competitively across all data sizes. Due to its compression, a Parquet file is much smaller than a JSON file with the same logical data content. Finally, Parquet is a columnar file format, which makes the Parquet-based DataFrame an excellent fit for column-based operations but slower on tasks that require access to the entire payload of each data record.

A unique characteristic that sets AFrame apart from other large-scale DataFrame libraries is its ability to operate on managed and indexed data. AFrame benefits from its AsterixDB backend in several ways. First, it can eliminate repetitive file scans during the DataFrame creation process since datasets have been ingested and stored on disk in AsterixDB. Second, it is able to operate on data larger than the available memory, seamlessly, without requiring additional effort. It thus eliminates the disconnect between data-intensive analytical tools (e.g., Pandas) and database management systems. Third, it eliminates issues that could arise from manually managing large amounts of data from various sources. Flat file storage requires effort to maintain and can be difficult to share between multiple users; modifying data in traditional storage can be prone to corruption because of a lack of transactional support. In addition, by having a distributed data management system as its backend, complex DataFrame operations that would otherwise execute inefficiently can be optimized by a database query optimizer. AsterixDB provides query plan optimization and indexing that enable AFrame to perform competitively, especially in terms of the total time evaluations (which arguably reflect the time from question to insight). 

%% file: conclusion.tex
\section{Conclusions \& Future Work}
In this work, we have shown the practicality of utilizing a distributed data management system to scale data scientists' familiar DataFrame operations to work against modern data at scale without requiring distributed data engineering expertise. We can also increase data analysts' productivity by optimizing their operations' execution times through lazy evaluation and database query optimization. AsterixDB also provides additional benefits to AFrame, such as the ability to utilize pre-trained models from packages such as Scikit-Learn (as-is) without requiring specialized large-scale machine learning skills. Finally, with AsterixDB's built-in social media data feeds, data scientists can operate on live datasets in the same way that they would work with static data.

In order to evaluate our initial AFrame prototype, we have also proposed a DataFrame benchmark for evaluating DataFrame performance on analytic operations. Our benchmark can be used in both single-node and distributed settings. Our experiments showed that AFrame can operate competitively in both settings. We have also demonstrated that optimizations can be crucial when dealing with data at scale. Our DataFrame benchmark, even at this early stage, can help data scientists better understand the performance of their workloads and understand distributed frameworks' tradeoffs.

Moving forward, we have a list of new functionality and improvements that we would like to implement for both AFrame and the DataFrame benchmark. For AFrame, while we have demonstrated its basic functionality and evaluated its performance against several similar frameworks, it is still in its early development stage. There are many more Pandas features to be implemented in order to better meet data scientists' requirements. We are also adding nested data handling and window functions to AFrame. We additionally plan to make AFrame less query language specific by abstracting its language layer. Currently, AFrame translates DataFrame operations into SQL++. By separating its language module from the DataFrame operation translation mechanism, we should also be able to deploy AFrame on other query-based data management systems (e.g., Postgres).

The DataFrame benchmark is preliminary work that has served a purpose by allowing us to evaluate the feasibility of AFrame for analytic operations and to compare its initial performance against other frameworks. However, the benchmark is a work in progress and needs more analytic operations to be included in order to evaluate other aspects of distributed DataFrames. We also intend to eventually add more frameworks (e.g., Dask~\cite{dask}) to our evaluation and to deploy them in a much larger distributed environment to gain more insight into the feasibility of supporting the DataFrame experience on large-scale distributed data.